\documentclass[12pt]{cernart}
\usepackage{epsfig}
\usepackage{times}
\usepackage{multirow}
\begin{document}



\setcounter{topnumber}{3}
\renewcommand{\topfraction}{0.999}
\renewcommand{\bottomfraction}{0.99}
\renewcommand{\textfraction}{0.0}
\setcounter{totalnumber}{6}

\def\thefootnote{\fnsymbol{footnote}}
\begin{titlepage}
%
\title{\large{
Inclusive Production of Charged Pions in p+p Collisions at 158 GeV/c
Beam Momentum
}}
%
\begin{Authlist}
%
%
%
%
%
%
\noindent
C.~Alt$^{8}$, T.~Anticic$^{17}$, B.~Baatar$^{7}$, D.~Barna$^{4}$,
J.~Bartke$^{5}$, L.~Betev$^{9}$, H.~Bia{\l}\-kowska$^{15}$,
C.~Blume$^{8}$,  B.~Boimska$^{15}$, M.~Botje$^{1}$, J.~Bracinik$^{3}$, 
P.~Bun\v{c}i\'{c}$^{9}$, V.~Cerny$^{3}$, P.~Christakoglou$^{2}$, O.~Chvala$^{12}$,
P.~Dinkelaker$^{8}$, J.~Dolejsi$^{12}$, V.~Eckardt$^{11}$, H.G.~Fischer$^{9}$,
D.~Flierl$^{8}$, Z.~Fodor$^{4}$, P.~Foka$^{6}$, V.~Friese$^{6}$, 
M.~Ga\'zdzicki$^{8,10}$, 
K.~Grebieszkow$^{16}$,
C.~H\"{o}hne$^{6}$, K.~Kadija$^{17}$, A.~Karev$^{11,9}$, M.~Kliemant$^{8}$, S.~Kniege$^{8}$,
V.I.~Kolesnikov$^{7}$, E.~Kornas$^{5}$, R.~Korus$^{10}$, M.~Kowalski$^{5}$, 
I.~Kraus$^{6}$, M.~Kreps$^{3}$, M.~van~Leeuwen$^{1}$, 
B.~Lungwitz$^{8}$, M.~Makariev$^{14}$, A.I.~Malakhov$^{7}$, 
M.~Mateev$^{14}$, G.L.~Melkumov$^{7}$, M.~Mitrovski$^{8}$, 
S.~Mr\'owczy\'nski$^{10}$, 
G.~P\'{a}lla$^{4}$, 
D.~Panayotov$^{14}$, A.~Petridis$^{2}$, 
R.~Renfordt$^{8}$,
M. Rybczy\'nski$^{10}$, A.~Rybicki$^{5,9}$, A.~Sandoval$^{6}$, N.~Schmitz$^{11}$, 
T.~Schuster$^{8}$, P.~Seyboth$^{11}$, F.~Sikl\'{e}r$^{4}$, 
E.~Skrzypczak$^{16}$,
G.~Stefanek$^{10}$, R.~Stock$^{8}$, H.~Str\"{o}bele$^{8}$, T.~Susa$^{17}$,
J.~Sziklai$^{4}$, P.~Szymanski$^{9,15}$,
V.~Trubnikov$^{15}$, D.~Varga$^{9}$, M.~Vassiliou$^{2}$,
G.I.~Veres$^{4}$, G.~Vesztergombi$^{4}$, D.~Vrani\'{c}$^{6}$, 
S.~Wenig$^{9,}$\footnote{Corresponding author: Siegfried.Wenig@cern.ch}, 
A.~Wetzler$^{8}$, Z.~W{\l}odarczyk$^{10}$, I.K.~Yoo$^{13}$
\vspace*{2mm}

\noindent
{\it (The NA49 Collaboration)}  \\
\vspace*{2mm}
\noindent
$^{1}$NIKHEF, Amsterdam, Netherlands. \\
$^{2}$Department of Physics, University of Athens, Athens, Greece.\\
$^{3}$Comenius University, Bratislava, Slovakia.\\
$^{4}$KFKI Research Institute for Particle and Nuclear Physics, Budapest, Hungary.\\
$^{5}$Institute of Nuclear Physics, Cracow, Poland.\\
$^{6}$Gesellschaft f\"{u}r Schwerionenforschung (GSI), Darmstadt, Germany.\\
$^{7}$Joint Institute for Nuclear Research, Dubna, Russia.\\
$^{8}$Fachbereich Physik der Universit\"{a}t, Frankfurt, Germany.\\
$^{9}$CERN, Geneva, Switzerland.\\
$^{10}$Institute of Physics \'Swi\c{e}tokrzyska Academy, Kielce, Poland.\\
$^{11}$Max-Planck-Institut f\"{u}r Physik, Munich, Germany.\\
$^{12}$Institute of Particle and Nuclear Physics, Charles University, Prague, Czech Republic.\\
$^{13}$Department of Physics, Pusan National University, Pusan, Republic of Korea.\\
$^{14}$Atomic Physics Department, Sofia University St. Kliment Ohridski, Sofia, Bulgaria.\\ 
$^{15}$Institute for Nuclear Studies, Warsaw, Poland.\\
$^{16}$Institute for Experimental Physics, University of Warsaw, Warsaw, Poland.\\
$^{17}$Rudjer Boskovic Institute, Zagreb, Croatia.\\
\end{Authlist}
%

%
%
%
\vspace*{2mm}
\begin{abstract}
\vspace{-3mm}
New results on the production of charged pions in p+p interactions
are presented. The data come from a sample of 4.8~million inelastic
events obtained with the NA49 detector at the CERN SPS at 158~GeV/c
beam momentum. Pions are identified by energy loss measurement in
a large TPC tracking system which covers a major fraction of the 
production phase space. Inclusive invariant cross sections are
given on a grid of nearly 300~bins per charge over intervals from 0 to
2~GeV/c in transverse momentum and from 0 to 0.85 in Feynman~x. The 
results are compared to existing data in overlapping energy ranges.
\end{abstract}

\cleardoublepage

\end{titlepage}

\section{Introduction
}
\vspace{3mm}
The NA49 collaboration has set out to explore an extended experimental
programme concerning non-perturbative hadronic interactions at SPS
energies. This programme covers elementary hadron--proton collisions
as well as hadron--nucleus and nucleus--nucleus reactions. It is aimed
at providing for each type of interaction large statistics data
samples obtained with the same detector layout which combines wide 
acceptance coverage with complete particle identification. It is 
therefore well suited for the comparison of the different
processes  and to a detailed scrutiny of the evolution from
elementary to nuclear hadronic phenomena. In the absence of reliable
theoretical predictions in the non-perturbative sector of QCD,
it is one of the main aims of this study to provide
the basis for a model independent approach to the underlying
production mechanisms. This approach has to rely on high quality
data sets which cover both the full phase space and a variety 
of projectile and target combinations.

The present paper addresses the inclusive production of charged pions
in p+p collisions.  In the multiparticle final states encountered at
SPS  energies, single  particle inclusive measurements cover
only the simplest  hypersurface of a complex multidimensional
phase space. Only moderate hopes may be nursed to learn enough from
such measurements alone to  experimentally constrain the
non-calculable sector of QCD, and additional  studies beyond the
inclusive surface constitute indeed an important part of the  NA49
programme.

However, as will be shown below, the experimental situation even in
the most simple case of inclusive cross sections is far from
being satisfactory. Notwithstanding a sizeable number of preceding
efforts especially in the SPS energy range which in part date back
several decades, it has not been possible to obtain sufficiently
precise and internally consistent data sets covering the whole
available phase space. Precision in this context may be defined
as the possibility to establish from the existing data an ensemble
of cross sections which will be consistent within well defined error
limits. From the physics point of view, this ensemble should permit
the study of the evolution of inclusive yields from elementary
to nuclear interactions. This evolution is characterized for pions by
deviations of typically some tens of percent in comparison to the
most straightforward superposition of elementary hadronic
interactions. Its interpretation and especially any claim
of connection with new physics phenomena has to rely completely
on comparison with elementary data. Due to the poverty of
available data sets in this sector, situations have arisen where
data from nucleus--nucleus collisions are more complete and
precise than the elementary reference.

It seems therefore mandatory and timely to attempt a new effort in 
this field with the aim of providing an improved elementary data base.

This paper is arranged as follows. In Section 2, the experimental
situation in the SPS energy range is described. Sections 3 to 6
present the data samples obtained by NA49 and a detailed discussion
of event cuts, particle identification, cross section normalization
and applied corrections. The final data are presented and tabulated 
in Sections 7 and 9. A detailed comparison to existing data including a
complete statistical analysis is given in Sections 8 and 9. A short
discussion of the results in relation to forthcoming supplementary
publications terminates the paper in Section 10.

\section{The Experimental Situation
}
\vspace{3mm}
A band of beam momenta extending from 100 to about 400~GeV/c
corresponding to an interval in $\sqrt{s}$ from 14 to 27~GeV may be
defined as SPS/Fermilab energy range. In this interval, quite
a number of experiments have published inclusive particle yields
[1--12]. 
They may be tentatively divided into three different categories.

The first class concerns bubble chamber applications, here mostly
restricted to relatively small data samples of order 10~000 events each
obtained with the Fermilab 30~inch bubble chamber. It is
characterized by the absence of particle identification as far as
pions are concerned.

The second class covers spectrometer measurements with small solid
angle devices measuring typically well delineated and restricted
ranges in production angle and featuring complete particle identification.

A third class contains large solid angle spectrometers like the
EHS and OMEGA facilities at CERN, with the restriction that very few, if any
results concerning inclusive data have been published from these experiments.

In the framework of the present publication the data without particle
identification and therefore with heavy assumptions about
kaon and baryon yields for pion extraction have been discarded from
comparison. 
   
We are interested in the available measurements of the double
differential cross section of identified pions
\begin{equation}
\frac{d^2\sigma}{dx_Fdp_T^2}
\end{equation}
\noindent
as a function of the phase space variables defined in
this paper as transverse momentum $p_T$ and reduced longitudinal
momentum
\begin{equation}
x_F = \frac{p_L}{\sqrt{s}/2}~~~,
\end{equation}
where $p_L$ denotes the longitudinal momentum component in the cms.

\begin{figure}[b]
\centering
\epsfig{file=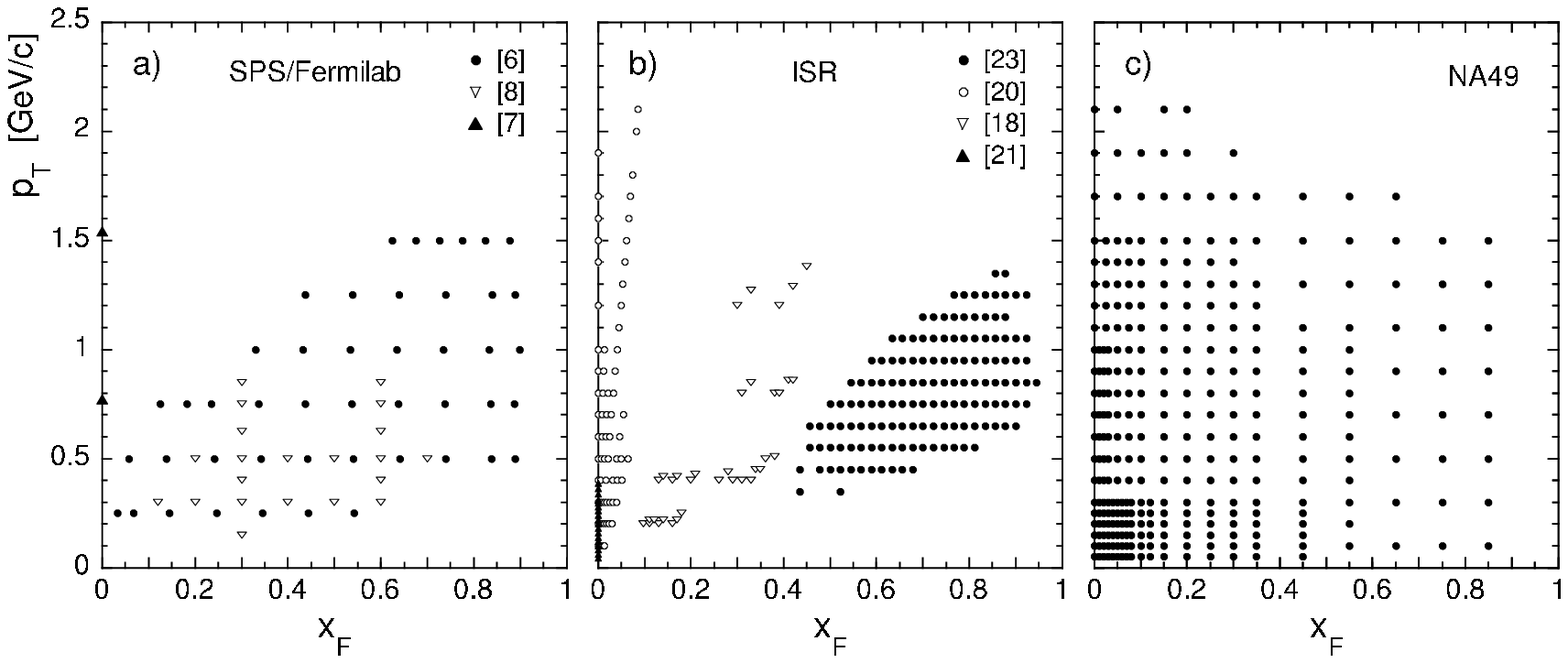,width=16cm}
\caption{Phase space coverage of existing data in the a) SPS/Fermilab and b) ISR energy range
compared to c) NA49. Each set of symbols represents a set of measurements.
}
\label{cov}
\end{figure}

A search for such data with pion identification practically eliminates
the first and third class. The phase space coverage of the remaining 
spectrometer experiments is characterized in Fig.~\ref{cov}a for the fixed target 
configurations. The very limited range of these experiments
is immediately apparent from this plot.
Data are scarce or lacking completely in the regions of $p_T$ below 0.3~GeV/c
and above 1~GeV/c as well as $x_F$ below 0.2. In addition it will be shown
in Section 8.2 below that the most copious data set of Johnson et al.~\cite{joh}
is afflicted with large systematic deviations which makes it unusable
for quantitative reference.

It has therefore been decided to also include some of the rather extensive data from
ISR experiments [13--23] into the comparison in the $\sqrt{s}$ range from 23
to 63~GeV as indicated in Fig.~\ref{cov}b. As a detailed study of $s$-dependence
over the complete kinematic plane is outside the scope of the 
present paper, only data at $x_F > 0.3$ have been used for the comparison. 
In this region Feynman scaling is expected to hold on a few percent level
whereas a more involved $s$-dependence is present at low $x_F$.

The phase space coverage of the NA49 apparatus is shown for comparison
in Fig.~\ref{cov}c. This acceptance allows for the first 
time a detailed study of the low $p_T$ region and is only limited by 
statistics at $p_T$ above 2~GeV/c. In addition, there is a restricted loss of 
acceptance at $x_F > 0.7$ in the lower $p_T$ range for $\pi^+$ due to the interaction trigger 
used (see Section 3.3).     

The main aim of the present paper is to contribute new data covering
the accessible phase space as densely and continuously as possible
in a single experiment. This aim needs first of all a very high
statistics event sample. In addition, high quality particle
identification has to be performed over the accessible range of
kinematic variables. As a third basic requirement the absolute 
normalization and the systematic error sources should be controlled on an
adequate level with respect to the statistical uncertainties.

\section{The NA49 Experiment
}
\begin{figure}[t]
\centering
\epsfig{file=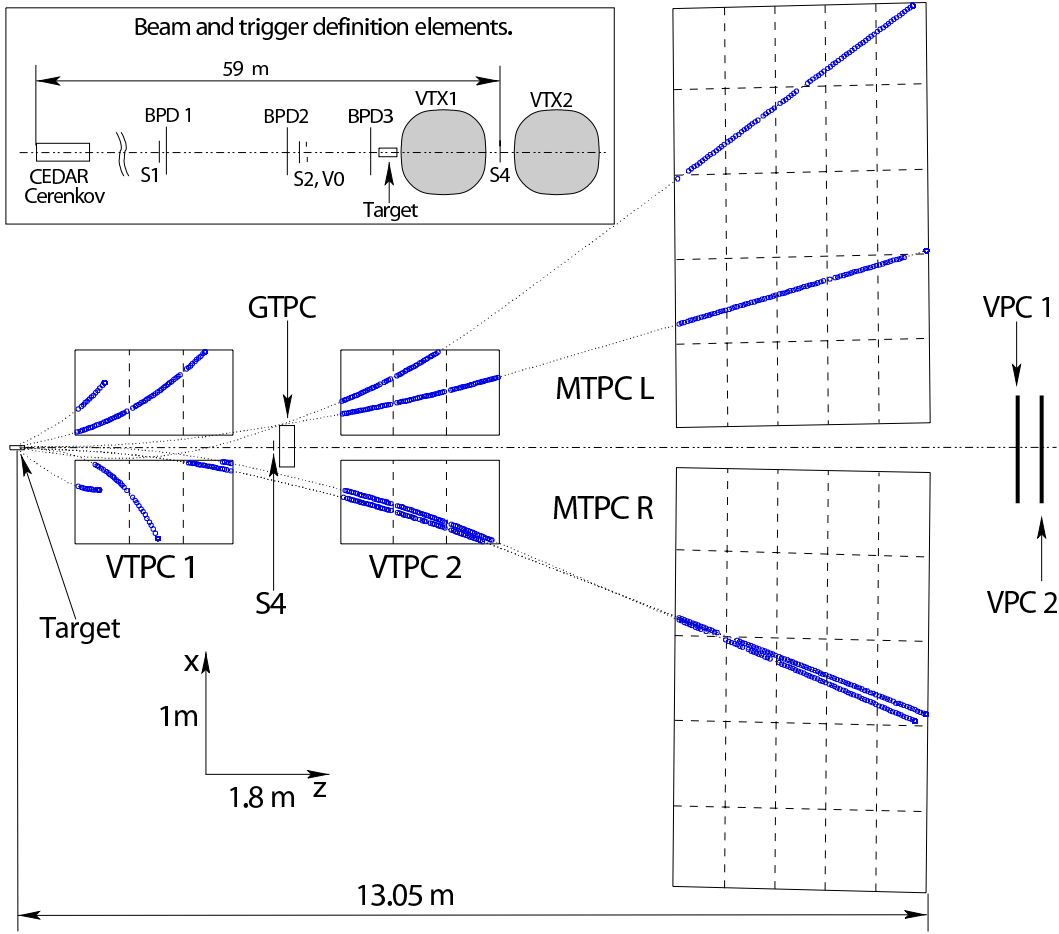,width=14cm}
\caption{NA49 detector layout and tracks of a typical mean multiplicity
p+p event. The open circles are the points registered in the TPC's, the dotted 
lines are the interpolation trajectories between the track segments and the 
extrapolations to the event vertex in the LH$_2$ target. The beam and trigger
definition counters are presented in the inset. Detector name abbreviations are 
explained in the text.
}
\label{exp}
\end{figure}

\vspace{3mm}
NA49 is a fixed target experiment situated in the H2 beam line at the
CERN SPS accelerator complex. It uses a set of large Time Projection 
Chambers (TPC) together with two large aperture Vertex Magnets (VTX1,2) for 
tracking and particle identification. 
A schematic view of the detector is
shown in Fig.~\ref{exp} with an overlay of tracks from a typical mean multiplicity
p+p event. The details of the detector layout, construction and performance
are described in \cite{nim}.

In order to introduce and explain the data taking and analysis policies
adopted in the present work on elementary hadronic cross sections,
some general features of the detector system have to be stressed:

\vspace{2mm}
\begin{itemize}
\item
The data flow from the $1.8 \cdot 10^5$ TPC electronics channels whose analog
outputs are digitized in 512 time buckets per channel produces event
sizes of about 1.5~Mbyte after zero suppression. This in turn limits
the total number of events that can be recorded, stored and processed.
The data sample analyzed in this paper corresponds to a raw data
volume of 10~Tbytes.
\item
The readout system which has been optimized for operation with heavy
ion collisions only allows a recording rate of 32 events
per accelerator cycle.
\item 
This readout rate can be saturated with rather modest beam intensities
of about $10^4$/s which keeps the rate of multiple events during
the 50~$\mu$s open time of the TPC tracker small enough to be readily 
eliminated off-line.
\item
Due to the limitations discussed above, the unbiased running on
a beam trigger alone with alternating full and empty target is   
not feasible. The experiment has therefore to be operated with an
interaction trigger which unavoidably introduces
a certain trigger bias. The corrections for this bias are discussed
in detail in Section~5.2 below.
\end{itemize}

\subsection{The Beam
}
\vspace{3mm}
The secondary hadron beam was produced by 400~GeV/c primary protons
impinging on a 10~cm long Be target. Secondary hadrons were selected
in a beam line set at 158~GeV/c momentum with a resolution of 
0.13\%. 
The particle composition of the beam was roughly 65\% protons, 30\% pions and
5\% kaons. Protons were identified using a CEDAR \cite{bov} Ring Cerenkov counter
(see insert Fig.~\ref{exp}) with a misidentification probability of less than $10^{-3}$. 
Typical beam intensities on target were $3 \cdot 10^4$ particles per 
extraction of 2.37~s.
Two scintillators (S1 and S2 in Fig.~\ref{exp}) provided beam definition 
and timing, together with a ring-shaped veto counter (V0) reducing the background 
from upstream interactions. 
Three two-plane proportional chambers (BPD1--3 in Fig.~\ref{exp}) with cathode
strip readout measured the projectile trajectory yielding 150~$\mu$m position
and 4.5~$\mu$rad angle resolution at the target position where the
beam profile had a full width at base of 6~mm horizontally and 4~mm
vertically.

\subsection{The Target
}
\begin{figure}
\centering
\epsfig{file=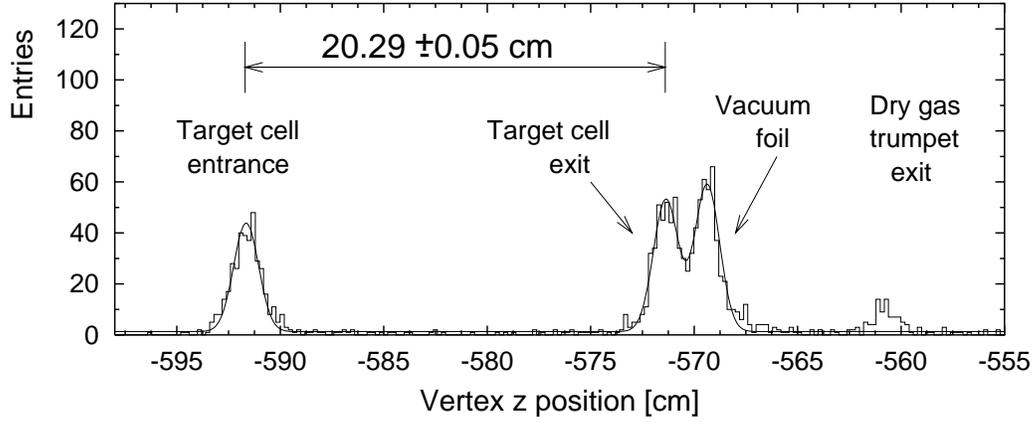,width=14cm}
\caption{Distribution of reconstructed vertex position in high multiplicity empty target 
events delivering the exact target length.
}
\label{tal}
\end{figure}

\vspace{3mm}
A liquid hydrogen target of 20.29~cm length (2.8\% interaction length) and
3~cm diameter placed 88.4~cm upstream of the first TPC (VTPC-1)
was used. The exact target length is determined from the distribution
of the reconstructed vertex positions in high multiplicity empty target 
events as shown in Fig.~\ref{tal}.
 
Runs with full and empty target were alternated. As the empty/full 
event ratio can be reduced from initially 18\% to 9\% after suitable 
offline cuts, and as the empty target rate has been shown to be 
treatable as a small correction to the full target data, see Section 5.1, 
the empty target running could be kept to a fraction of a few percent of the
total data taking.

The target was filled with para-Hydrogen obtained in a closed-loop 
liquefaction system which was operated at 75~mbar overpressure with respect
to atmosphere. This results, at the mean atmospheric pressure of 965~mbar
measured over the running periods of the experiment, in a density of
$\rho_H =0.0707$~g/cm$^3$. 
In addition, the cross section determination has to take into account the 
density of gaseous hydrogen $\rho_{ET}$ present in the empty target runs. 
This density is obtained from the empty over full target ratio of high multiplicity 
events observed in a small fiducial volume around the target center.
The density ratio $\rho_{ET}/\rho_H$ turns out to be 0.5\% which indicates
an increase of the average temperature in the empty target by about
$40^{\circ}$K above liquid temperature. The boiling rate in full-target
condition has been verified to present a negligible modification of the 
liquid density.

\subsection{The Trigger
}
\vspace{3mm}
Beam protons were selected by the coincidence 
C$\cdot$S1$\cdot$S2$\cdot$$\overline{\mbox{V0}}$ yielding the
beam rate $R_{beam}$. The interaction trigger
was defined by a small scintillation counter (S4 in Fig.~\ref{exp}) in 
anti-coincidence with the beam. This counter of 2~cm 
diameter was placed on the beam trajectory 380~cm downstream of the 
target, between the two Vertex Magnets. The trigger condition
C$\cdot$S1$\cdot$S2$\cdot$$\overline{\mbox{V0}}\cdot \overline{\mbox{S4}}$
delivered the interaction rates $R_{FT}$ and $R_{ET}$
for full and empty target operation, respectively. The corresponding trigger 
cross section $\sigma_{trig}$ is calculated according to the formula
\begin{equation}
\sigma_{trig}=
\frac{P^H}{\rho_{H} \cdot l \cdot N_{A}/A}~~,
\end{equation}
where $l$, $N_{A}$, $A$ and $\rho_{H}$ denote, respectively, the target length, the
Avogadro constant, the atomic number and the liquid hydrogen target density.
The interaction probability in hydrogen $P^H$ has to be extracted
from the measured rates $R_{FT}$, $R_{ET}$ and $R_{beam}$. 
Taking into account the exponential beam attenuation in the target,
the reduction of beam intensity due to interactions upstream of the target,
the reduction of the downstream interaction probability in full target operation,
and the gaseous hydrogen content of the empty target cell as discussed above,
it is determined by the relation
\begin{equation}
P^H =
\frac{R_{FT}-R_{ET}}{R_{beam}} \cdot ( 1 + \frac{R_{FT}-R_{ET}}{2 \cdot R_{beam}} +
\frac{R_{ET}}{R_{beam}} + \frac{\rho_{ET}}{\rho_H} )~~~.
\end{equation}
In this expression higher order terms are neglected. The resulting trigger cross
section, averaged over three running periods in the years 1999, 2000 and
2002, amounts to 28.23~mb. From this value the total inelastic cross section as
measured by NA49 is derived using a detailed Monte Carlo calculation
which takes into account the measured inclusive distributions of protons,
kaons and pions as well as the contribution from elastic scattering in order to
determine the loss of events due to produced particles hitting S4. 
The different components resulting from this calculation are presented in
Table~\ref{tri}.
    
\begin{table}[h]
\begin{center}
\begin{tabular}{|l|r|}
\hline
$\sigma_{trig}$                           &  28.23 mb     \\
loss from p                               &   3.98 mb     \\
loss from $\pi$, K                        &   0.33 mb     \\
contribution from $\sigma_{el}$           &  -1.08 mb     \\
\hline
predicted $\sigma_{inel}$                 &  31.46 mb     \\
\hline
\hline
literature value                          &  31.78 mb     \\
\hline
\end{tabular}
\end{center}
\vspace{-2mm}
\caption{Contributions derived by a detailed Monte Carlo calculation to the
determination of the inelastic cross section $\sigma_{inel}$.
}
\label{tri}
\end{table}

It appears that the extracted inelastic cross section compares with the
literature value of 31.78~mb \cite{car} to within one percent. The total trigger
loss concerning inelastic events amounts to 14.4\%. The rejected events
come mostly from target diffraction (about 2.5~mb) and from non-
diffractive events containing forward charged particles at high $x_F$ which hit S4.
In order to account for this event rejection, a topology-dependent correction
to the inclusive cross sections is applied, as described in Section 5.2.

\subsection{Tracking and Event Reconstruction
}
\vspace{3mm}
The tracking system of the NA49 detector comprises a set of large volume
Time Projection Chambers covering a total volume of about 50~m$^3$.
Two of them (VTPC-1 and VTPC-2 in Fig.~\ref{exp}) are placed inside superconducting
Vertex Magnets with a combined bending power of 9~Tm. The magnets define by
their aperture the phase space available for tracking. 
Two larger TPC's (MTPC-R and MTPC-L in Fig.~\ref{exp}) are positioned downstream of the 
magnets in order to extend the acceptance to larger momenta and to provide sufficient 
track length for precise particle identification via ionization energy loss $dE/dx$
measurement.
The use of the same detector for the complete range of interactions
implies a separation of the sensitive TPC volumes with respect to the
beam trajectory. Otherwise, in the case of heavy ion collisions non-interacting
beam particles would create excessive chamber loads and the density of
secondary tracks would be prohibitively high. The corresponding acceptance
loss for low $p_T$ particles in the $x_F$ range above 0.5 has been remedied by
the introduction of a small TPC on the beam line in between the two
magnets (GTPC in Fig.~\ref{exp}) which, in combination with two strip-readout
proportional chambers (VPC-1 and VPC-2 in Fig.~\ref{exp}) ensures tracking acceptance
up to the kinematic limit. As the target is placed upstream of the magnets
there is a gradual reduction of acceptance at low $p_T$ in the backward 
hemisphere which for pions starts to be effective for $x_F < -0.05$.    
The overall acceptance region for tracking of pions in the forward $x_F/p_T$ plane 
has been shown in Fig.~\ref{cov}c.

The event reconstruction proceeds through several steps. Firstly,
all charged track candidates leaving at least 8 space points (clusters)
in the TPC system are pattern recognized and momentum fitted. This step
includes the formation of global tracks from track segments visible in
different TPC's (see Fig.~\ref{exp}). Secondly, a primary event vertex
is fitted using all global tracks found in the event together with the measured 
beam track. 
After a successful vertex reconstruction, a second path of momentum fitting is 
performed also including tracks which are only recorded in the downstream TPC's outside
the magnetic field.
In this final momentum fit, the determined vertex point is used as an additional 
measured point on the track. All tracks with an acceptable $\chi^2$ of the fit are 
retained for further analysis.

\subsection{Event Selection
}
\vspace{3mm}
Two steps of event selection are introduced in order to clean up the
event sample and to reduce the empty target background. 

\begin{figure}[h]
\centering
\epsfig{file=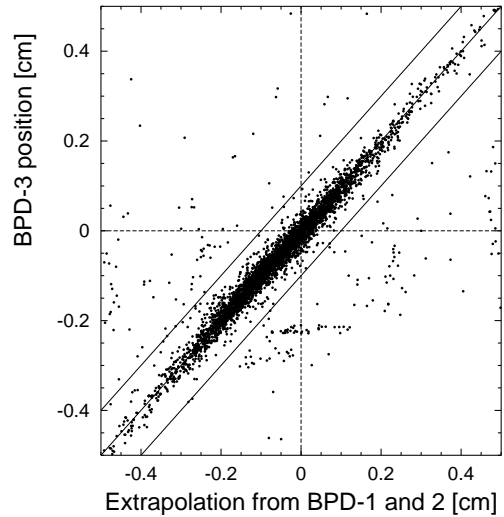,width=7cm}
\caption{Beam position at BPD-3 versus extrapolation of beams from BPD-1 and BPD-2 to BPD-3.
Only events falling between the two lines are accepted.
}
\label{bpd}
\end{figure}

Firstly, cuts on the beam position close to the target are performed.
The position of the incoming beam particle is registered in three
Beam Position Detectors (BPD1-3 in Fig.~\ref{exp}). The three corresponding
measurements are required to be well defined and collinear in both measured 
transverse coordinates by imposing that the extrapolation from BPD-1 and BPD-2
to BPD-3 coincides with the beam position measured in BPD-3, as demonstrated 
in Fig.~\ref{bpd}. A further, less restrictive cut is imposed on the 
beam profile. 
These cuts are bias-free as the measurement takes place before the interaction.

\begin{figure}
\centering
\epsfig{file=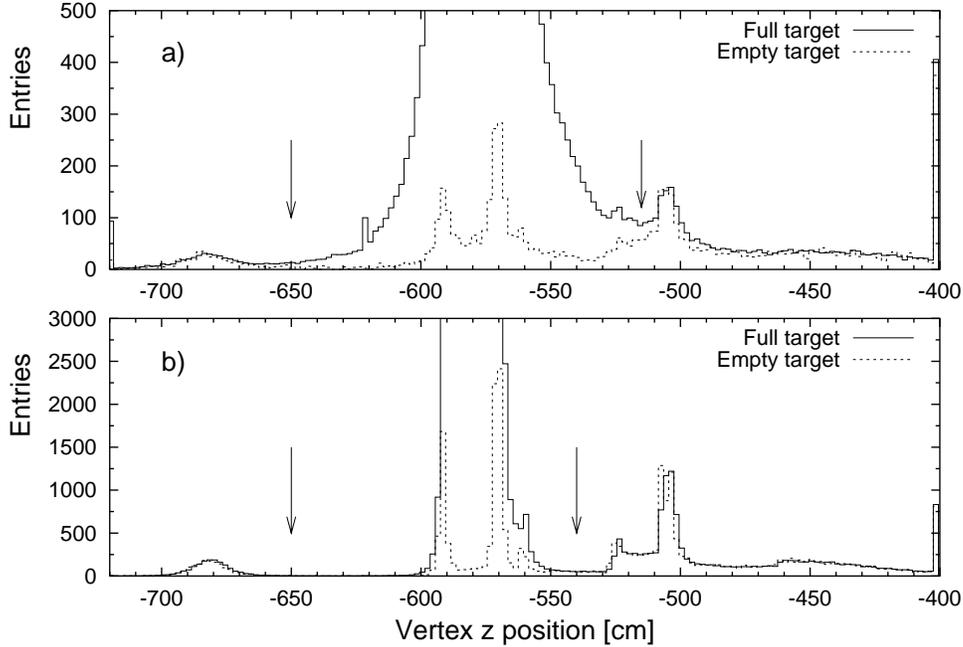,angle=-90,width=13cm}
\caption{Normalized vertex distributions from full and empty target events
with selected track multiplicity a) one and b) five and more.
The imposed vertex cuts are indicated by arrows.
}
\label{vtz}
\end{figure}

Secondly, the interaction vertex is constrained to a fiducial
region around the target position by applying a cut on the longitudinal coordinate of
the reconstructed vertex position. Though this cut is very efficient
in reducing the empty target contribution, it must be carefully designed
as the precision of the reconstructed vertex position depends strongly
on the event configuration. 
The longitudinal vertex cut is therefore performed depending on the track multiplicity
and on the track topology. Short or very small laboratory angle tracks
are not entering the track sample determining the vertex and the cut boundaries are 
placed so that no event from the target is rejected. This is exemplified in the
normalized vertex distributions from full and empty target shown in Fig.~\ref{vtz}a and b.
For the 5\% of the target events with at least one reconstructed track, for which the 
reconstruction software does not give a reliable longitudinal vertex position, the target
center is used.

The combined event selection described above reduces the relative empty 
target yield from 18\% to 9\% retaining 85\% of the liquid hydrogen target events.
This final sample corresponds to $4.8 \cdot 10^6$ events which were obtained
in three running periods between the years 1999 and 2002, as shown in
Table~\ref{evsamp}.

\begin{table}[b]
\begin{center}
\begin{tabular}{|c|cc|cc|}
\hline
       & \multicolumn{2}{c|}{Events taken}  & \multicolumn{2}{c|}{Events after cuts}  \\
Year   & Full target & Empty target & Full target & Empty target                      \\
\hline
1999   &  1 211 k  &  41.2 k  &    906 k  &   13.7 k    \\
2000   &  2 648 k  &  47.8 k  &  2 049 k  &   16.9 k    \\
2002   &  2 508 k  &  69.0 k  &  1 814 k  &   21.8 k    \\
\hline
Total  &  6 367 k  & 158.0 k  &  4 769 k  &   52.4 k    \\
\hline
\end{tabular}
\end{center}
\vspace{-2mm}
\caption{Data samples analyzed.
}
\label{evsamp}
\end{table}

\subsection{Track Selection
}
\vspace{3mm}
A track in the NA49 detector is defined by the ensemble of clusters
(points) in three dimensions that a charged particle leaves
in the effective volume of the TPC system. These clusters have a typical
spacing in track direction of about 3~cm in the VTPC's and 4~cm in
the MTPC regions. The tracking in a TPC environment has several decisive
advantages which make the track selection a relatively straight-forward and
safe task:

\vspace{2mm}
\begin{itemize}
\item
The local efficiency for cluster formation is practically 100\% 
  with a loss rate below the permille level. This is guaranteed 
  by the choice of operation point of the readout proportional chambers 
  and by the high reliability of the readout electronics which has a 
  fraction of less than $10^{-3}$ of missing or malfunctioning channels.
\item
The total number of points expected on each track can be reliably 
  predicted by detector simulation with the exception of some edge
  regions for the VTPC's due to magnetic field $E \times B$ effects.
\item
Regions of 100\% acceptance can be readily defined in each kinematic
  bin $\Delta x_F$, $\Delta p_T$, and the azimuthal angle wedge 
  $\Delta \Phi$ by inspecting the distribution of points
  per track in comparison to the expected value. In practice this is
  achieved by adjusting $\Delta \Phi$ such that this distribution does
  not show tails beyond a well-defined average. This gives at the same
  time an experimental handle to stay away from edge regions which show
  a drop of the number of points.
\item
The only possibility of track losses or a reduction of track length 
  is due to weak decays or hadronic interaction in the detector gas.
  The policy adopted in the present analysis is to allow a track to
  be shorter than expected if it has a minimum length and if the lost
  points are concentrated at the end of the track. It has been verified
  by eye-scans on such shortened tracks that either a decay (presence
  of one additional downstream track at an angle to the primary track)
  or a hadronic interaction (several additional tracks emerging from
  a defined interaction vertex) is present.
\end{itemize}
\vspace{2mm}

\begin{figure}[h]
\centering
\epsfig{file=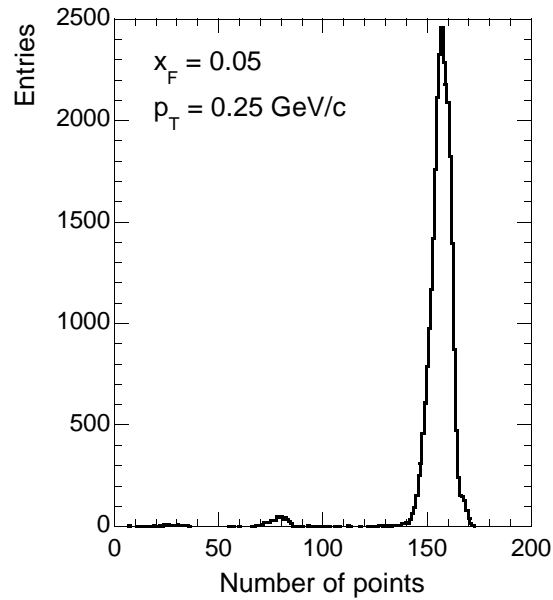,width=15cm}
\caption{Distribution of the number of measured points in a typical analysis bin at 
$x_F=0.05$ and $p_T=0.25$~GeV/c.
}
\label{pointnum}
\end{figure}

In practice each track entering the analysis sample has to have at least
30 points in order to ensure a minimum quality of particle identification
(see Section~4). This corresponds to detected track lengths in excess of 90~cm.
The only exception to this criterion is in the extreme forward direction
where the GTPC/VPC combination has only 9 space points and where no
particle identification via $dE/dx$ is feasible. An example of a point 
number distribution in a typical analysis bin
is given in Fig.~\ref{pointnum}.

The corresponding tracks span 3 TPC's (VTPC-1,VTPC-2,MTPC) with an expected
number of about 160~clusters. 2.15\% of all tracks are found in two distinct
accumulations around 80 and 30 points, corresponding to tracks
detected up to the end of VTPC-2 (1.7\%) and VTPC-1 (0.45\%), respectively.
This has to be compared to the expected fraction from nuclear interactions, 
1.2\% and 0.3\%, confirming that the reconstruction software does not introduce short 
tracks, and therefore potential background, of unexplained sources.
The small fraction of tracks falling below the 30~point
cut is corrected for by the absorption correction described
in Section~5.4.

The track selection criteria defined above have been controlled by extensive
eye-scans. This method is made very efficient in the low multiplicity 
environment of p+p collisions by the photograph-like picture built up
by the measured space points and by the excellent pattern recognition
capability of the human eye. Based on extensive studies with special 
care for potentially problematic areas (high $p_T$, edge regions of acceptance,
short tracks) the tracking efficiency is found to be 100\% with an upper
error limit of 0.5\% below $x_F = 0.3$ rising to less than 2\% close to the
kinematic boundary approached by the GTPC/VPC combinations.

\subsection{Acceptance Coverage, Binning and Statistical Errors
}
\begin{figure}[h]
\centering
\epsfig{file=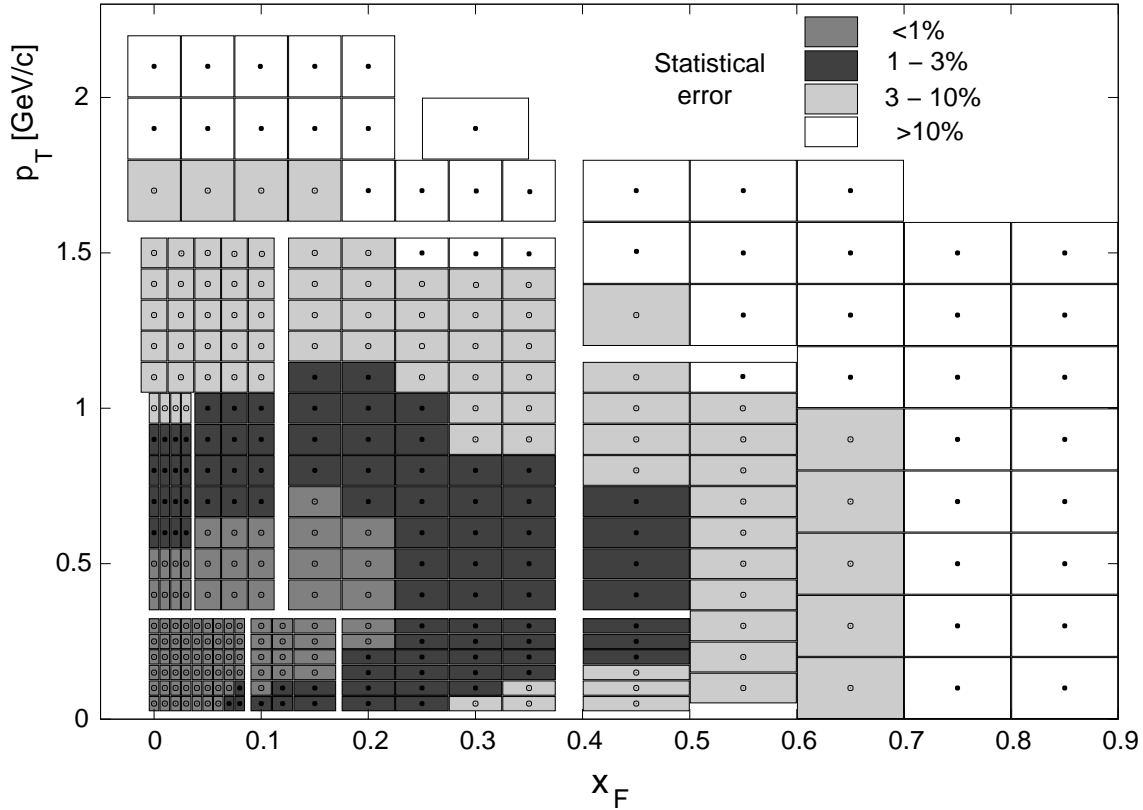,width=15cm}
\caption{Binning scheme in $x_F$ and $p_T$ together with information about
the statistical error.
}
\label{binning}
\end{figure}

\vspace{3mm}
The event sample defined above contains a total of about 28~million pions.
In order to cover the available acceptance in an optimum fashion, a
binning scheme presented in Fig.~\ref{binning} has been chosen in the variables
$x_F$ and $p_T$. There are several aspects determining this choice:

\vspace{2mm}
\begin{itemize}
\item
Optimum exploitation of the available statistics
\item
Definition of bin centers at user-friendly and consistent values of 
  $x_F$ and $p_T$
\item 
Compliance with the structure of the inclusive cross sections
\item
Sufficiently small bins in $p_{tot}$ for optimum extraction of ionization
  energy loss for particle identification
\item
Avoidance of overlaps and minimization of lost regions
\item
Optimization of bin sizes for minimum binning effects and corresponding
  corrections
\end{itemize}
\vspace{2mm}

The resulting statistical precision per bin is also indicated in Fig.~\ref{binning}.
This precision is superior or equal to all other existing measurements
in the SPS/Fermilab and ISR energy regions with the exception of a few points
at large $p_T$ and/or large $x_F$.

\section{Particle Identification
}
\vspace{3mm}
\subsection{Identification Method, Parametrizations and Performance
}
\begin{figure}[h]
\centering
\epsfig{figure=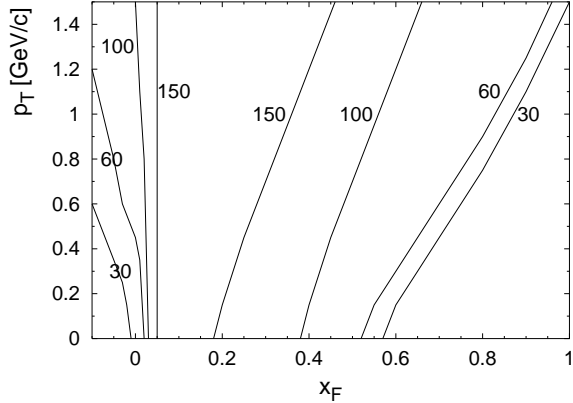,angle=-90,width=8cm} 
\caption{Lines for constant number of $dE/dx$ samples per track in the $x_F$/$p_T$ plane
(restricted $\Delta \Phi$ range of selected tracks).
}
\label{dedx-1}
\end{figure}

\vspace{3mm}
The NA49 detector offers a powerful combination of tracking and
particle identification via ionization energy loss ($dE/dx$) measurement
in the TPC system. This system combines four large TPC volumes and
features track lengths of between 1~m and 6~m in the kinematic region
for pions covered in this publication.
Each track is sampled by readout pads of 2.8 and 4~cm length in the
Vertex and Main TPC's, respectively. 
This results in the pattern of sample numbers $N_s$ per track as a function 
of $x_F$ and $p_T$ shown in Fig.~\ref{dedx-1} .   

The $dE/dx$ measurement is achieved by forming a truncated mean 
of the 50\% smallest charge deposits sampled along each track. 
The truncation transforms the Landau distribution of the samples
into a Gaussian distribution of the mean per track if the number of 
samples stays above about 30. Under this condition the relative 
resolution of the energy loss measurement can be parametrized as 

\begin{equation}
\frac{\sigma(N_s,dE/dx)}{dE/dx}=\sigma_0 \frac{1}{N_s^{\beta}} (dE/dx)^{\alpha}
\label{dedx-s}
\end{equation}

\noindent
where the dependence of the truncated mean on the number of samples
$dE/dx(N_s)$ and the parameters $\sigma_0$, $\alpha$  and $\beta$ are determined 
experimentally. To this end, for each of the 62 readout sectors of
the TPC system, tracks are binned in laboratory momentum $p_{lab}$ and a preliminary 
truncated mean distribution is formed with the samples found
outside the sector under study. Sharp cuts are performed on this 
preliminary $dE/dx$ measurement in order to separate electrons, pions,
kaons and protons in each momentum bin. Samples from different, identified
tracks in the given sector are then combined into truncated means with an arbitrary
number of samples in order to determine the dependences defined above.

\begin{figure}[h]
\centering
\epsfig{figure=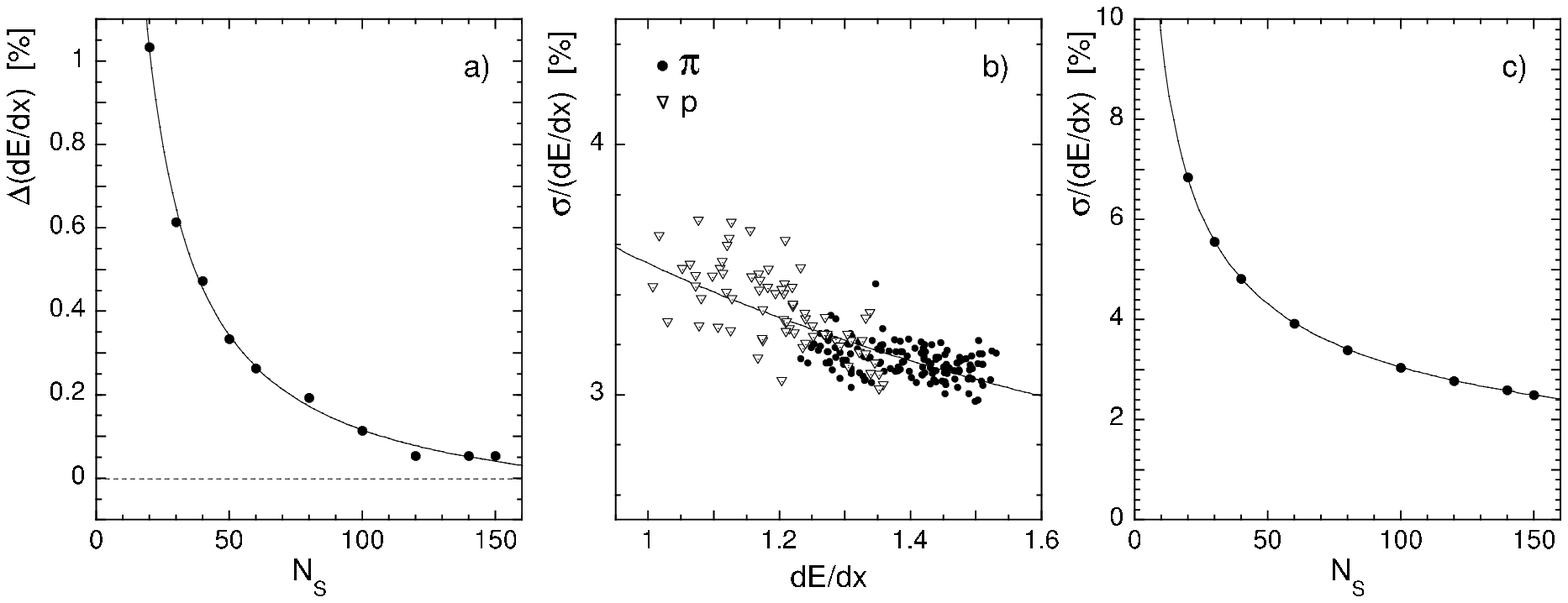,width=15cm}
\caption{a) Percentage deviation of the mean $dE/dx(N_s)$ from $dE/dx(\infty)$
as a function of $N_s$; resolution as a function of b) $dE/dx$ and c) $N_s$.
All 3 panels are for MTPC.
}
\label{dedx-2}
\end{figure}

The dependence of the $dE/dx$ measurement on the number of samples, $dE/dx(N_s)$ is
shown in Fig.~\ref{dedx-2}a. The dependences of the resolution on $dE/dx$ and $N_s$ are
presented in Fig.~\ref{dedx-2}b and \ref{dedx-2}c.
The parameters $\alpha$ and $\beta$ are fitted to

\begin{center}
$\alpha = - 0.39 \pm 0.03$
\hspace{1cm}
and 
\hspace{1cm}
$\beta  = 0.50 \pm 0.01$~~~.
\end{center}

The NA49 TPC system employs different gases based on Ne in the VTPC 
and Ar in the MTPC. The function $dE/dx(N_s)$ and the parameters $\alpha$ 
and $\beta$ are found to be independent of the type of gas. Also the 
parameter $\sigma_0$ referred to the same pad length is the same for 
Ne and Ar. This has been shown by Lehraus et al. in 1982 \cite{leh} and 
can be understood by the different behaviour of secondary ionization for 
different noble gases \cite{fis}. The only difference of resolution observed 
in the Vertex chambers is due to the shorter pad length with a 
dependence which is again given by the parameter $\alpha$. The resulting 
values are

\begin{center}
$\sigma_0^{\mbox{\small{Vertex}}} = 0.41$
\hspace{1cm}
and
\hspace{1cm}
$\sigma_0^{\mbox{\small{Main}}} = 0.352$~~~.
\end{center}

The measured truncated means have a nonlinear relationship to the Bethe
Bloch function which describes the primary ionization loss and has
a different dependence on particle velocity ($\beta \gamma$) in the two gases.
This difference which is of the order of 0-5\% for different regions of
$\beta \gamma$ has to be taken into account. Using the above methodology
the energy loss functions are independently determined
for the two gases. In order to combine samples from the VTPC and MTPC
on the same track two independent truncated means are formed. The value
from the MTPC is transformed to the corresponding value of
the VTPC using a linear transformation. The weighted average of these
numbers, taking the respective resolutions into account, results in
the final $dE/dx$ measurement.

It has to be mentioned that before the formation of truncated means a number 
of corrections have to
be applied to each ionization sample. This concerns detailed time dependence 
extending over the complete data taking period of several years including
pressure dependence up to second order terms, and corrections for track
angles, effects of magnetic field and
drift length dependences principally induced by the electronics threshold  
cut combined with electron diffusion in the gas.

\begin{figure}[h]
\centering
\epsfig{figure=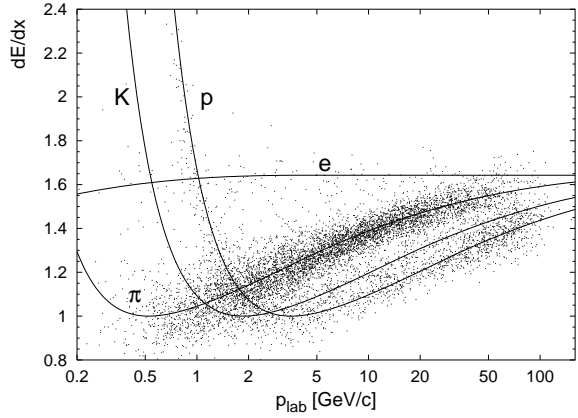,angle=-90,width=8cm}
\caption{Energy loss $dE/dx$ with respect to minimum ionization as a function 
of track momentum $p_{lab}$.
}
\label{dedx-3}
\end{figure}

The resulting relative resolution is typically on the 3-4\% level over
most of the phase space covered by this experiment with tails up to
8\% at low $x_F$ and $p_T$ and at large $x_F$ due to the decrease of sample
numbers. The scatter plot of $dE/dx$ values (referred to minimum ionization)
versus track momentum shown in Fig.~\ref{dedx-3} gives an impression of the performance
with respect to the necessity to separate particles in the region of the
relativistic rise. 
The lines shown in Fig.~\ref{dedx-3} represent the mean response of the detector to 
the different types of particle derived from the measured energy loss.

\subsection{Fit Procedure and Yield Determination
}
\begin{figure}[h]
\centering
\epsfig{figure=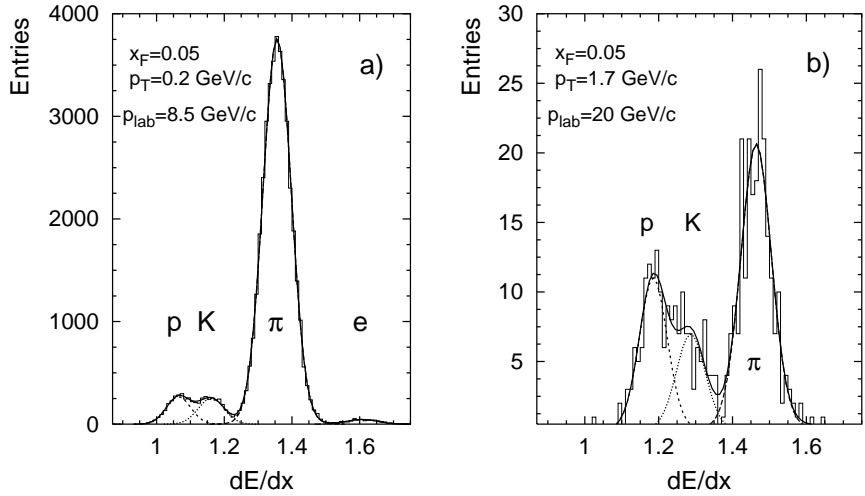,angle=-90,width=12cm}
\caption{$dE/dx$ distribution for two different $x_F/p_T$ bins. The line represents
the fit.
}
\label{dedx-4}
\end{figure}

\vspace{3mm}
The particle identification, i.e. the determination of the yields of
individual particle species is achieved by a fit to the $dE/dx$
distribution for a small region of momentum defined by the analysis
bin. 
This distribution is a superposition of Gaussians with variances following
for each particle type the parametrization specified in Formula \ref{dedx-s} above, 
and thus taking proper account of the variation of $N_s$ over the bin. 
Two typical examples are shown in Fig.~\ref{dedx-4}.

\begin{figure}[t]
\centering
\epsfig{figure=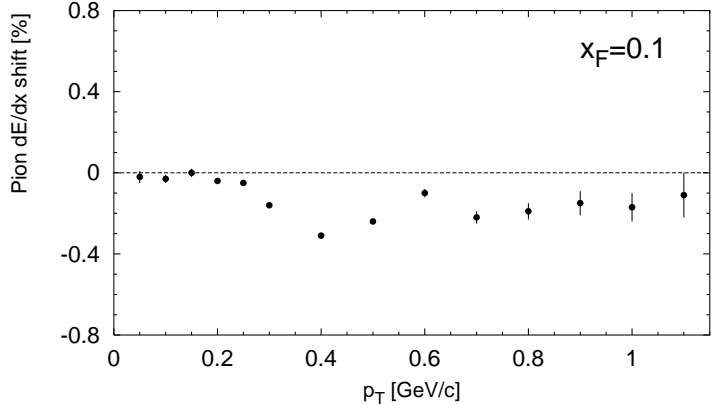,angle=-90,width=10cm}
\caption{Deviation of the measured mean $dE/dx$ from the predicted one as a 
function of $p_T$ for pions.
}
\label{dedx-5}
\vspace{-2mm}
\end{figure}

The fit has in principle nine parameters: four particle yields, four
mean $dE/dx$ values and the $\sigma_0$ parameter describing the width of
the individual distributions. The last five parameters would be  
determined exactly by the energy loss function and the resolution function
(Formula \ref{dedx-s}) if these quantities could be absolutely predicted.
Due to the complexity of the primary and secondary ionization
processes this prediction is, however, not possible on the
level of precision needed here. The mean energy loss which is
a unique function of $p_{lab}/m = \beta \gamma$ (see Fig.~10) has therefore
to be described by a multi-parameter approximation which 
must be expected to show deviations from the measured response. 
Detector related imperfections in the elaboration of 
local calibration, magnetic field effects and pulse formation
introduce additional deviations which in general violate the
$\beta \gamma$ scaling of the elementary process. The resulting
pattern of displacements has a smooth dependence on the kinematic
variables
%
%
as exemplified in Fig.~\ref{dedx-5}, where the
deviations are shown as a function of $p_T$ at a fixed $x_F$ for pions.

Indeed, in order to keep the systematic error of the yield extraction well 
below the statistical error in a bin, these shifts have to be and can 
be determined on a sub-percent level by performing complete nine parameter 
fits. The fit procedure minimizes the $\chi^2$ over the complete $dE/dx$
distribution in each bin. For the $\chi^2$ definition different 
prescriptions representing the degree to which the fit is 
considered reliable have been tested. Comparison of the results from these
different prescriptions reveals consistency within the expected statistical
errors, confirming that there are no visible tails or shape distortions
affecting the yield extraction. The statistical error of the fit is 
calculated from the covariance matrix of the nine parameters using the
standard $\chi^2$ definition. For the determination of the pion yield
which is dominating over most of the measured phase space, the error turns out
to be equal to the square-root of the number of pions in each bin with the exception 
of the far forward region ($x_F > 0.3$) discussed in Section ~4.5 below.
This means that the fitting method itself does not introduce any further 
systematic uncertainty to be added to the purely statistical fluctuations of the number 
of pions in each bin.

As each of the 589 $x_F/p_T$ bins is fitted individually without
imposing external constraints, extensive checks of the fit outputs are
performed. In fact, the results have been extracted by three people with two
different programs and the consistency has been confirmed in about 300 mutual
cross-check bins.

\subsection{The Region of $dE/dx$ Crossing
}
\vspace{3mm}
Below $p_{lab}$ of about 3~GeV/c the energy loss functions of
pions, kaons and protons approach each other in the so-called cross-over
region, see Fig.~\ref{dedx-3}. This, together with a significant reduction of track
length in the same region, see Fig.~\ref{dedx-1}, prohibits independent pion
identification. On the other hand, as pions at $x_F=p_T=0$ have a 
$p_{lab}$ of 1.3~GeV/c, this region contains most of the inclusive pion yield
and is therefore essential to complete the data set.

In order to make the measurement possible, a reflection technique
already used in bubble chamber experiments \cite{zab} without any means of
particle identification in this region is employed. At the laboratory
momentum which corresponds to $x_F=p_T=0$ for pions, protons are at
$x_F^{\mbox{\small{p}}} \simeq -0.3$. 
The proton yield can therefore be extracted from forward/backward 
symmetry by fitting protons in the symmetric bin at 
$x_F^{\mbox{\small{p}}} \simeq +0.3$. 
The corresponding reflection can be applied for kaons.

\begin{figure}[h]
\centering
\epsfig{figure=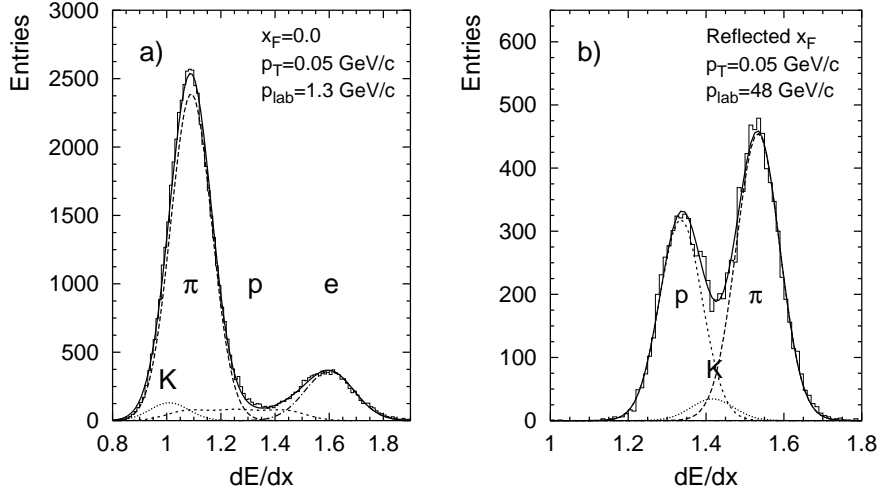,angle=-90,width=12cm}
\caption{$dE/dx$ distribution in a) the pion bin ($p=1.3$~GeV/c) and 
b) the corresponding proton reflected bin ($p=48$~GeV/c).
}
\label{dedx-6}
\end{figure}

In practice, the complete pion analysis bin is reflected with proton
(or kaon) assumption. 
The $dE/dx$ distributions for the original pion bin and the corresponding
proton-reflected bin is shown in Fig.~\ref{dedx-6}.

The reflection technique is used for $p_T < 0.3$~GeV/c at $x_F = 0$ and
up to $x_F =  0.02$ at $p_T = 0.05$~GeV/c. The estimated systematic error
from this method is below 1\% since the proton and kaon 
contributions are on the 5-10\% level in this kinematic area.
The consistency of the method has been tested in bins where both the normal
extraction and the reflection method are usable, and its reliability
was confirmed.

\subsection{Electrons at High Momenta
}
\begin{figure}[h]
\centering
\epsfig{figure=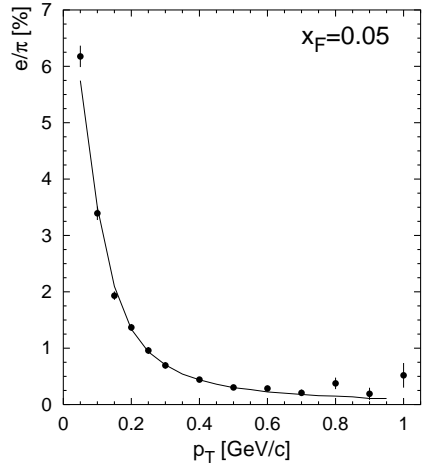,angle=-90,width=6cm}
\caption{e/$\pi$ ratio as a function of $p_T$ at $x_F=0.05$. 
The line represents the Monte Carlo result.
}
\label{dedx-7}
\end{figure}

\vspace{3mm}
Above $p_{lab}$ of about 40~GeV/c the energy loss
of pions approaches the relativistic plateau occupied by the electrons,
rendering independent extraction difficult. This difficulty is
enhanced by the fact that the e/$\pi$ ratio becomes very small, typically
of the order of a few permille. In order to be able to control this
small contribution, a Monte Carlo simulation has been constructed using
$\pi^0$ as the main source of electrons by gamma conversion and
Dalitz decay. The $\pi^0$ cross section is obtained as the average
of the measured charged pions and the combined conversion probability
is adjusted from the measurement of the e/$\pi$ ratio at $x_F=0.05$ as
shown in Fig.~\ref{dedx-7}.
In the high $x_F$ region, the ratio is then constrained to the value
predicted by the simulation and the small excess over the prediction obtained
from the fit added to the pion yield.

\subsection{Constraints from K/$\pi$ and p/$\pi$ Ratios at High $x_F$
}
\vspace{3mm}
The region of $x_F$ above about 0.3 presents specific problems for the
independent extraction of kaons necessary to obtain bias-free pion
cross sections. This is due to the fact that the mean ionizations of the
different particle types approach each other towards saturation, and that 
the average track length steadily decreases.
Furthermore, the cross sections fall steeply with $x_F$, which progressively reduces 
the available statistics per bin.
 
\begin{figure}[h]
\centering
\epsfig{figure=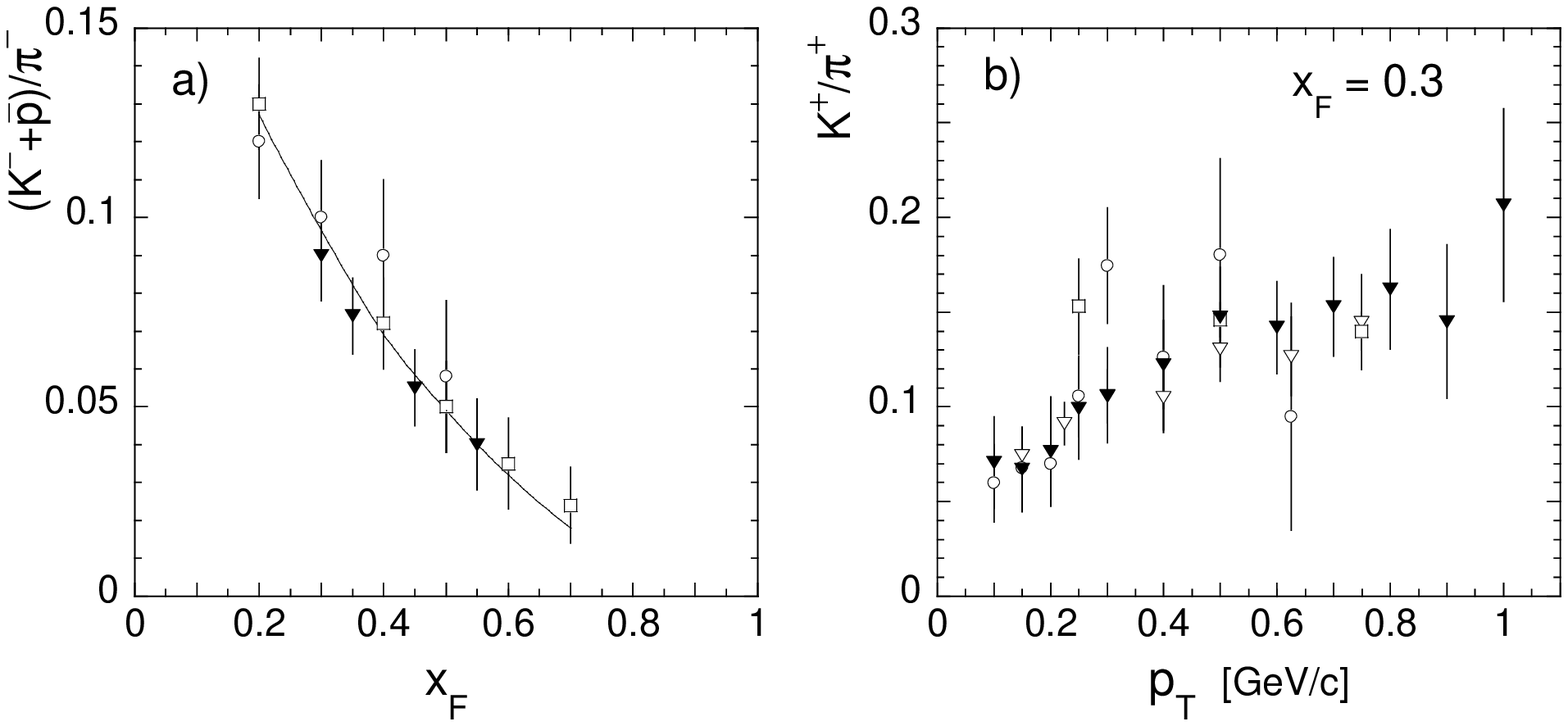,width=14cm}
\vspace*{-2mm}
\caption{a) (K$^-$+$\overline{\mbox{p}}$)/$\pi^-$ ratio as a function of $x_F$ measured by
\cite{bre} (open circles) and \cite{joh} (open squares) compared to NA49 (full triangles), 
b) K$^+$/$\pi^+$ ratio as a function of $p_T$ measured by
\cite{bre} (open circles and triangles) and \cite{joh} (open squares) compared to NA49 
(full triangles).
}
\label{dedx-8}
\end{figure}

The combined (K$^-$+$\overline{\mbox{p}}$)/$\pi^-$ ratio is rapidly
falling with $x_F$ for all $p_T$ values. This $p_T$ averaged ratio as obtained 
from the NA49 data is shown in Fig.~\ref{dedx-8}a as a function of $x_F$. 
As the $dE/dx$ fits become rather unreliable for values of this ratio below a few
percent and at low statistics, the measurement is complemented by values 
from other experiments \cite{joh, bre} also shown in Fig.~\ref{dedx-8}a and extrapolated smoothly 
to zero at $x_F \simeq 0.85$.


For K$^+$ extraction, the situation becomes difficult already above $x_F \simeq 0.3$ due
to the preponderant p component. This leads to problems with the
independent determination of the kaon shift with respect to the
energy loss function as described in Section~4.2 above. In fact the
fit tends to find unphysical local minima of $\chi^2$ corresponding
to large displacements of the K$^+$. In this area the K$^+$ shifts have
to be constrained by using the values obtained for pions and 
protons. The resulting K$^+$/$\pi^+$ ratios are again compared to data from
other experiments \cite{joh, bre} which agree well with the results from NA49 at 
$x_F=0.3$ presented in Fig.~\ref{dedx-8}b.
This confirmation of consistency is necessary, as for higher $x_F$ values the
K$^+/\pi^+$ ratio was constrained to that from other measurements.
The rapid decrease of the $\pi^+$/p ratio to values below 5\% at $x_F > 0.55$
imposes, together with the slightly increasing K$^+$/$\pi^+$ ratio in this
region, a limit on reliable $\pi^+$ extraction at $x_F = 0.55$. 

\section{Evaluation of Invariant Cross Sections and Corrections
}
\vspace{3mm}
The invariant inclusive cross section is defined as
\begin{equation}
   f(x_F,p_T) = E(x_F,p_T) \cdot \frac{d^3\sigma}{dp^3} (x_F,p_T)
\end{equation}
where $dp^3$ is the infinitesimal volume element in three dimensional momentum
space.

This cross section is approximated by the measured quantity
\begin{equation}
  f_{meas}(x_F,p_T,\Delta p^3) = 
  E(x_F,p_T,\Delta p^3) \cdot \frac{\sigma_{trig}}{N_{ev}} \cdot 
  \frac{\Delta n(x_F,p_T,\Delta p^3)}{\Delta p^3}~~~,
\end{equation}
where 
$\Delta p^3$ is the finite volume element defined by the experimental bin
width, $\sigma_{trig}$ the trigger cross section discussed in Section~3.3,
$N_{ev}$ the number of events originating from the liquid hydrogen target, and
$\Delta n$ the number of identified pions from the target measured
in the bin $\Delta p^3$. In general and for arbitrary functional shapes of
$f(x_F,p_T)$, the measured quantity $f_{meas}$ depends on the bin width $\Delta p^3$
via $E$ and $\Delta n$.

Several steps of normalization and correction are necessary in order to
make $f_{meas}$ approach $f(x_F,p_T)$:

\vspace{2mm}
\begin{itemize}
\item
The numbers $N_{ev}$ and $\Delta n$ have to be determined from the measured
  full and empty target yields.
\item
The fact that $\sigma_{trig}$ is not equal to $\sigma_{inel}$ necessitates
  a trigger bias correction.
\item
$\Delta n$ has to be corrected for re-interaction of produced particles
  in the target volume, for weak decay and absorption of pions on their
  way through the detector, and for feed-down from weak decay of strange
  particles.
\item
The finite volume element $\Delta p^3$ has to be replaced by the infinitesimal
  $dp^3$ which requires a binning correction.
\end{itemize}
\vspace{2mm}

This list defines seven corrections which will be discussed in turn below.

\subsection{Empty Target Correction
}
\vspace{3mm}
The particle yield $\Delta n/N_{ev}$ can in principle be determined from separate
yield determinations in full and empty target conditions for each bin with the formula
%
%
$$ \left( \frac{\Delta n}{N_{ev}} \right)^{FT-ET} = \frac{1}{1-\epsilon} \left( \left( \frac{\Delta
n}{N_{ev}} \right)^{FT} - \epsilon \left( \frac{\Delta n}{N_{ev}} \right)^{ET} \right)~~~, $$
where $\epsilon=R_{ET}/R_{FT}$.

\begin{figure}[h]
\centering
\epsfig{file=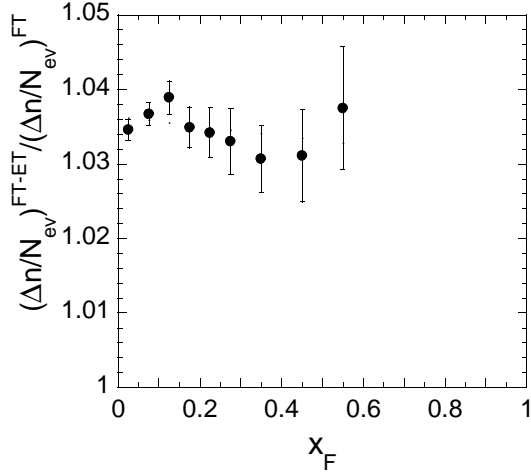,width=14cm}
\caption{Correction factor applied to account for the  empty target contribution 
as a function of $x_F$ for the average of $\pi^+$ and $\pi^-$.
}
\label{etcor}
\end{figure}

In practice the bin-by-bin subtraction would necessitate a large enough 
event sample from empty
target to comply with the statistical precision of the full target data
and in addition would give identification problems in the regions of
low cross sections where fits to the energy loss distributions anyhow
become critical. Given the constraints in data taking of the NA49
experiment and the sizeable reduction of the empty target rate, both evoked 
and described in Section~3, a more efficient approach is used. 
For the achieved empty/full target event ratio of 9\%,
the empty target contribution in terms of track numbers in a given bin
is in fact only about 4-5\% due to the much larger yield of empty ("zero
prong") events in this sample. In addition the empty target events 
are mostly of the type p+C and p+air and produce pion yields in the
forward (projectile) hemisphere which are very similar to p+p collisions.
These points make it possible to extract the cross sections from full
target runs alone and to treat the empty target contribution as a small
correction. 
Detailed studies have shown that this correction, which is about
3--4\%, is the same for $\pi^+$ and $\pi^-$, has no measurable
$p_T$ dependence, and only a slight variation with $x_F$. The resulting
correction, defined as the ratio of the cross section measured with
the proper full-empty subtraction to the one extracted from full target data
only, is shown in Fig.~\ref{etcor}.

\subsection{Trigger Bias Correction
}
\begin{figure}[h]
\centering
\epsfig{file=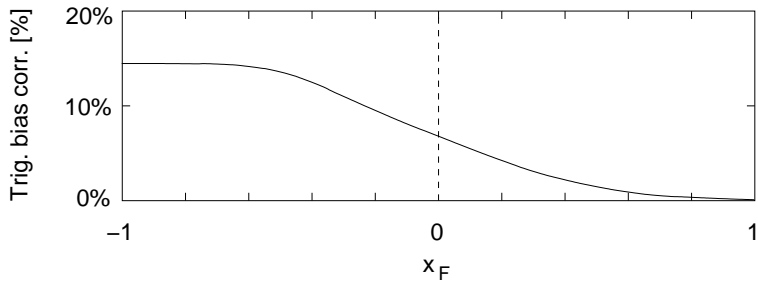,width=10cm}
\caption{Qualitative expectation of trigger bias correction as a function of $x_F$.
}
\label{s4exp}
\end{figure}

\vspace{3mm}
As explained in Section~3.3 the interaction trigger defined by the 
scintillator S4 accepts only 85.6\% of the total inelastic cross section.
This bias will reflect into the measured cross sections via the expression
\begin{equation}
        f_{meas} \sim \sigma_{trig} \cdot \frac{\Delta n}{N_{ev}}
\end{equation}
in a non-trivial way. It will be zero for all event topologies with no
hit of S4, i.e. for the far forward region where -- once a particle
is detected there -- no further particle can reach S4 by energy-momentum
conservation. On the other hand, as it has been shown that particle yields
completely decouple between the forward and the backward hemispheres \cite{bob},
the loss of 14.4\% should be entirely felt in the backward region. Therefore a
dependence on $x_F$ as illustrated in Fig.~\ref{s4exp} may be expected,
whereas short range correlations should modify this prediction in the forward
hemisphere.

Detailed correction tables are obtained experimentally by increasing
the diameter of the S4 counter off-line and
extrapolating the observed change in cross section to surface zero. The
method resembles the technique used in transmission experiments and 
relies entirely on measured quantities. This is important to note as
no event generator codes can be expected to be reliable at the level
of precision required here. The systematic error from this method is dominated 
by the statistical error of the evaluation, typically a factor of three 
smaller than the statistical error of the extracted data. 

\begin{figure}[h]
\centering
\epsfig{file=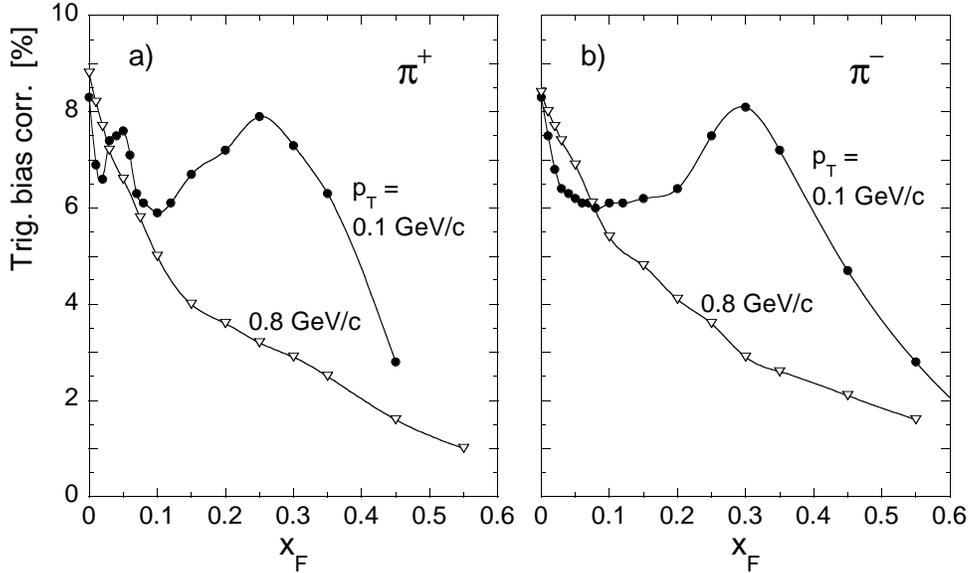,width=14cm}
\caption{Trigger bias correction as a function of $x_F$ at various $p_T$
for a) $\pi^+$ and b) $\pi^-$.
}
\label{s4}
\end{figure}

The resulting trigger bias correction over the kinematic range
of the experiment as presented in Fig.~\ref{s4}a and b follows indeed the 
expectation for $p_T$ values above about 0.6~GeV/c. For smaller $p_T$ a significant
micro-structure in the $x_F$ dependence appears which is a reflection of
hadronic two-body correlations driven by resonance decay (see also the
discussion in Section~10).

\subsection{Re-interaction in the Liquid Hydrogen Target
}
\begin{figure}[h]
\centering
\epsfig{file=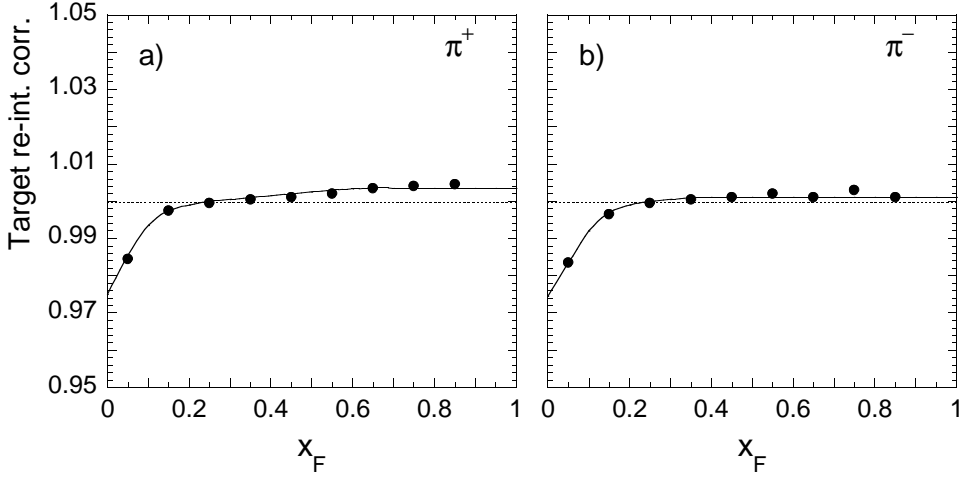,width=14cm}
\caption{Correction factor applied accounting for the target re-interaction 
of produced particles.
}
\label{tr}
\end{figure}

\vspace{3mm}
The produced particles will travel through the target material downstream
of the primary interaction. Given the 2.8\% interaction length of the
target for protons, a secondary interaction will occur in about 1.4\%
of the cases for produced hadrons. The correction for this effect is 
evaluated using the PYTHIA event generator \cite{sjo}, assuming that all daughter particles
from these secondary interactions are reconstructed at the primary vertex. 
As hadronic interactions will also produce pions, not only make them vanish,
the correction factor is $>1.0$ in the high $x_F$ region and $<1.0$ in the low
$x_F$ region where pion production dominates. The $x_F$ dependence of the 
correction is shown in Fig.~\ref{tr}.

\subsection{Absorption in Detector Material
}
\begin{figure}[h]
\centering
\epsfig{file=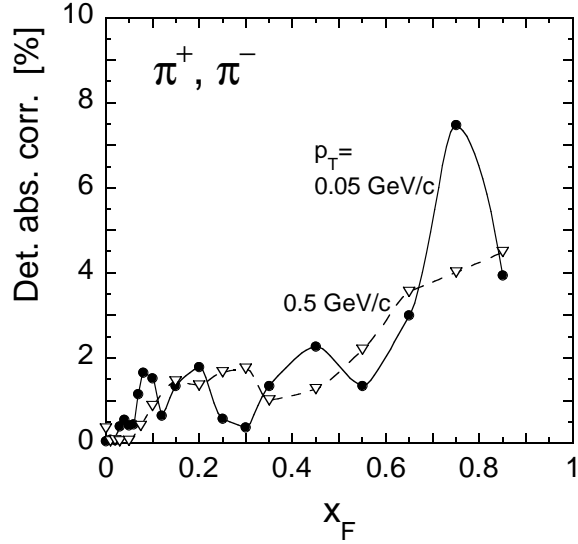,width=16cm}
\caption{Correction due to absorption of produced pions in the downstream
detector material.
}
\label{da}
\end{figure}

\vspace{3mm}
The correction for pions interacting in the downstream material of the
detector is determined using the GEANT simulation of the NA49
detector. Based on eye-scans it is assumed that all primary pions
undergoing hadronic interactions before detection are lost. This
assumption largely simplifies the analysis, and given the small value
of the correction even in the critical regions of phase space, it introduces
only a small systematic error. The absorption correction as a function
of $x_F$ at two $p_T$ values is shown in Fig.~\ref{da}.
At low $p_T$ the $x_F$ dependence exhibits multiple maxima which correspond
to the position of the ceramic support tubes of the TPC field cages \cite{nim}. 

\subsection{Pion Weak Decay
} 
\begin{figure}[h]
\centering
\epsfig{file=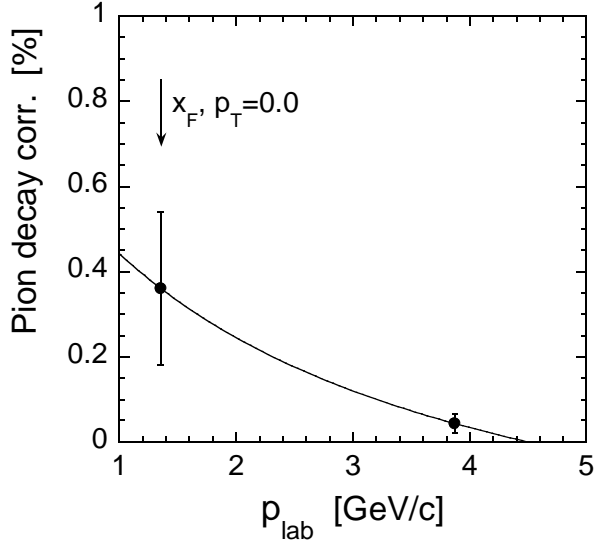,width=16cm}
\caption{Correction (close to $1/p$) due to the weak decay of pions.
}
\label{pd}
\end{figure}

\vspace{3mm}
Due to the sizeable decay length of pions in the weak channel 
$\pi \rightarrow \mu+\nu_{\mu}$
only a small fraction of pions will decay on their way through the
detector, ranging from a maximum of 3.6\% at the smallest detected 
$p_{lab}$ at $x_F$ and $p_T=0.0$ to less than 0.5\% for $p_{lab} > 10$~GeV/c. The
bulk of these decays will cause neither a loss of the particle track
nor a misidentification problem. The dip angle of the muon track which is
most critical for its reconstruction at the primary vertex deviates
less than $1^{\circ}$ from the pion direction at the lowest energy, and
the muon takes on average about 80\% of the pion momentum. Muons from
decays in front of the detector in the field-free or stray field regions
are therefore reconstructed at vertex. The same applies for decays 
downstream of VTPC-2. The only sample contributing to track losses 
corresponds to decays inside the magnetic field before the primary
track has left the necessary 30 points for analysis and where the 
secondary track escapes reconstruction. 
The tracking inefficiency for this sample, which amounts to 1.5\% of
pions at the lowest detected momentum, is determined to about
20\% $\pm$ 10\%, by making use of the total fraction of tracks with
less than about 30 points in the low momentum range. This low
inefficiency was confirmed by detailed eye-scans, showing that
secondary interactions or kaon decays are the primary source of these
short tracks, and not pion weak decays.
The resulting correction decreases rapidly with increasing
laboratory momentum from a maximum value of 0.36\% at $x_F$ and $p_T=0$ as shown 
in Fig.~\ref{pd}.

As far as particle identification is concerned, muons appear shifted 
upwards by about one standard deviation from the pion position in the 
energy loss distribution. This small deviation is absorbed in the
independent fit of position and width of the pion peak in each bin.

\subsection{Feed-down to Pions from Weak Decays of Strange Particles
}
\vspace{3mm}
It may in principle be a matter of discussion whether decay pions from
weak decays of strange particles should be subtracted or counted into
the total sample. In bubble chamber experiments with usually a small
detector size, most decays escape detection via too long decay lengths.
For decays inside the fiducial volume, secondary vertices are readily 
detected and the corresponding decay particles eliminated from the sample.
For counter experiments the situation is less clear. In fixed target
geometry the detectors are long enough to see a sizeable fraction of the
decay daughters and vertex fitting is usually not precise enough to
eliminate all secondary vertices. In the early collider experiments
at the ISR one can start with the assumption that almost all decay
products are counted into the track sample. In fact none of the experiments
compared to the present data, see Section~8, has attempted a feed-down
correction. As feed-down pions are mostly concentrated at low $p_T$ and low $x_F$
the case is saved by the fact that these experiments usually have no
acceptance in these regions.

In view of this situation and as the present experiment has full acceptance
in the critical areas of phase space, a complete feed-down subtraction is
performed by considering all relevant sources, i.e. K$^0_s, \Lambda, 
\Sigma^+, \Sigma^-, \bar{\Lambda}$. The main source of systematic uncertainty
is the very limited knowledge of the corresponding production cross
sections.

The correction is determined in three steps. First, the double-differential
parent distributions are adjusted to existing data. With this input, in the
second step, a decay Monte Carlo is used to predict yields of daughter
particles in the $x_F/p_T$ bins of the experiment. These yields are, in the
third step, folded with the reconstruction efficiency which is obtained
from a detailed NA49 detector simulation using complete generated events
containing the appropriate strange hadrons.

\begin{figure}[h]
\centering
\epsfig{file=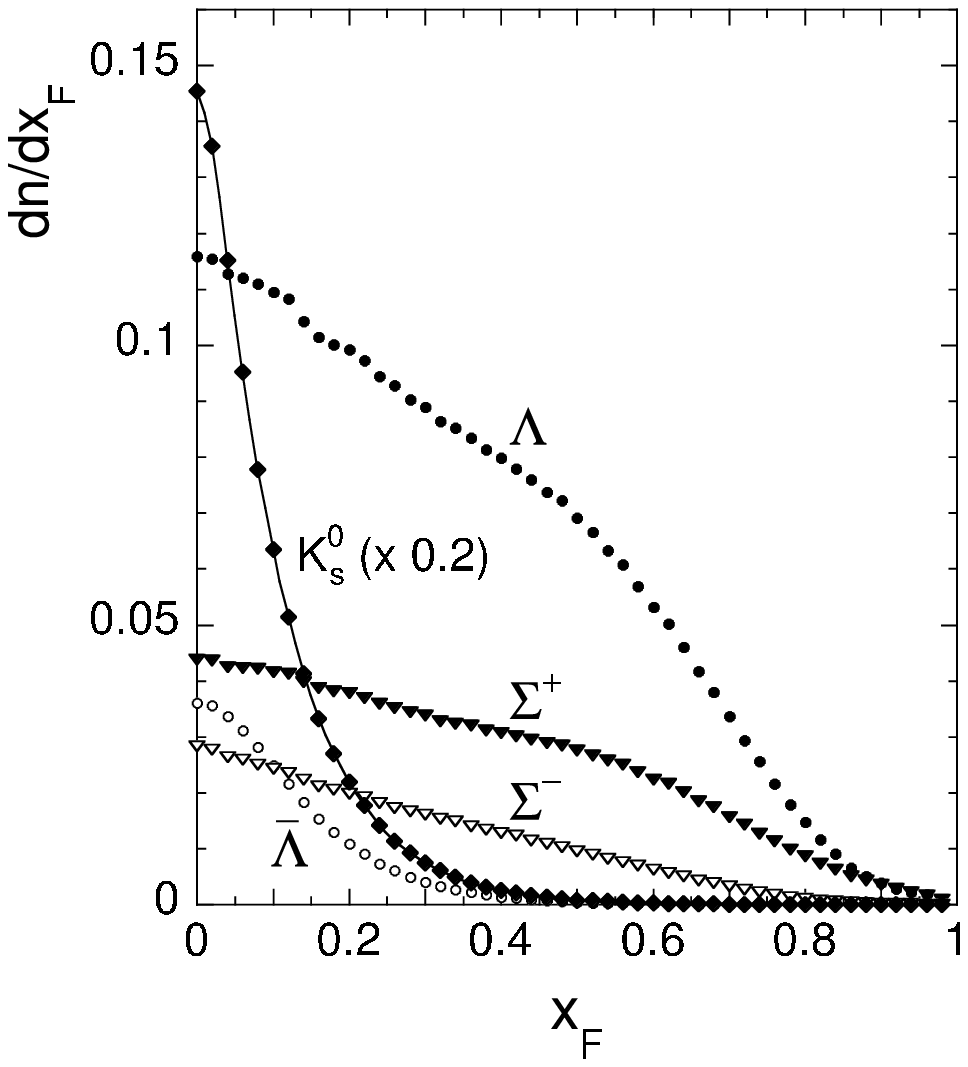,width=7cm}
\caption{$p_T$ integrated density distribution $dn/dx_F$ of parent particles
contributing to the feed-down correction.
}
\label{fd-dndxf}
\end{figure}

The $p_T$ integrated density distributions of parent particles are shown in
Fig.~\ref{fd-dndxf} as a function of $x_F$. The corresponding $p_T$ distributions are
extracted for K$^0_s$ from averaged charged kaon data and for $\Lambda$ from
a combined set of bubble chamber, NA49 and ISR data \cite{fdd}.

The reconstruction efficiency for the daughter pions, determined from
a full GEANT simulation of the NA49 detector using VENUS for the input
distributions, is defined as the ratio of reconstructed vertex tracks in a
given analysis bin to the daughter tracks emitted to the same
bin. As such, this ratio can reach values
above 1.0. It has been verified that the efficiency is only
depending weakly on the input parent-particle distribution, and can
therefore be used as a multiplicative factor.

\begin{figure}[h]
\centering
\epsfig{file=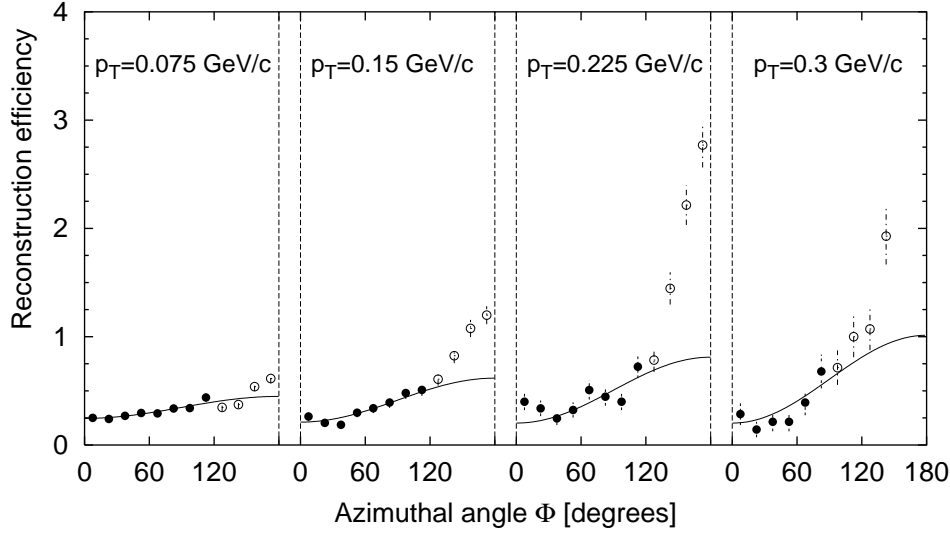,angle=-90,width=13cm}
\caption{Reconstruction efficiency for pions with $p_{lab}=3.2$~GeV/c from $\Lambda$ decay
as a function of $\Phi$ for various $p_T$ values. 
The $\Phi$ region chosen for pion extraction is indicated by full symbols, 
where $\Phi=0$ is in the horizontal plane.
}
\label{fd-eff}
\end{figure}

The average reconstruction efficiency reaches levels of up to 50\% 
in the low $x_F$ region. It is rather independent of $p_T$ but shows a strong variation
with the azimuthal angle which is due to biases in the fitting of 
decay products to the primary vertex. This is exemplified in Fig.~\ref{fd-eff}
which shows the azimuthal dependence for Lambda decay to pions with 
$p_{lab}=3.2$~GeV/c in different bins of $p_T$. This observation
leads to the restriction of the $\Phi$ window in bins which have otherwise
full acceptance for vertex tracks (see Section 3.6). 

\begin{figure}[h]
\centering
\epsfig{file=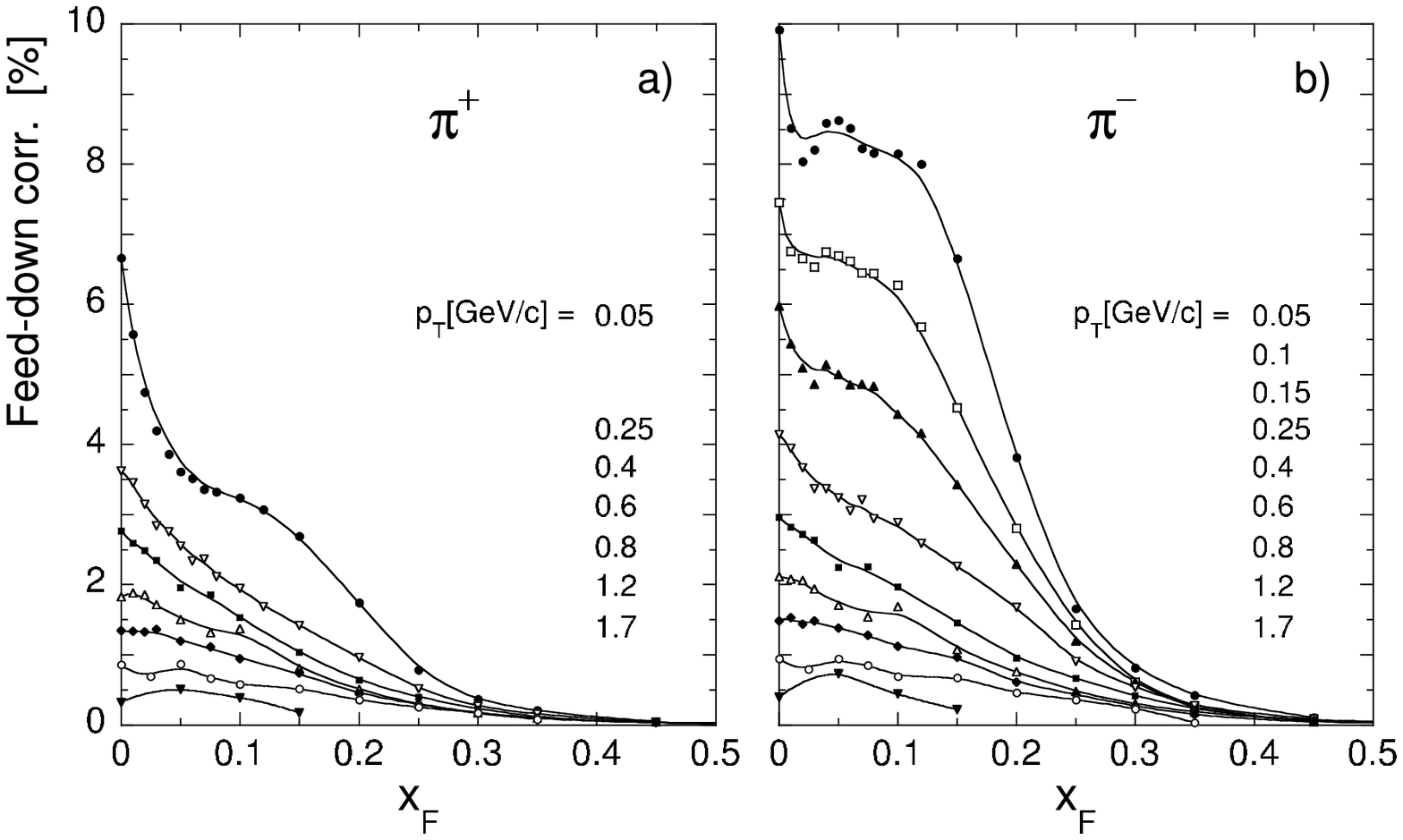,width=14cm}
\caption{Correction of feed-down to pions from weak decays for a) $\pi^+$ and b) $\pi^-$.
}
\label{fdd}
\end{figure}

The resulting total feed-down correction for $\pi^+$ and $\pi^-$ is presented in 
Fig.~\ref{fdd} as a function of $x_F$ for various $p_T$ bins. 
Apparently this correction is very sizeable especially at low $p_T$ and
extends rather far up in $x_F$ with a complex overall structure.
Therefore it is no question that great care has to be taken in comparing
experiments with undefined feed-down treatment in these areas.

\subsection{Binning Correction
}
\vspace{3mm}
The extraction of invariant cross sections has by necessity to be performed
in finite bins of phase space. The measured yield is therefore the integral
of the particle density over the phase space volume element. As the density
distribution as a function of the  azimuthal angle is flat by symmetry, the 
problem reduces to a determination of binning corrections in $x_F$ and $p_T$.

Approximating the variation of particle density $\rho$ in the coordinate $t$ by an 
expansion in local derivatives,
$$ \rho(t) \approx \rho(t_0) + \rho'(t_0) (t-t_0) + \rho''(t_0) \frac{(t-t_0)^2}{2} $$
the measured value corresponding to the bin center $t_0$ is
$$ \rho_{meas}(t_0)= \frac{1}{\Delta} \int_{t_0-\Delta/2}^{t_0+\Delta/2} \rho(t) \textrm{d} t 
\approx \rho(t_0) + \frac{1}{24} \rho''(t_0) \Delta^2 $$
up to second order terms, where $\Delta$ is the bin width. Hence the difference
between the real particle density at $t_0$ and its measured value is proportional to the
second derivative of the density function and to the square of the bin
width. This approximation holds if the difference does not exceed
the level of a few percent.

For the functional forms discussed here the second derivative may be 
approximated from neighbouring data points by
$$ \rho''(t_0) \approx \left( \left( { \frac{\Delta_1 \rho(t_2) + \Delta_2 \rho(t_1)}{\Delta_1 + 
\Delta_2}}\right) - \rho(t_0) \right) { \frac{2}{\Delta_1 \Delta_2}}~~,$$
where $\Delta_1=t_1-t_0$ and $\Delta_2=t_0-t_2$.

The generalization of the method to the case of double differential
cross sections is straightforward and it can be shown that the correction
can be determined independently in the two coordinates.

The above consideration defines the binning correction, which can be
evaluated for all data points if statistics permits an estimation of
the second derivative. The statistical uncertainty induced by the
correction, determined by the error of neighbouring points, is about a
factor of 10 lower than the error of the data points itself. The
direct application of the correction makes it model or parametrization
independent.

\begin{figure}[h]
\centering
\epsfig{file=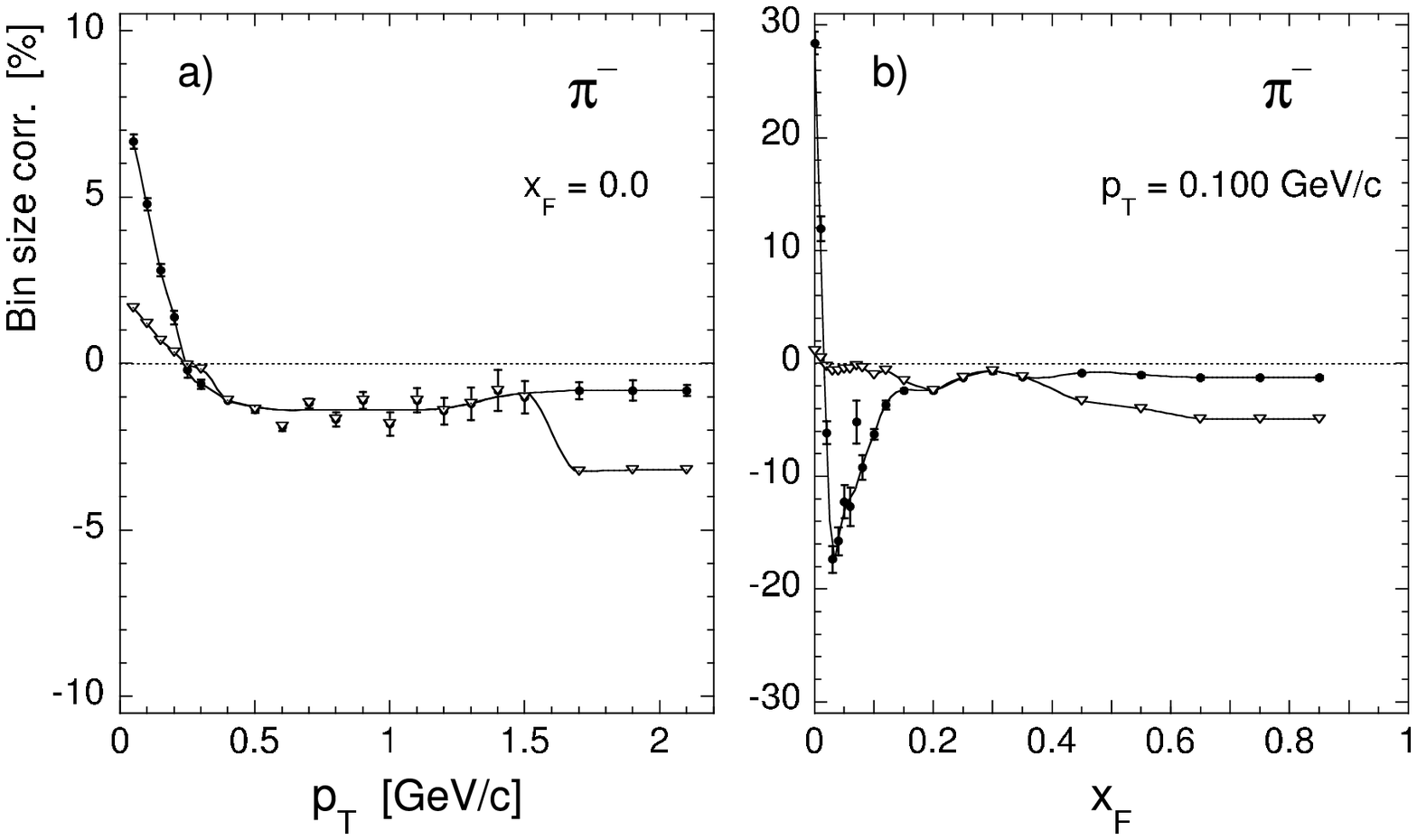,width=15cm}
\caption{Correction due to the binning in a) $p_T$ and b) $x_F$. Full circles 
represent the correction for a fixed bin of $\Delta p_T=0.1$~GeV/c and $\Delta x_F=0.05$,
respectively; open triangles describe the correction for the bin sizes actually used.
}
\label{bs}
\end{figure}

The correction is on the 1--4\% level as shown in Fig.~\ref{bs} for some typical
$p_T$ and $x_F$ dependences. In fact, one of the important aspects in the 
elaboration of the binning scheme, see Section 3.7, is to keep this 
correction low in the high statistics regions in order to capture
the structures of the density function in an optimum fashion.
The remaining systematic error from neglecting higher terms is below
0.5\% as verified by Monte Carlo studies. 

\section{Systematic Errors
}
\vspace{3mm}
In addition to the overall normalization uncertainty, the corrections
discussed in Section~5 introduce systematic errors which are
estimated by allowing appropriate error bands for the underlying
physics inputs. The corresponding values are given in Table~\ref{tsyserr}. By summing
up all contributions, an upper limit of 4.8\% for the total systematic uncertainty can be
claimed.

\begin{table}[h]
\begin{center}
\begin{tabular}{|l|r|}
\hline
Normalization                                &   1.5~\%      \\
\hline
Tracking efficiency                          &   0.5~\%      \\
\hline
Trigger bias                                 &   0.5~\%      \\
\hline
Feed-down                                     &   0.5--1.5~\%      \\
\hline
Detector absorption                          &               \\
Pion decay $\pi \rightarrow \mu+\nu_{\mu}$   &   0.5~\%      \\
Re-interaction in target                      &               \\
\hline
Binning                                      &   0.3~\%      \\
\hline
\hline
Total (upper limit)                          &   4.8~\%      \\
\hline
Total (quadratic sum)                        &   2.0~\%      \\
\hline
\end{tabular}
\end{center}
\vspace{-2mm}
\caption{Summary of systematic errors.
}
\label{tsyserr}
\end{table}

It should be noted that in the high $x_F$ region where certain assumptions
have to be made on K$^+/\pi^+$ and K$^-/\pi^-$ ratios for the purpose of particle
identification, additional systematic uncertainties of between 1 and
3\% have to be added.

\begin{figure}[b]
\centering
\epsfig{file=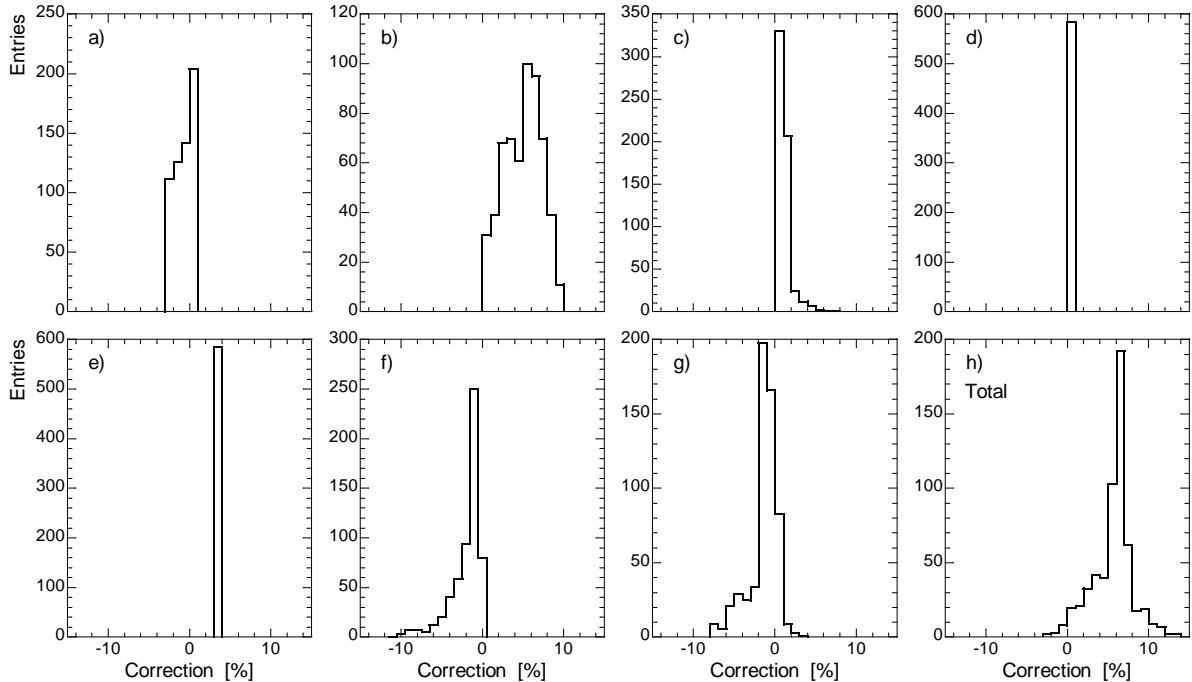,width=16cm}
\caption{Distribution of correction for a) target re-interaction, b) trigger bias,
c) absorption in detector material, d) pion decay, e) empty target contribution,
f) feed-down, g) binning, and h) resulting total correction.
}
\label{fsyserr}
\end{figure}

Further information on the error sources is contained in Fig.~\ref{fsyserr}
which gives the  distributions of each single correction
and of the resulting total correction for the 589 data points of this 
experiment. It appears that the individual corrections compensate to
a large extent in the overall result so that only few points had
to be corrected by more than 10\%. This might give some support 
to the estimation of the total systematic error as the sum of the 
squares to only 2\%. This low value has to be regarded with some 
caution since the authors are the first to recognize that this way 
of minimizing possible systematic uncertainties has been in the past, one 
of the major difficulties in the quantitative comparison of hadronic cross sections.
Finally, only comparison between different independent measurements
can help to understand the problem of systematic error propagation.
This comparison will be attempted in Sections~8 and 9 below.

\section{Results
}
\vspace{3mm}
The following chapter presents the double differential invariant cross 
sections for charged pions resulting from the event sample and from
the extraction and correction procedures discussed above. The basic numerical 
information is summarized in data tables. The completeness and the small 
statistical and systematic errors of these data allow the identification of a complex 
structure in their dependence on the kinematic variables which cannot 
be parametrized by straight-forward arithmetic expressions. A numerical 
interpolation scheme has therefore been developed. 
The resulting curves are presented 
together with the data points in $p_T$ distributions at fixed $x_F$, in $x_F$ 
distributions at fixed $p_T$, and in corresponding $\pi^+/\pi^-$ distributions. 
Finally, the cross sections are also shown in their dependence on rapidity $y$
and transverse mass $m_T$.

\subsection{Data Tables
}
\vspace{3mm}
The following Tables summarize the double differential invariant cross sections
as a function of $x_F$ and $p_T$ for the binning scheme discussed in Section
3.7 above. This scheme results in 281 measured values for $\pi^+$ (Table~\ref{tpi+}) 
and 308 values for $\pi^-$ (Table~\ref{tpi-}) production, the difference for 
the two charges being
due to the limitation of $\pi^+$ identification in the region above $x_F=0.55$  
(see Section 4.5).


\begin{table}[p]
\renewcommand{\tabcolsep}{0.10pc} 
\renewcommand{\arraystretch}{1.05} 
\begin{center}
{\scriptsize
\begin{tabular}{|l|cc|cc|cc|cc|cc|cc|cc|cc|cc|cc|}
\hline
\multicolumn{21}{|c|}{$f(x_F,p_T)~~~~\Delta f$}               \\
\hline
\raisebox{-0.2ex}{$p_T$}$\backslash$\raisebox{0.2ex}{$x_F$} 
& \multicolumn{2}{c|}{0.0}  & \multicolumn{2}{c|}{0.01}
& \multicolumn{2}{c|}{0.02} & \multicolumn{2}{c|}{0.025}
& \multicolumn{2}{c|}{0.03} & \multicolumn{2}{c|}{0.04}
& \multicolumn{2}{c|}{0.05} & \multicolumn{2}{c|}{0.06}
& \multicolumn{2}{c|}{0.07} & \multicolumn{2}{c|}{0.075}                          \\
\hline
0.050
& 62.87    &   0.61 & 62.79    &   0.66 & 59.44    &   0.65 &          &        & 51.53    &   0.63
& 43.29    &   0.63 & 37.64    &   0.75 & 32.64    &   0.86 & 29.78    &   0.98 &          &         \\
\hline
0.100
& 59.66    &   0.58 & 58.91    &   0.62 & 54.55    &   0.48 &          &        & 51.49    &   0.54
& 45.19    &   0.54 & 41.09    &   0.63 & 36.46    &   0.73 & 32.92    &   0.81 &          &         \\
\hline
0.150
& 51.20    &   0.55 & 50.87    &   0.58 & 47.98    &   0.55 &          &        & 45.66    &   0.58
& 42.21    &   0.48 & 39.68    &   0.54 & 36.39    &   0.59 & 34.55    &   0.67 &          &         \\
\hline
0.200
& 41.42    &   0.65 & 40.78    &   0.49 & 39.58    &   0.51 &          &        & 37.37    &   0.56
& 35.33    &   0.54 & 33.69    &   0.50 & 31.49    &   0.58 & 29.16    &   0.63 &          &         \\
\hline
0.250
& 32.00    &   0.57 & 31.93    &   0.53 & 30.47    &   0.55 &          &        & 29.47    &   0.61
& 27.77    &   0.53 & 26.54    &   0.60 & 25.04    &   0.68 & 23.80    &   0.63 &          &         \\
\hline
0.300
& 24.11    &   0.49 & 23.89    &   0.62 & 23.03    &   0.60 &          &        & 22.00    &   0.67
& 21.35    &   0.69 & 20.37    &   0.62 & 19.44    &   0.71 & 18.71    &   0.68 &          &         \\
\hline
0.400
& 13.20    &   0.57 & 13.21    &   0.57 & 12.76    &   0.58 &          &        & 12.11    &   0.61
&          &        & 11.56    &   0.43 &          &        &          &        & 10.89    &   0.35  \\
\hline
0.500
&  7.212   &   0.77 &  7.297   &   0.78 &  7.046   &   0.80 &          &        &  6.690   &   0.85
&          &        &  6.423   &   0.59 &          &        &          &        &  6.033   &   0.62  \\
\hline
0.600
&  4.102   &   1.02 &  3.920   &   1.07 &  3.923   &   1.06 &          &        &  3.723   &   1.11
&          &        &  3.585   &   0.77 &          &        &          &        &  3.356   &   0.79  \\
\hline
0.700
&  2.219   &   1.41 &  2.233   &   1.40 &  2.170   &   1.43 &          &        &  2.027   &   1.52
&          &        &  2.007   &   1.01 &          &        &          &        &  1.924   &   1.11  \\
\hline
0.800
&  1.213   &   1.86 &  1.259   &   1.87 &  1.183   &   1.89 &          &        &  1.148   &   1.94
&          &        &  1.156   &   1.30 &          &        &          &        &  1.082   &   1.38  \\
\hline
0.900
&  0.689   &   2.49 &  0.681   &   2.51 &  0.702   &   2.43 &          &        &  0.692   &   2.47
&          &        &  0.652   &   1.96 &          &        &          &        &  0.607   &   1.93  \\
\hline
1.000
&  0.386   &   3.31 &  0.397   &   3.22 &  0.383   &   3.33 &          &        &  0.383   &   3.32
&          &        &  0.379   &   2.54 &          &        &          &        &  0.368   &   2.62  \\
\hline
1.100
&  0.248   &   2.60 &          &        &          &        &  0.236   &   3.48 &          &       
&          &        &  0.223   &   3.35 &          &        &          &        &  0.219   &   3.42  \\
\hline
1.200
&  0.136   &   3.54 &          &        &          &        &  0.135   &   4.57 &          &       
&          &        &  0.120   &   4.32 &          &        &          &        &  0.128   &   4.17  \\
\hline
1.300
&  0.0797  &   4.56 &          &        &          &        &  0.0750  &   6.07 &          &       
&          &        &  0.0729  &   5.55 &          &        &          &        &  0.0734  &   5.67  \\
\hline
1.400
&  0.0482  &   5.99 &          &        &          &        &  0.0448  &   8.04 &          &       
&          &        &  0.0443  &   6.97 &          &        &          &        &  0.0473  &   6.58  \\
\hline
1.500
&  0.0296  &   7.61 &          &        &          &        &  0.0277  &   9.96 &          &       
&          &        &  0.0302  &   8.45 &          &        &          &        &  0.0266  &   9.20  \\
\hline
1.700
&  0.0122  &   6.14 &          &        &          &        &          &        &          &       
&          &        &  0.0115  &   6.76 &          &        &          &        &          &         \\
\hline
1.900
&  0.00406 &  10.2  &          &        &          &        &          &        &          &       
&          &        &  0.00373 &  11.9 &          &        &          &        &          &         \\
\hline
2.100
&  0.00172 &  15.4  &          &        &          &        &          &        &          &       
&          &        &  0.00149 &  18.8 &          &        &          &        &          &         \\
\hline
\hline
\raisebox{-0.2ex}{$p_T$}$\backslash$\raisebox{0.2ex}{$x_F$} 
& \multicolumn{2}{c|}{0.08} & \multicolumn{2}{c|}{0.1}
& \multicolumn{2}{c|}{0.12} & \multicolumn{2}{c|}{0.15}
& \multicolumn{2}{c|}{0.2}  & \multicolumn{2}{c|}{0.25}
& \multicolumn{2}{c|}{0.3}  & \multicolumn{2}{c|}{0.35}
& \multicolumn{2}{c|}{0.45} & \multicolumn{2}{c|}{0.55}                          \\
\hline
0.050
& 27.71    &   1.07 & 24.05    &   0.92 & 22.01    &   1.04 & 19.13    &   0.92 & 17.35    &   1.08
& 15.80    &   1.57 & 11.80    &   2.06 &  7.690   &   2.75 &  3.720   &   3.18 &          &         \\
\hline
0.100
& 29.30    &   0.91 & 25.11    &   0.77 & 22.98    &   0.89 & 20.40    &   0.78 & 16.72    &   0.93
& 14.59    &   1.35 & 11.23    &   1.81 &  6.811   &   2.11 &  3.734   &   2.23 &          &         \\
\hline
0.150
& 30.72    &   0.75 & 26.17    &   0.63 & 23.42    &   0.75 & 20.15    &   0.65 & 16.16    &   0.78
& 12.87    &   1.16 &  9.319   &   1.68 &  5.970   &   1.82 &  3.266   &   1.94 &          &         \\
\hline
0.200
& 27.80    &   0.67 & 24.38    &   0.59 & 21.57    &   0.67 & 18.40    &   0.59 & 13.64    &   0.73
& 10.84    &   1.10 &  7.535   &   1.57 &  5.169   &   1.67 &  2.812   &   1.82 &  1.683   &   2.32  \\
\hline
0.250
& 22.94    &   0.66 & 20.16    &   0.59 & 18.06    &   0.67 & 15.33    &   0.57 & 11.59    &   0.71
&  8.601   &   1.17 &  6.478   &   1.55 &  4.354   &   1.62 &  2.435   &   1.75 &          &         \\
\hline
0.300
& 18.18    &   0.71 & 16.19    &   0.60 & 14.57    &   0.67 & 12.59    &   0.58 &  9.413   &   0.72
&  7.164   &   1.14 &  5.146   &   1.57 &  3.659   &   1.61 &  1.970   &   1.79 &  1.198   &   2.18  \\
\hline
0.400
&          &        & 10.00    &   0.42 &          &        &  8.047   &   0.40 &  6.189   &   0.61
&  4.754   &   0.85 &  3.514   &   1.16 &  2.590   &   1.18 &  1.388   &   1.29 &  0.853   &   3.58  \\
\hline
0.500
&          &        &  5.749   &   0.50 &          &        &  5.026   &   0.46 &  4.040   &   0.68
&  3.204   &   0.93 &  2.367   &   1.25 &  1.812   &   1.28 &  0.949   &   1.37 &  0.538   &   3.94  \\
\hline
0.600
&          &        &  3.186   &   0.64 &          &        &  2.874   &   0.65 &  2.561   &   0.77
&  2.071   &   1.06 &  1.681   &   1.38 &  1.313   &   1.38 &  0.660   &   1.49 &  0.349   &   4.59  \\
\hline
0.700
&          &        &  1.770   &   1.02 &          &        &  1.623   &   0.83 &  1.452   &   1.02
&  1.272   &   1.25 &  1.006   &   1.63 &  0.863   &   1.58 &  0.482   &   1.67 &  0.227   &   5.29  \\
\hline
0.800
&          &        &  1.027   &   1.45 &          &        &  0.905   &   1.04 &  0.806   &   1.29
&  0.697   &   1.58 &  0.608   &   1.98 &  0.490   &   1.96 &  0.305   &   1.98 &  0.148   &   6.20  \\
\hline
0.900
&          &        &  0.576   &   2.03 &          &        &  0.516   &   1.35 &  0.471    &  1.63 
&  0.398   &   2.01 &  0.304   &   2.64 &  0.281   &   2.47 &  0.188   &   2.39 &  0.0956   &  7.24  \\
\hline
1.000
&          &        &  0.356   &   2.40 &          &        &  0.298   &   1.71 &  0.252   &   2.14
&  0.219   &   2.56 &  0.181   &   3.27 &  0.158   &   3.14 &  0.101   &   3.10 &  0.0592  &   8.74  \\
\hline
1.100
&          &        &  0.203   &   3.67 &          &        &  0.176   &   2.41 &  0.155   &   2.72
&  0.130   &   3.18 &  0.113   &   3.98 &  0.0802  &   4.16 &  0.0514  &   4.07 &  0.0326  &   7.89  \\
\hline
1.200
&          &        &  0.125   &   4.27 &          &        &  0.106   &   3.36 &  0.0876  &   3.49
&  0.0739  &   4.11 &  0.0555  &   5.46 &  0.0497  &   5.09 &  0.0290  &   5.27 &          &         \\
\hline
1.300
&          &        &  0.0701  &   6.07 &          &        &  0.0653  &   4.54 &  0.0530  &   4.47
&  0.0482  &   5.16 &  0.0405  &   6.20 &  0.0274  &   6.74 &  0.0189  &   6.38 &  0.00724 &  15.1   \\
\hline
1.400
&          &        &  0.0416  &   7.65 &          &        &  0.0405  &   5.59 &  0.0295  &   7.45
&  0.0246  &   6.72 &  0.0234  &   7.92 &  0.0169  &   8.28 &  0.0116  &   7.74 &          &         \\
\hline
1.500
&          &        &  0.0290  &   9.23 &          &        &  0.0258  &   7.49 &  0.0204  &   7.79
&  0.0179  &   9.45 &  0.0138  &   4.93 &  0.0110  &   7.01 &  0.00612 &   8.78 &  0.00214 &  26.7   \\
\hline
1.700
&          &        &  0.00981 &   7.56 &          &        &  0.0114  &   7.78 &  0.00720 &   9.78
&  0.00615 &  10.5  &  0.00527 &   7.69 &  0.00448 &  10.5  &  0.00315 &  11.8  &          &         \\
\hline
1.900
&          &        &  0.00339 &  12.7  &          &        &  0.00374 &  13.9  &  0.00323 &  13.7
&          &        &  0.00273 &  10.3  &  0.00209 &  17.7  &          &        &          &         \\
\hline
2.100
&          &        &          &        &          &        &  0.00215 &  17.3  &  0.00151 &  22.6
&          &        &          &        &  0.00110 &  24.5  &          &        &          &         \\
\hline
\end{tabular}
}
\end{center}
\vspace{-2mm}
\caption{Double differential invariant cross section $f(x_F,p_T)$ [mb/(GeV$^2$/c$^3$)] for $\pi^+$
produced in p+p interactions at 158~GeV/c. The statistical uncertainty $\Delta f$ is given in \%.
}
\label{tpi+}
\end{table}


\begin{table}[p]
\renewcommand{\tabcolsep}{0.10pc} 
\renewcommand{\arraystretch}{1.05} 
\begin{center}
{\scriptsize
\begin{tabular}{|l|cc|cc|cc|cc|cc|cc|cc|cc|cc|cc|}
\hline
\multicolumn{21}{|c|}{$f(x_F,p_T)~~~~\Delta f$}               \\
\hline
\raisebox{-0.2ex}{$p_T$}$\backslash$\raisebox{0.2ex}{$x_F$} 
& \multicolumn{2}{c|}{0.0}  & \multicolumn{2}{c|}{0.01}
& \multicolumn{2}{c|}{0.02} & \multicolumn{2}{c|}{0.025}
& \multicolumn{2}{c|}{0.03} & \multicolumn{2}{c|}{0.04}
& \multicolumn{2}{c|}{0.05} & \multicolumn{2}{c|}{0.06}
& \multicolumn{2}{c|}{0.07} & \multicolumn{2}{c|}{0.075}                          \\
\hline
0.050
& 59.19    &   0.52 & 55.36    &   0.57 & 49.58    &   0.69 &          &        & 41.35    &   0.70
& 35.72    &   0.67 & 30.52    &   0.81 & 26.76    &   0.93 & 23.79    &   1.05 &          &         \\
\hline
0.100
& 54.40    &   0.48 & 52.21    &   0.52 & 47.69    &   0.49 &          &        & 41.32    &   0.57
& 35.95    &   0.59 & 31.55    &   0.71 & 27.53    &   0.81 & 24.76    &   0.91 &          &         \\
\hline
0.150
& 46.74    &   0.47 & 45.73    &   0.48 & 41.85    &   0.57 &          &        & 37.78    &   0.63
& 34.21    &   0.51 & 30.42    &   0.61 & 27.50    &   0.68 & 24.47    &   0.77 &          &         \\
\hline
0.200
& 37.85    &   0.53 & 36.91    &   0.51 & 34.65    &   0.55 &          &        & 31.96    &   0.62
& 29.26    &   0.57 & 26.45    &   0.55 & 24.03    &   0.65 & 22.16    &   0.71 &          &         \\
\hline
0.250
& 29.00    &   0.59 & 28.74    &   0.55 & 26.68    &   0.59 &          &        & 25.41    &   0.66
& 23.56    &   0.58 & 21.70    &   0.68 & 19.90    &   0.74 & 18.51    &   0.69 &          &         \\
\hline
0.300
& 22.02    &   0.68 & 22.25    &   0.67 & 20.62    &   0.67 &          &        & 19.51    &   0.70
& 18.26    &   0.75 & 16.87    &   0.69 & 15.98    &   0.79 & 14.91    &   0.71 &          &         \\
\hline
0.400
& 12.28    &   0.61 & 12.05    &   0.62 & 11.63    &   0.55 &          &        & 10.75    &   0.59
&          &        & 10.05    &   0.45 &          &        &          &        &  8.853   &   0.39  \\
\hline
0.500
&  6.673   &   0.80 &  6.371   &   0.83 &  6.296   &   0.84 &          &        &  5.946   &   0.91
&          &        &  5.697   &   0.63 &          &        &          &        &  5.069   &   0.67  \\
\hline
0.600
&  3.498   &   1.10 &  3.542   &   1.11 &  3.507   &   1.12 &          &        &  3.261   &   1.20
&          &        &  3.147   &   0.81 &          &        &          &        &  2.849   &   0.86  \\
\hline
0.700
&  2.016   &   1.44 &  1.959   &   1.47 &  1.893   &   1.51 &          &        &  1.852   &   1.55
&          &        &  1.705   &   1.06 &          &        &          &        &  1.616   &   1.19  \\
\hline
0.800
&  1.098   &   1.95 &  1.089   &   1.96 &  1.101   &   1.94 &          &        &  1.049   &   2.03
&          &        &  0.982   &   1.41 &          &        &          &        &  0.940   &   1.48  \\
\hline
0.900
&  0.641   &   2.51 &  0.616   &   2.59 &  0.622   &   2.57 &          &        &  0.570   &   2.80
&          &        &  0.567   &   2.08 &          &        &          &        &  0.529   &   2.02  \\
\hline
1.000
&  0.346   &   3.37 &  0.372   &   3.43 &  0.341   &   3.43 &          &        &  0.336   &   3.48
&          &        &  0.318   &   2.68 &          &        &          &        &  0.307   &   2.79  \\
\hline
1.100
&  0.209   &   3.09 &          &        &          &        &  0.215   &   3.64 &          &       
&          &        &  0.190   &   3.35 &          &        &          &        &  0.181   &   3.53  \\
\hline
1.200
&  0.122   &   3.67 &          &        &          &        &  0.117   &   4.91 &          &       
&          &        &  0.111   &   4.52 &          &        &          &        &  0.0982  &   5.38  \\
\hline
1.300
&  0.0784  &   4.64 &          &        &          &        &  0.0686  &   6.49 &          &       
&          &        &  0.0597  &   6.04 &          &        &          &        &  0.0582  &   6.39  \\
\hline
1.400
&  0.0469  &   5.91 &          &        &          &        &  0.0411  &   8.23 &          &       
&          &        &  0.0372  &   7.76 &          &        &          &        &  0.0329  &   8.41  \\
\hline
1.500
&  0.0242  &   8.33 &          &        &          &        &  0.0262  &  10.5  &          &       
&          &        &  0.0232  &   9.59 &          &        &          &        &  0.0212  &  10.5   \\
\hline
1.700
&  0.0102  &   6.25 &          &        &          &        &          &        &          &       
&          &        &  0.00785 &   8.09 &          &        &          &        &          &         \\
\hline
1.900
&  0.00399 &  10.3  &          &        &          &        &          &        &          &       
&          &        &  0.00344 &  12.2  &          &        &          &        &          &         \\
\hline
2.100
&  0.00124 &  27.4  &          &        &          &        &          &        &          &       
&          &        &  0.00162 &  17.7  &          &        &          &        &          &         \\
\hline
\hline
\raisebox{-0.2ex}{$p_T$}$\backslash$\raisebox{0.2ex}{$x_F$} 
& \multicolumn{2}{c|}{0.08} & \multicolumn{2}{c|}{0.1}
& \multicolumn{2}{c|}{0.12} & \multicolumn{2}{c|}{0.15}
& \multicolumn{2}{c|}{0.2}  & \multicolumn{2}{c|}{0.25}
& \multicolumn{2}{c|}{0.3}  & \multicolumn{2}{c|}{0.35}
& \multicolumn{2}{c|}{0.45} & \multicolumn{2}{c|}{0.55}                          \\
\hline
0.050
& 21.27    &   1.19 & 17.41    &   1.04 & 14.35    &   1.26 & 11.31    &   1.16 &  7.564   &   1.55
&  6.064   &   2.36 &  4.669   &   3.32 &  3.007   &   4.44 &  1.418   &   5.09 &          &         \\
\hline
0.100
& 21.39    &   1.06 & 17.17    &   0.90 & 14.49    &   1.08 & 11.40    &   1.02 &  7.652   &   1.33
&  5.802   &   1.97 &  4.429   &   2.94 &  2.964   &   3.17 &  1.494   &   3.52 &  0.528   &   5.91  \\
\hline
0.150
& 21.66    &   0.87 & 17.62    &   0.78 & 14.45    &   0.96 & 11.23    &   0.84 &  7.367   &   1.11
&  5.539   &   1.63 &  3.971   &   2.51 &  2.712   &   2.66 &  1.300   &   3.07 &          &         \\
\hline
0.200
& 20.30    &   0.79 & 16.41    &   0.71 & 13.62    &   0.87 & 10.63    &   0.76 &  6.743   &   1.00
&  4.839   &   1.51 &  3.453   &   2.30 &  2.191   &   2.57 &  1.191   &   2.79 &  0.550   &   4.20  \\
\hline
0.250
& 17.28    &   0.80 & 14.32    &   0.67 & 12.03    &   0.81 &  9.195   &   0.73 &  5.986   &   0.95
&  4.334   &   1.52 &  2.872   &   2.31 &  2.152   &   2.30 &  0.978   &   2.71 &          &         \\
\hline
0.300
& 13.89    &   0.77 & 12.27    &   0.70 & 10.08    &   0.75 &  8.226   &   0.71 &  5.279   &   0.92
&  3.795   &   1.47 &  2.538   &   2.21 &  1.689   &   2.37 &  0.875   &   2.28 &  0.432   &   3.90  \\
\hline
0.400
&          &        &  7.674   &   0.48 &          &        &  5.567   &   0.48 &  3.892   &   0.76
&  2.741   &   1.06 &  1.914   &   1.57 &  1.286   &   1.66 &  0.637   &   2.30 &  0.267   &   4.23  \\
\hline
0.500
&          &        &  4.524   &   0.57 &          &        &  3.485   &   0.54 &  2.646   &   0.83
&  1.908   &   1.14 &  1.366   &   1.66 &  0.927   &   1.75 &  0.429   &   2.49 &  0.194   &   4.46  \\
\hline
0.600
&          &        &  2.564   &   0.71 &          &        &  2.133   &   0.75 &  1.670   &   0.96
&  1.258   &   1.28 &  0.958   &   1.82 &  0.676   &   1.89 &  0.305   &   2.68 &  0.133   &   4.88  \\
\hline
0.700
&          &        &  1.492   &   1.15 &          &        &  1.225   &   0.92 &  1.010   &   1.20
&  0.788   &   1.50 &  0.600   &   2.13 &  0.441   &   2.18 &  0.211   &   2.99 &  0.0821  &   5.73  \\
\hline
0.800
&          &        &  0.867   &   1.60 &          &        &  0.686   &   1.18 &  0.590   &   1.49
&  0.474   &   1.84 &  0.364   &   2.55 &  0.265   &   2.63 &  0.135   &   3.55 &  0.0627  &   6.15  \\
\hline
0.900
&          &        &  0.472   &   2.25 &          &        &  0.401   &   1.48 &  0.328   &   1.90
&  0.270   &   2.29 &  0.226   &   3.07 &  0.165   &   3.15 &  0.0840  &   4.22 &  0.0345  &   7.78  \\
\hline
1.000
&          &        &  0.264   &   2.81 &          &        &  0.234   &   1.88 &  0.181   &   2.45
&  0.159   &   2.86 &  0.133   &   3.80 &  0.0971  &   3.89 &  0.0524  &   5.34 &  0.0185  &  10.1   \\
\hline
1.100
&          &        &  0.174   &   3.68 &          &        &  0.139   &   2.69 &  0.115   &   2.99
&  0.0880  &   3.71 &  0.0779  &   4.77 &  0.0553  &   4.90 &  0.0315  &   5.41 &  0.0126  &   8.19  \\
\hline
1.200
&          &        &  0.105   &   5.10 &          &        &  0.0803  &   3.65 &  0.0680  &   3.78
&  0.0514  &   4.67 &  0.0421  &   6.38 &  0.0332  &   6.14 &          &        &          &         \\
\hline
1.300
&          &        &  0.0606  &   6.51 &          &        &  0.0499  &   4.99 &  0.0401  &   4.80
&  0.0332  &   5.63 &  0.0247  &   7.93 &  0.0187  &   7.93 &  0.0103  &   6.39 &  0.00463 &  12.6   \\
\hline
1.400
&          &        &  0.0420  &   7.67 &          &        &  0.0320  &   6.09 &  0.0225  &   7.27
&  0.0186  &   7.35 &  0.0136  &  10.7  &          &        &          &        &          &         \\
\hline
1.500
&          &        &  0.0201  &  11.1  &          &        &  0.0196  &   8.76 &  0.0131  &   9.13
&  0.0137  &  10.5  &  0.00879 &   6.22 &  0.00676 &   6.13 &  0.00439 &   8.92 &  0.00104 &  24.9   \\
\hline
1.700
&          &        &  0.00868 &   8.05 &          &        &  0.00885 &   8.78 &  0.00488 &  11.1
&  0.00466 &  11.2  &  0.00338 &   9.64 &  0.00276 &   9.43 &  0.00127 &  17.6  &  0.00052 &  33.2   \\
\hline
1.900
&          &        &  0.00254 &  14.6  &          &        &  0.00255 &  16.5  &  0.00197 &  17.0 
&          &        &  0.00114 &  16.5  &          &        &          &        &          &         \\
\hline
2.100
&          &        &          &        &          &        &  0.00071 &  30.9  &  0.00101 &  23.3 
&          &        &          &        &          &        &          &        &          &         \\
\hline
\hline
\raisebox{-0.2ex}{$p_T$}$\backslash$\raisebox{0.2ex}{$x_F$} 
& \multicolumn{2}{c|}{0.65} & \multicolumn{2}{c|}{0.75}
& \multicolumn{2}{c|}{0.85} & \multicolumn{2}{c|}{}
& \multicolumn{2}{c|}{}     & \multicolumn{2}{c|}{}
& \multicolumn{2}{c|}{}     & \multicolumn{2}{c|}{}
& \multicolumn{2}{c|}{}     & \multicolumn{2}{c|}{}                          \\
\hline
0.100
&  0.221   &   7.4  &  0.0588  &  15.8  &  0.0259  &  25.0  &          &        &          &       
&          &        &          &        &          &        &          &        &          &         \\
\hline
0.300
&  0.149   &   5.1  &  0.0409  &  10.6  &  0.0111  &  21.8  &          &        &          &       
&          &        &          &        &          &        &          &        &          &         \\
\hline
0.500
&  0.0582  &   6.3  &  0.0199  &  11.5  &  0.00421 &  26.7  &          &        &          &       
&          &        &          &        &          &        &          &        &          &         \\
\hline
0.700
&  0.0282  &   7.9  &  0.00742 &  16.0  &  0.00236 &  30.0  &          &        &          &       
&          &        &          &        &          &        &          &        &          &         \\
\hline
0.900
&  0.0142  &   9.5  &  0.00250 &  24.3  &  0.00116 &  37.6  &          &        &          &       
&          &        &          &        &          &        &          &        &          &         \\
\hline
1.100
&  0.00517 &  14.3  &  0.00185 &  25.8  &  0.00014 & 100    &          &        &          &       
&          &        &          &        &          &        &          &        &          &         \\
\hline
1.300
&  0.00161 &  23.5  &  0.00074 &  37.3  &  0.00012 & 100    &          &        &          &       
&          &        &          &        &          &        &          &        &          &         \\
\hline
1.500
&  0.00048 &  41.7  &  0.00019 &  68.4  &          &        &          &        &          &       
&          &        &          &        &          &        &          &        &          &         \\
\hline
1.700
&  0.00014 &  71.4  &          &        &          &        &          &        &          &      
&          &        &          &        &          &        &          &        &          &         \\
\hline
\end{tabular}
}
\end{center}
\vspace{-2mm}
\caption{Double differential invariant cross section $f(x_F,p_T)$ [mb/(GeV$^2$/c$^3$)] for $\pi^-$
produced in p+p interactions at 158~GeV/c. The statistical uncertainty $\Delta f$ is given in \%.
}
\label{tpi-}
\end{table}

The detailed structure of the data tables results from the attempt to cover
as completely as possible the available phase space with bins which
comply with the statistical errors of the data spanning five orders of
magnitude in cross section.

\subsection{Interpolation Scheme
}
\begin{figure}[t]
\centering
\epsfig{file=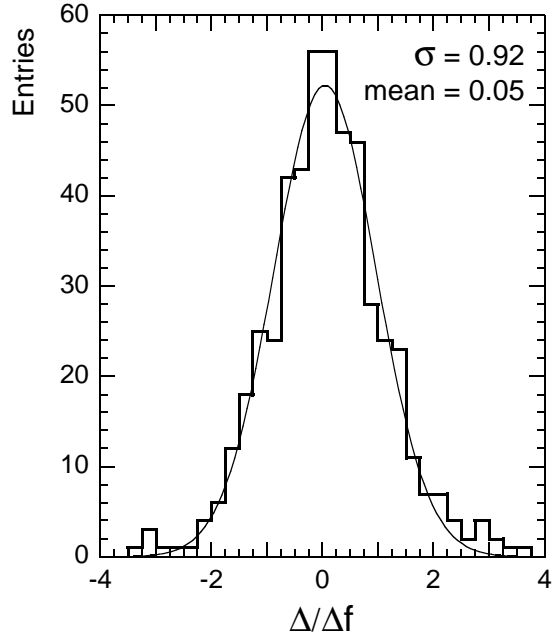,width=16cm}
\caption{Histogramme of the differences $\Delta$ between the measured invariant cross sections 
and the corresponding interpolated values ($\pi^+$ and $\pi^-$ combined) 
divided by the experimental uncertainty $\Delta f$ of the data points.
}
\label{ipol-stat}
\end{figure}

\vspace{3mm}
Although the data binning has been chosen to correspond to well defined
bin centers in $x_F$ and $p_T$, it is desirable to also offer an interpolation
scheme which produces smooth overall $x_F$ and $p_T$ dependences
for the internal extension of the data to non-measured intermediate
values and for comparison with other data measured at different $x_F$ and 
$p_T$ values. Any attempt at such numerical interpolation meets with the
problem that the dense phase space coverage combined with the small
statistical errors of the data reveals a microstructure both in the $p_T$
and $x_F$ dependences which makes a description with arithmetic
parametrizations if not impossible, then at least incompatible with the
data quality. A multistep manual interpolation method is therefore
selected which relies on local continuity of the cross sections
both in the $p_T$ and $x_F$ variables and which complies with the statistical
accuracy of the data points. The final result can be controlled by evaluating the
distribution of the differences between data and interpolated values, divided 
by the statistical error of each data point. This 
distribution should be a Gaussian centered at zero and with variance 
unity if the interpolation scheme does not introduce a bias. 
As shown in Fig.~\ref{ipol-stat} this is indeed the case for the 589 data points of
this experiment.

In the Figures shown in the subsequent Sections, the interpolation
for $\pi^+$ is extended beyond the region of identification, $x_F > 0.55$, 
by using data from experiments at ISR energies. This is discussed and 
justified in Section 8.2.

\subsection{Dependence of Cross Section on $p_T$ and $x_F$
}
\begin{figure}[p]
\centering
\epsfig{file=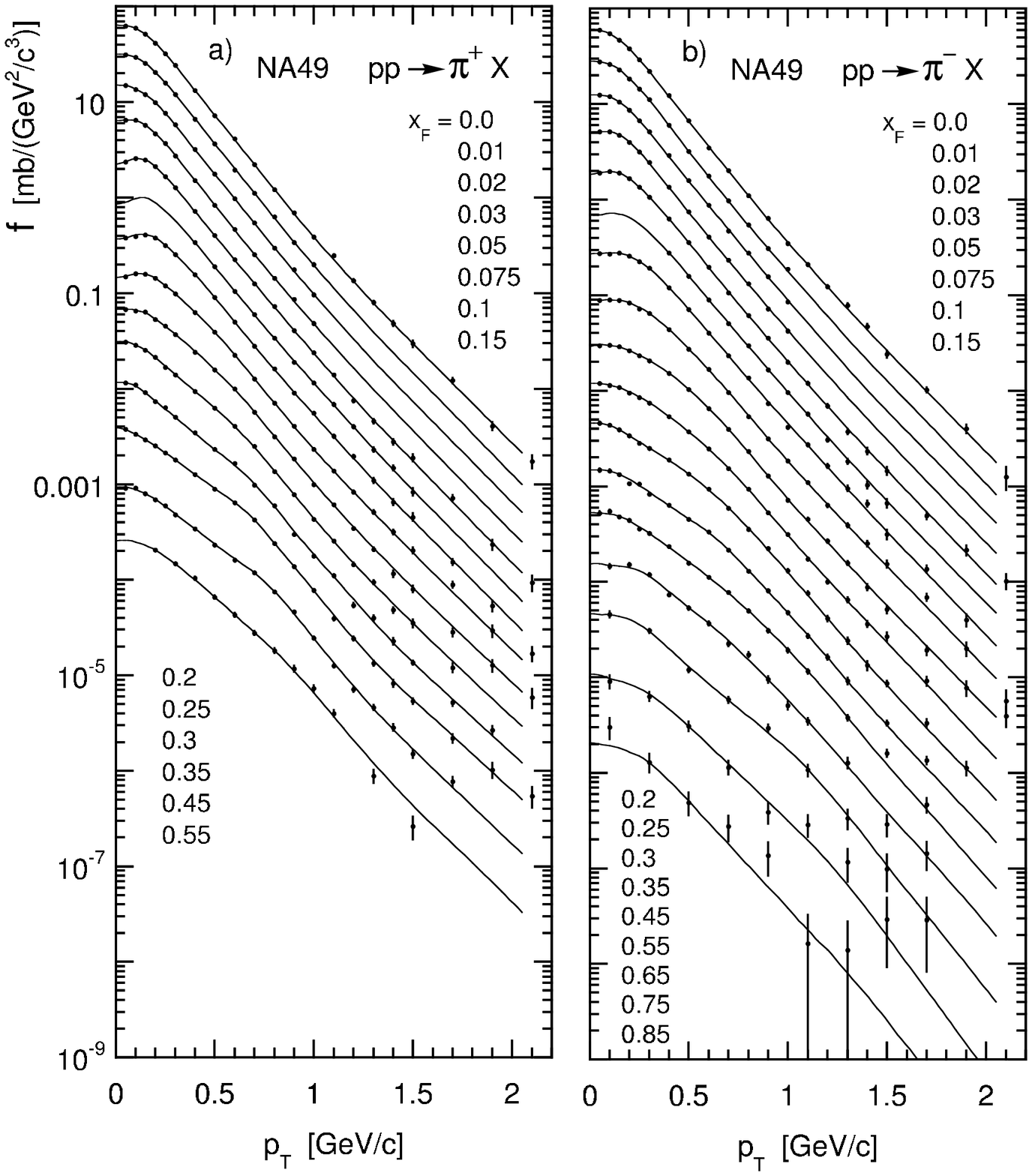,width=16cm}
\caption{Invariant cross section as a function of $p_T$ at fixed $x_F$ for a) $\pi^+$ 
and b) $\pi^-$ produced in p+p collisions at 158~GeV/c. 
Data and lines are multiplied successively by 0.5 for $\pi^+$ and by 0.5 up to $x_F=0.35$ 
and 0.75 for $x_F\geq0.45$ for $\pi^-$ to allow for a better separation.
}
\label{ccs-pt}
\end{figure}

\vspace{3mm}
The general behaviour of the invariant cross section as a function of $p_T$
at fixed $x_F$ is presented in Fig.~\ref{ccs-pt}. In order to clearly bring out
the shape evolution and to avoid overlapping of the interpolated curves
and error bars, the values at subsequent $x_F$ values are multiplied
by factors of typically 0.5 as given in the figure caption.

\begin{figure}[p]
\centering
\epsfig{file=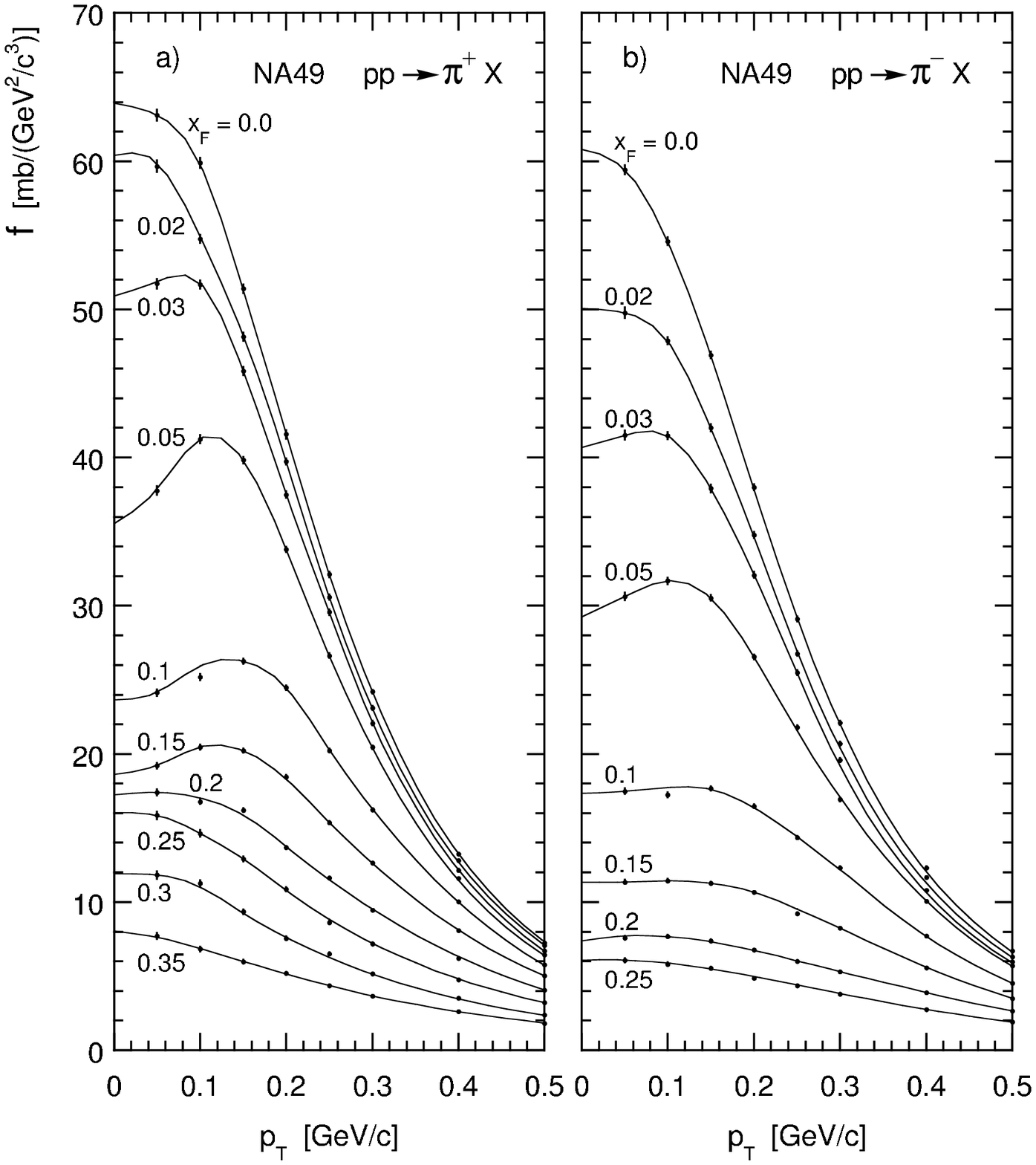,width=16cm}
\caption{Invariant cross section as a function of $p_T$ at fixed $x_F$ for a) $\pi^+$ 
and b) $\pi^-$ produced in p+p collisions at 158~GeV/c. 
The behaviour in the low $p_T$ region is emphasized by using a linear scale.
}
\label{ccs-pt-li}
\end{figure}

The observed $p_T$ dependence is definitely not describable by a simple 
overall exponential parametrization at any value of $x_F$, unless one wants to 
introduce local exponential slope parameters for small regions of $p_T$.
In addition, distinct structures are found at low $p_T$ and $x_F < 0.3$ and 
in the $p_T$ range from 0.5 to 1 GeV/c for $x_F > 0.25$. The structure at low $p_T$
is shown in more detail in the linear plots of Fig.~\ref{ccs-pt-li} where a local maximum
at $p_T \simeq 0.15$~GeV/c is seen to develop in the region $0.03 < x_F < 0.2$.
This maximum is less pronounced for $\pi^-$ than for $\pi^+$.

\begin{figure}[p]
\centering
\epsfig{file=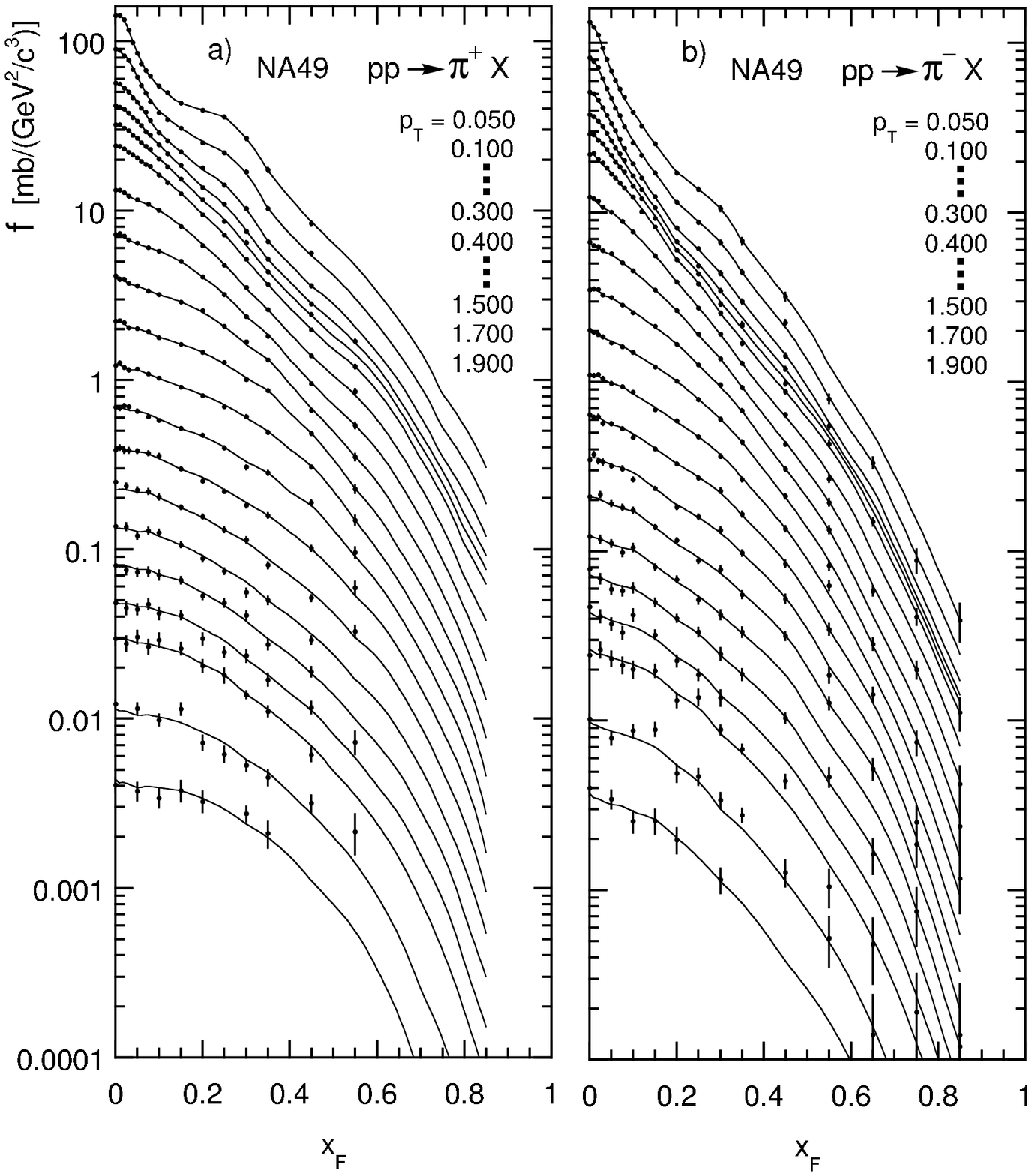,width=16cm}
\caption{Invariant cross section as a function of $x_F$ at fixed $p_T$ for a) $\pi^+$ 
and b) $\pi^-$ produced in p+p collisions at 158~GeV/c. 
The steps in $p_T$ are 50~MeV/c up to $p_T=0.300$~GeV/c, then 100~MeV/c up to 1.5~GeV/c 
and finally 200~MeV/c.
Data and lines for $p_T=0.05, 0.1, 0.15$~GeV/c are multiplied by 2.25, 1.5, 1.1, respectively, 
to allow for a better separation.
}
\label{ccs-xf}
\end{figure}

Corresponding $x_F$ distributions at fixed values of $p_T$ are presented in
Fig.~\ref{ccs-xf}. Again in order to clarify the overall shape development, the 
curves for the three lowest $p_T$ values 0.050, 0.100 and 0.150 GeV/c are multiplied
by different factors specified in the figure caption.

\begin{figure}[p]
\centering
\epsfig{file=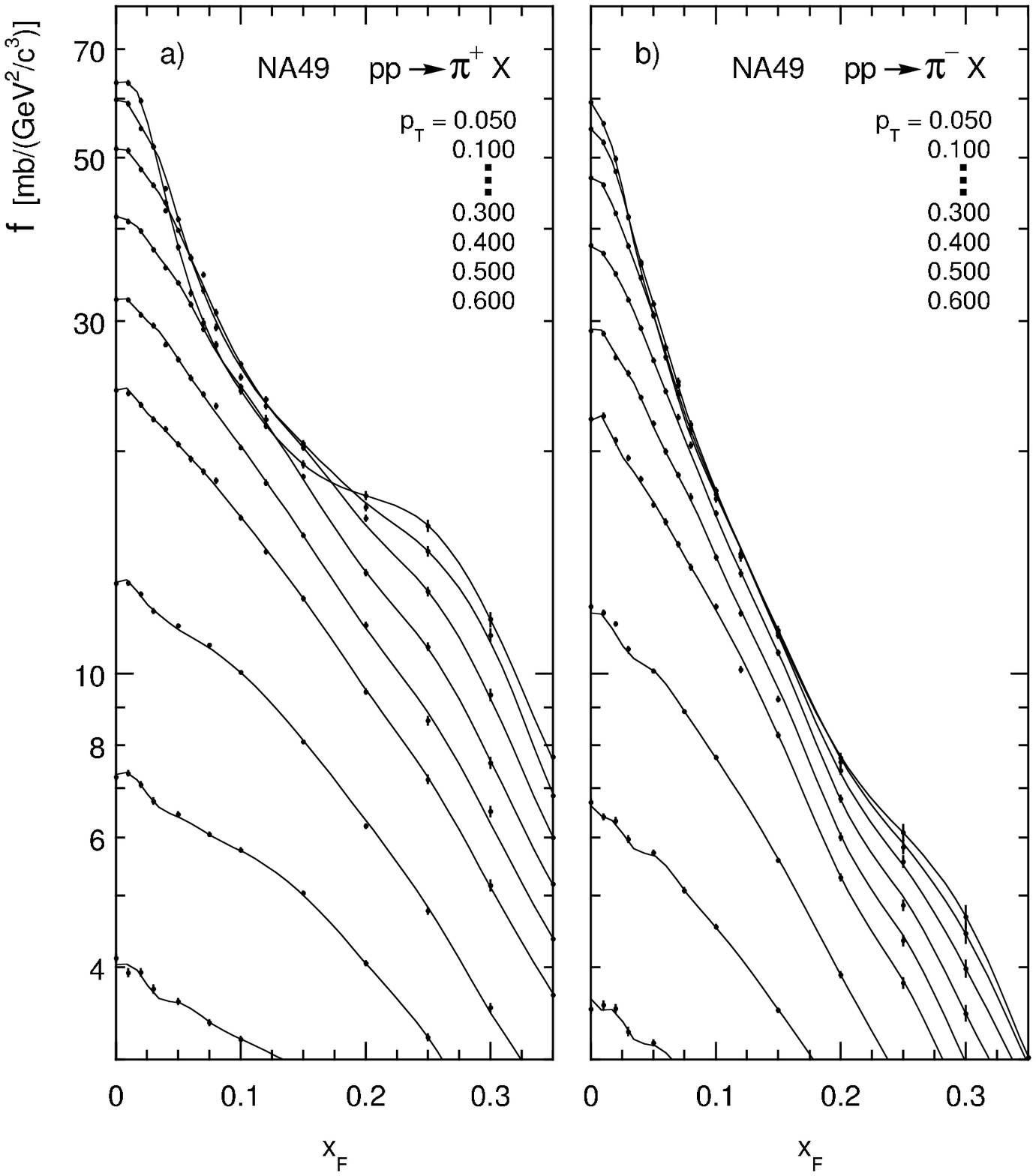,width=16cm}
\caption{Invariant cross section as a function of $x_F$ at fixed $p_T$ for a) $\pi^+$ 
and b) $\pi^-$ produced in p+p collisions at 158~GeV/c. 
The low $x_F$ and low $p_T$ region is magnified in order to demonstrate the complicated behaviour
of the invariant cross section in this region. Note that data and lines are not multiplied by 
factors in contrast to Fig.~\ref{ccs-xf}.
}
\label{ccs-xf-ma}
\end{figure}

The low $p_T$ region is magnified in the plots of Fig.~\ref{ccs-xf-ma} where the measured
cross sections are displayed without multiplication. A complicated
cross-over pattern with several inflexion points is evident 
for $p_T$ values below 0.3~GeV/c which is again less developed for $\pi^-$ than 
for $\pi^+$.

\subsection{$\pi^+/\pi^-$ Ratios
}
\begin{figure}[h]
\centering
\epsfig{file=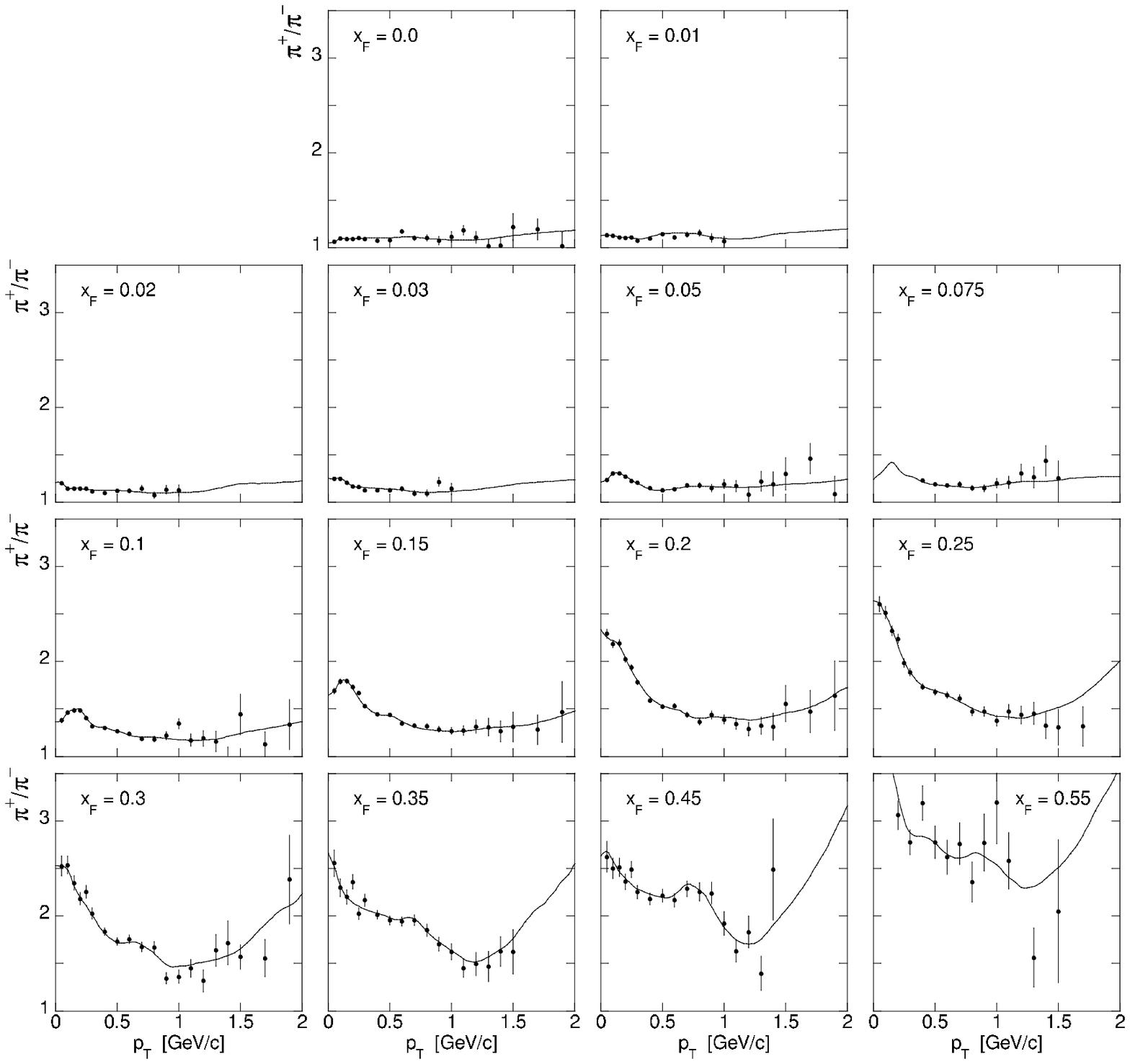,width=16cm}
\caption{Ratio of invariant cross section for $\pi^+$ and $\pi^-$ as a function of $p_T$ 
at fixed $x_F$. 
}
\label{pirat-pt}
\end{figure}

\begin{figure}[t]
\centering
\epsfig{file=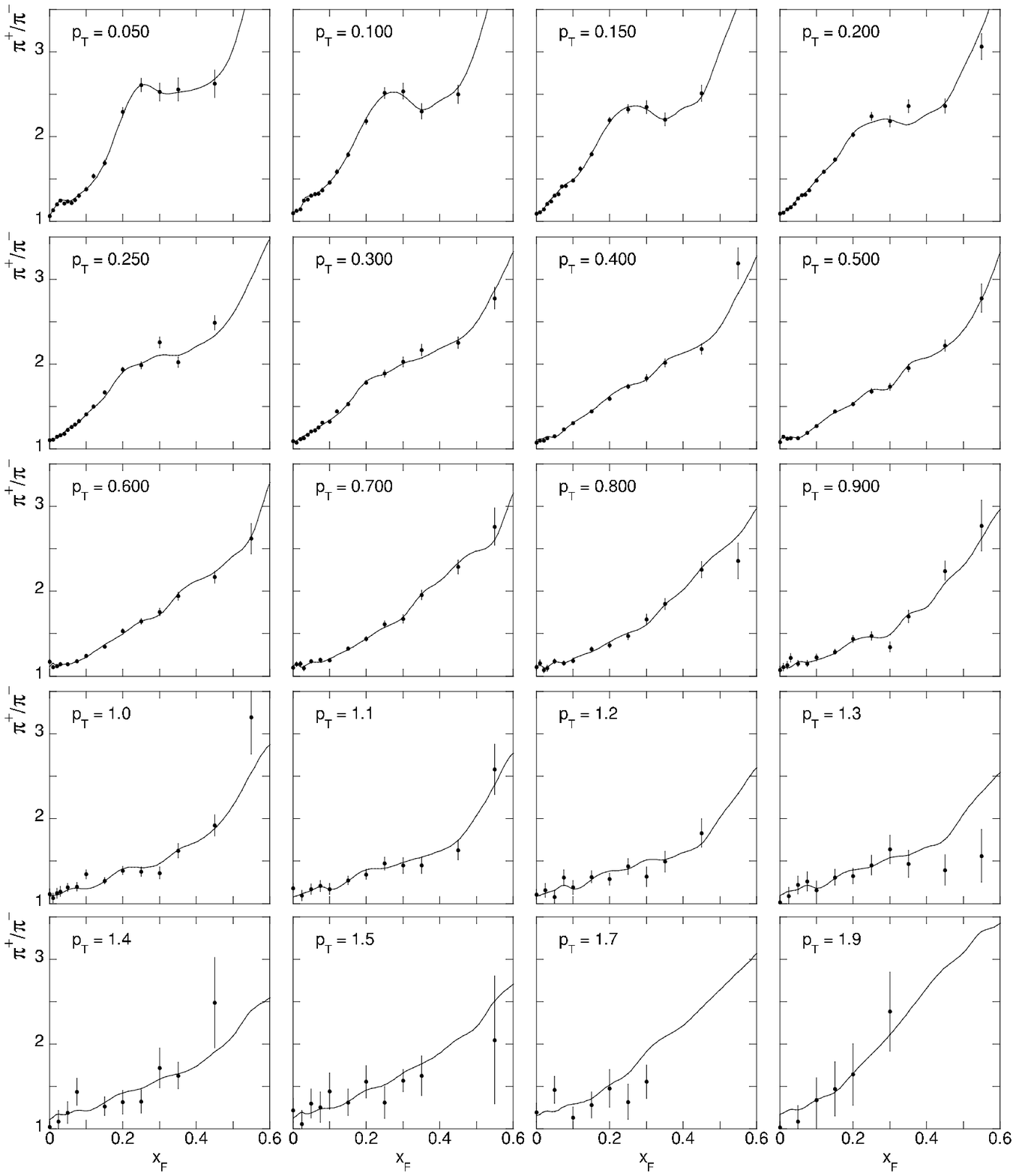,width=16cm}
\caption{Ratio of invariant cross section for $\pi^+$ and $\pi^-$ as a function of $x_F$ 
at fixed $p_T$. 
}
\label{pirat-xf}
\end{figure}

\vspace{3mm}
The distinct differences in the distributions of $\pi^+$ and $\pi^-$ over
phase space result in a complex pattern of the corresponding $\pi^+/\pi^-$
ratios. This is demonstrated in Figs.~\ref{pirat-pt} and \ref{pirat-xf} where the 
$p_T$ dependence for fixed $x_F$ and the $x_F$ dependence for fixed $p_T$ are shown 
in separate panels.

The well-known increase of the ratios with $x_F$ up to values of about
3 at $x_F \simeq 0.5$ is evident (see also Fig.~\ref{pirat-bja}); 
but for the first time with these data, well
defined structures at low ($<0.3$~GeV/c), intermediate ($0.5-0.7$~GeV/c) and
high $p_T$ ($>1.5$~GeV/c) as well as $x_F$ around 0.25 become visible.

\subsection{Rapidity and Transverse Mass Distributions
}
\begin{figure}[p]
\centering
\epsfig{file=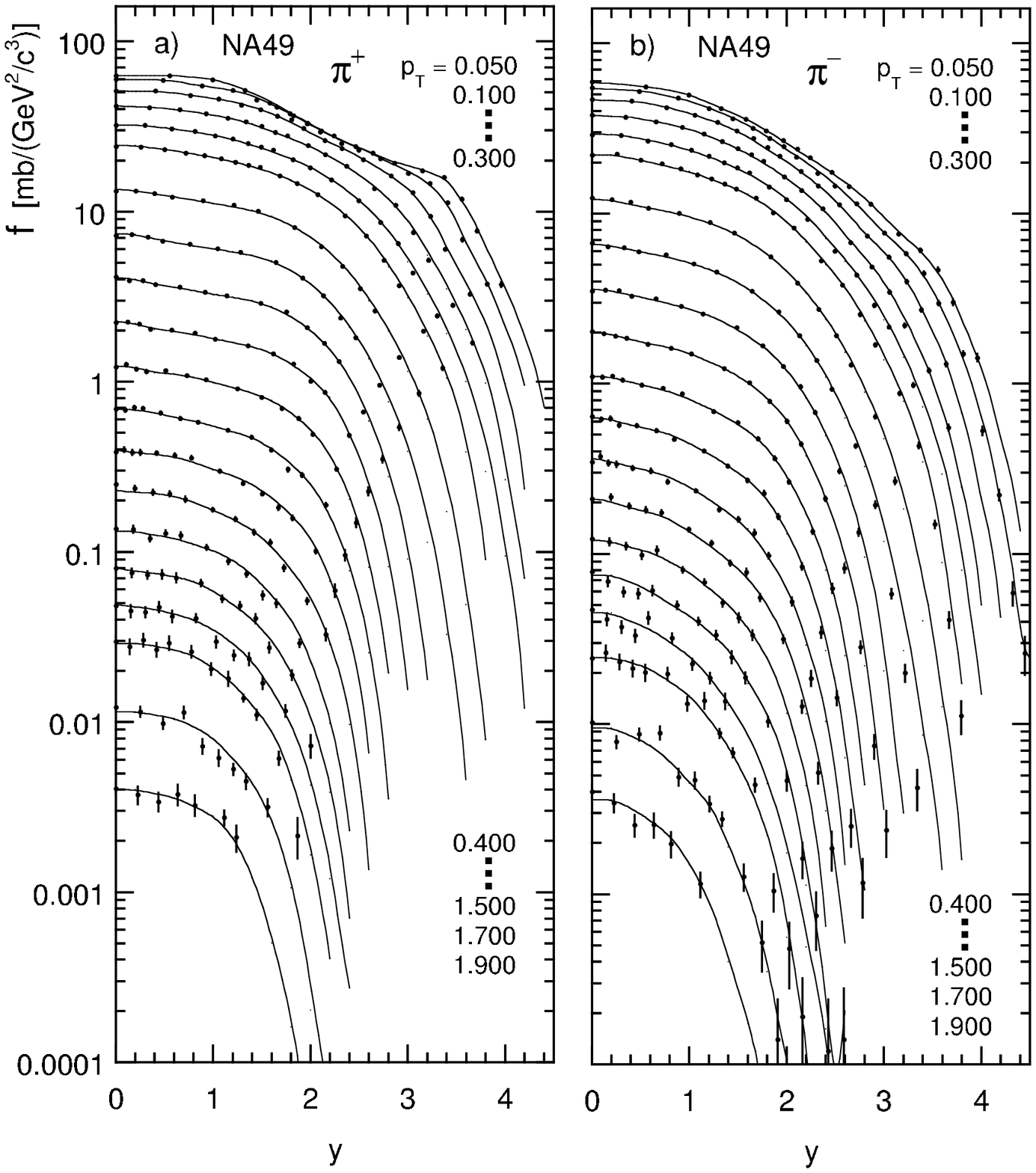,width=16cm}
\caption{Invariant cross section as a function of $y$ at fixed $p_T$ for a) $\pi^+$ 
and b) $\pi^-$ produced in p+p collisions at 158~GeV/c. 
The steps in $p_T$ are 50~MeV/c up to $p_T=0.300$~GeV/c, then 100~MeV/c up to 1.5~GeV/c 
and finally 200~MeV/c.
}
\label{ccs-y}
\end{figure}

\vspace{3mm}
It is customary to view double differential invariant cross sections 
also on the phase space subsurface of $y$ and $p_T$ although these two
variables are not orthogonal in momentum space. The data of 
Tables~\ref{tpi+} and \ref{tpi-} are
readily transformed into $y$ distributions at fixed values of $p_T$
and shown in Fig.~\ref{ccs-y}.

\begin{figure}[p]
\centering
\epsfig{file=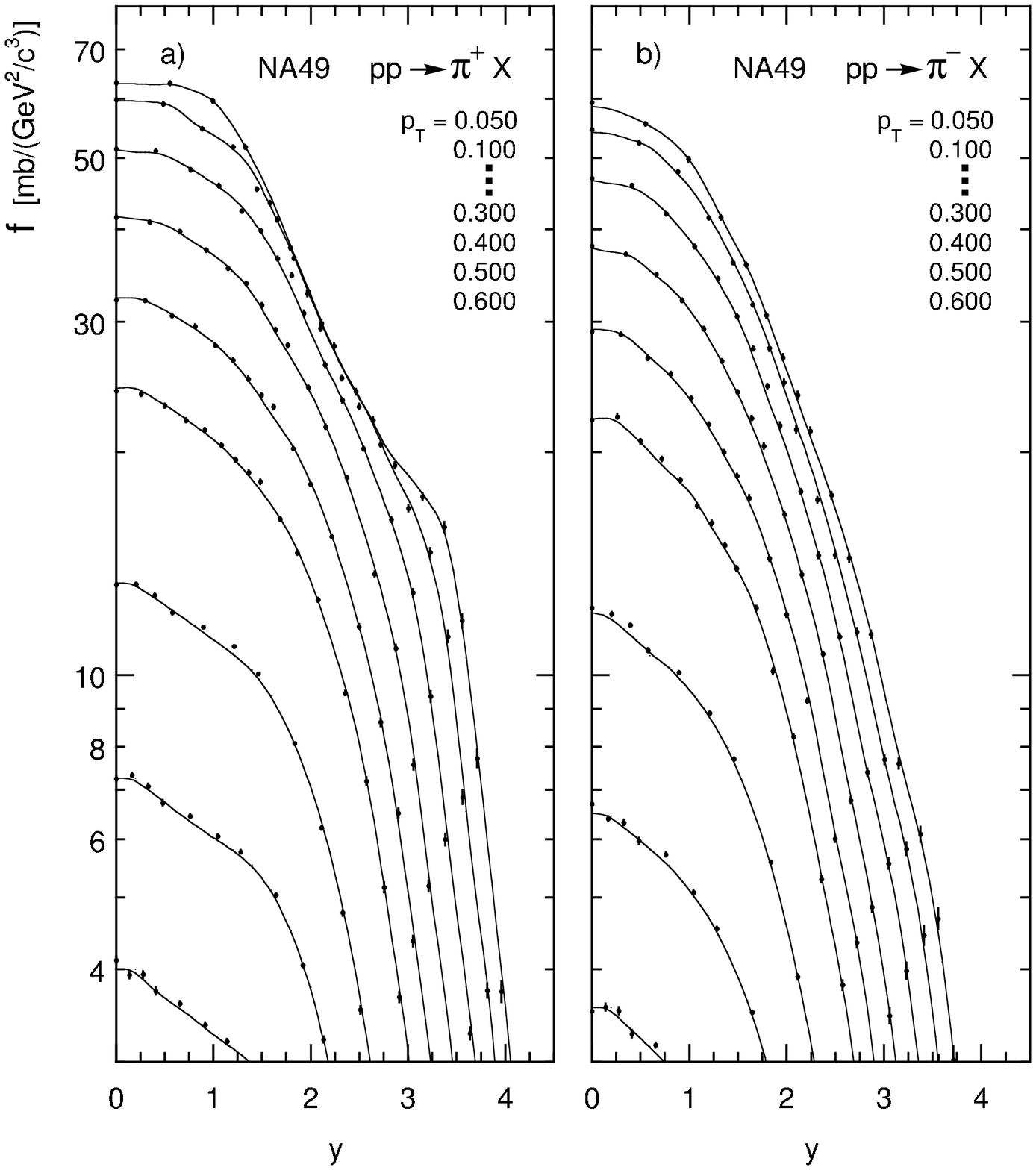,width=16cm}
\caption{Invariant cross section as a function of $y$ at fixed $p_T$ for a) $\pi^+$ 
and b) $\pi^-$ produced in p+p collisions at 158~GeV/c. 
The low $p_T$ region is magnified in order to demonstrate that the complicated behaviour
of the invariant cross section in $x_F$ is also apparent in $y$.
}
\label{ccs-y-ma}
\end{figure}

At fixed $p_T$ the $y$ distribution corresponds essentially to a distorted
longitudinal momentum distribution. The structures observed as a function
of $x_F$ will therefore also appear at the corresponding $y$ values which
is indeed the case. This is visible in more detail in the magnified $y$
distributions for $p_T < 0.5$~GeV/c shown in Fig.~\ref{ccs-y-ma}. With increasing $p_T$ these
structures happen to be less pronounced as compared to the $x_F$ distributions
(Fig.~\ref{ccs-xf-ma}) due to the different projection of phase space.

\begin{figure}[h]
\centering
\epsfig{file=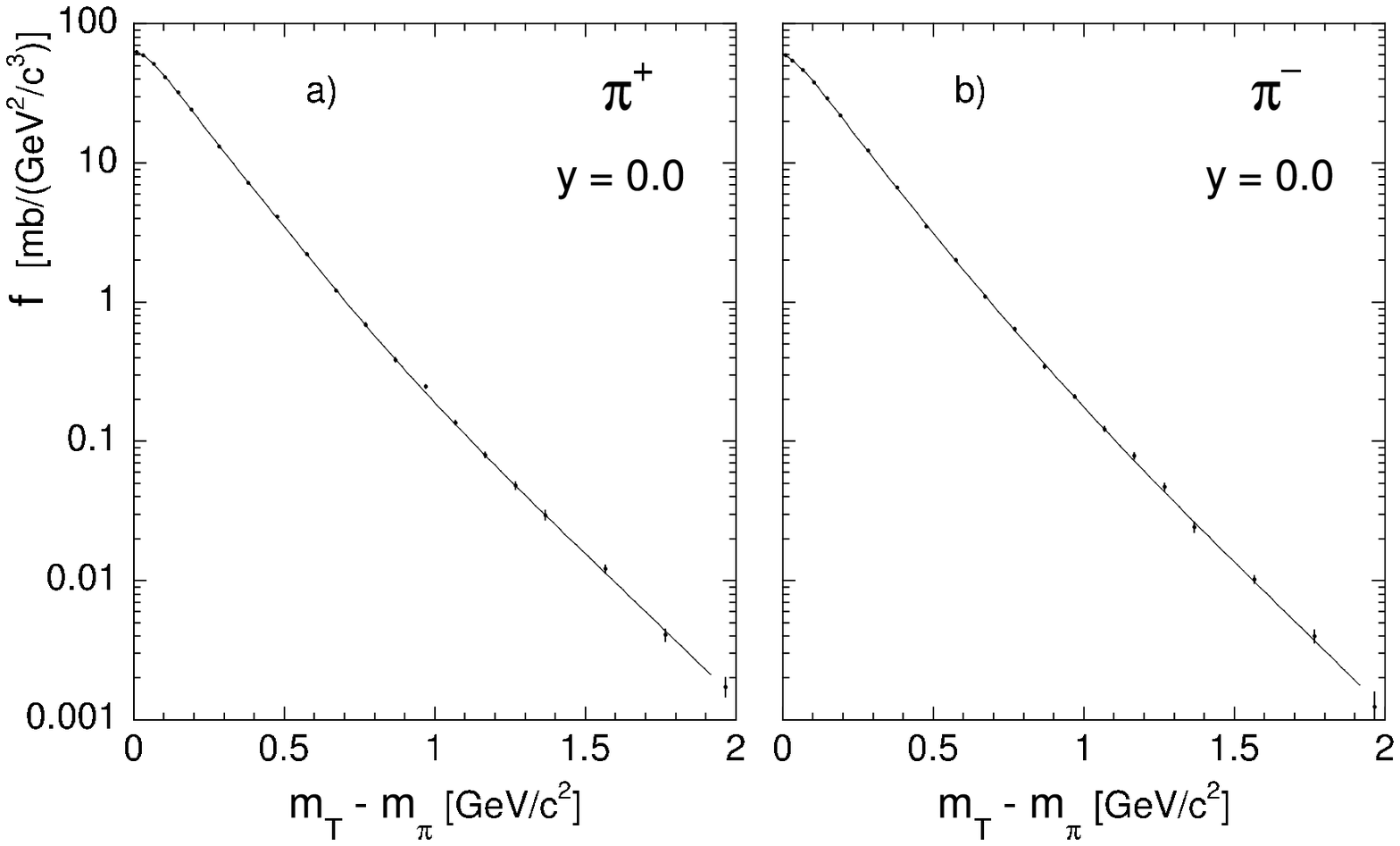,width=14cm}
\caption{Invariant cross section as a function of $m_T - m_{\pi}$ for a) $\pi^+$ and
b) $\pi^-$ produced at $y=0.0$
}
\label{ccs-mt}
\end{figure}

Another frequently used variable is the transverse mass $m_T=\sqrt{m_{\pi}^2+p_T^2}$. 
In fact, the shape of the $m_T$ distributions is often claimed to be close to exponential with
an inverse slope parameter $T$ as predicted for particle emission from a thermal source. 
In Fig.~\ref{ccs-mt} the invariant
cross section is presented as a function of $m_T - m_{\pi}$ for $\pi^+$ and $\pi^-$ produced at
$y=0.0$. 

\begin{figure}[b]
\centering
\epsfig{file=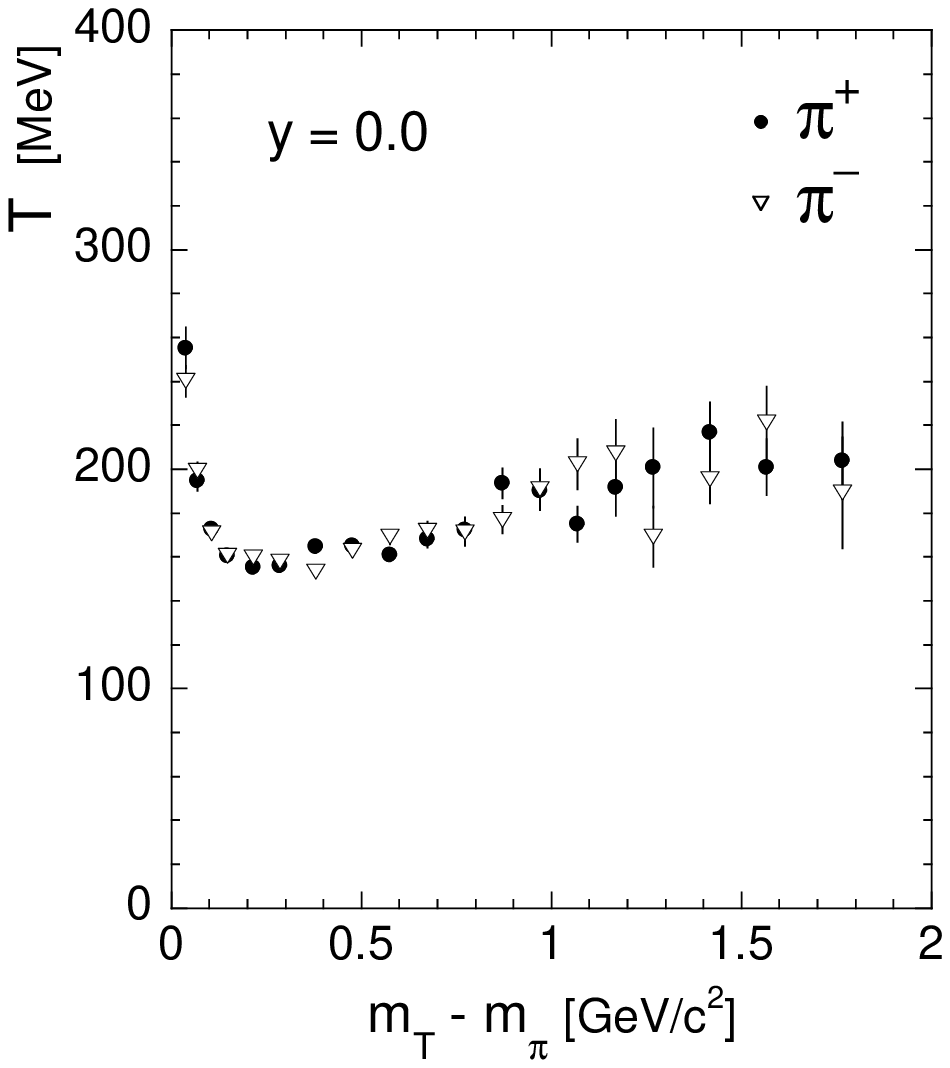,width=15cm}
\caption{Local slope of the $m_T$ distribution as a function of $m_T - m_{\pi}$ for $\pi^+$ and
$\pi^-$.
}
\label{mt-slope}
\end{figure}

The quality of the data allows to demonstrate that the shape of the $m_T$ distributions
cannot be described by a single exponential and makes the determination of the local
inverse slope $T(m_T-m_{\pi})$ as a function of transverse mass possible. As shown
in  Fig.~\ref{mt-slope} there is, both for $\pi^+$ and for $\pi^-$, a smooth and
continuous variation of $T$ with $m_T - m_{\pi}$ around a minimum of
$T = 0.155$~GeV at $m_T - m_{\pi} \simeq 0.250$~GeV/c$^2$ extending to values in excess
of $T = 0.200$~GeV both at low and and high $m_T - m_{\pi}$.

\section{Comparison to other Data
}
\vspace{3mm}
\subsection{Comparison at SPS/Fermilab Energies
}
\begin{figure}[b]
\centering
\epsfig{file=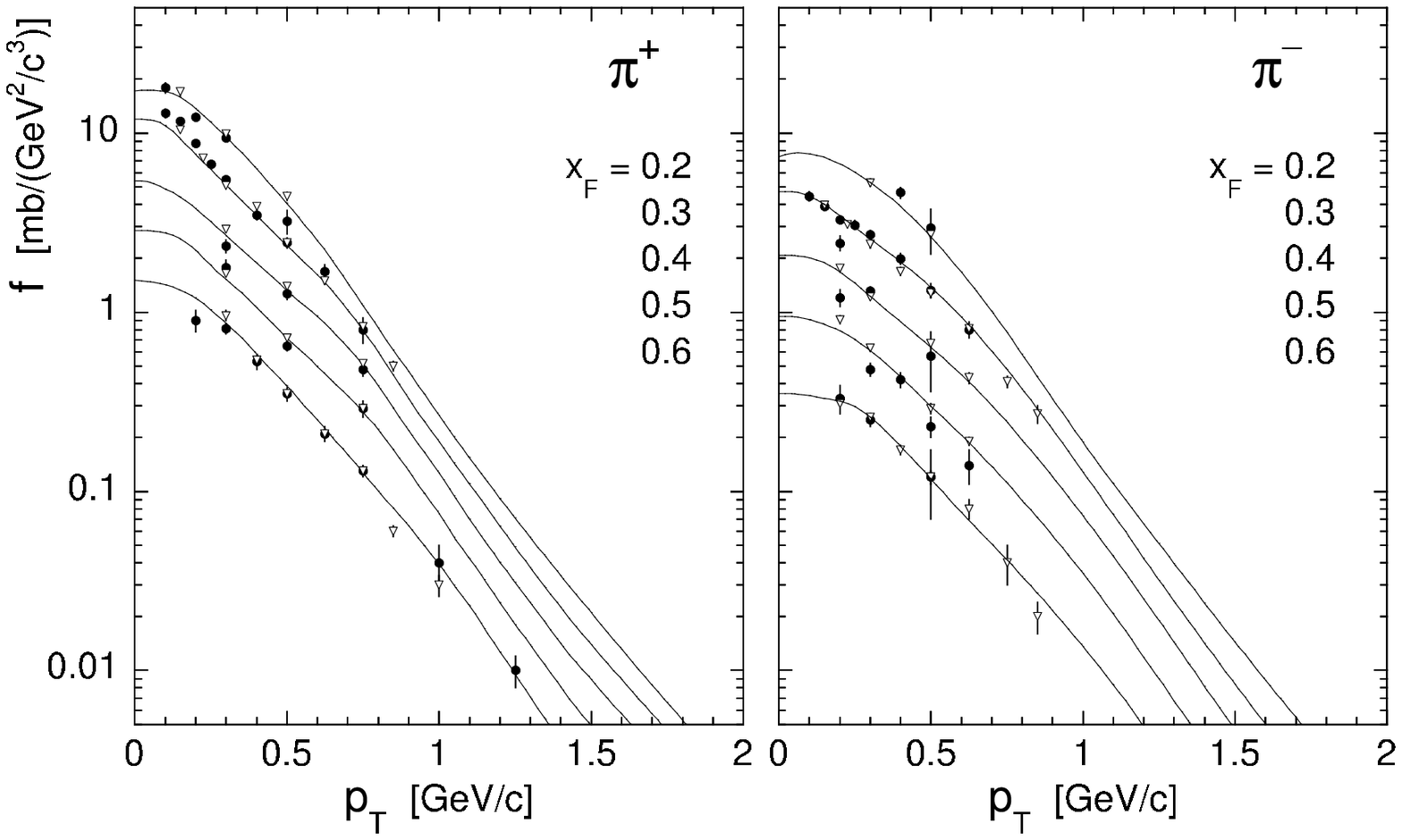,width=16cm}
\caption{Comparison of invariant cross section as a function of $p_T$ at fixed $x_F$ from
NA49 (lines) with measurements from \cite{bre} at 100 (full circles) and 175~GeV/c (open triangles).
}
\label{pi-bre-pt}
\end{figure}

\begin{figure}
\centering
\epsfig{file=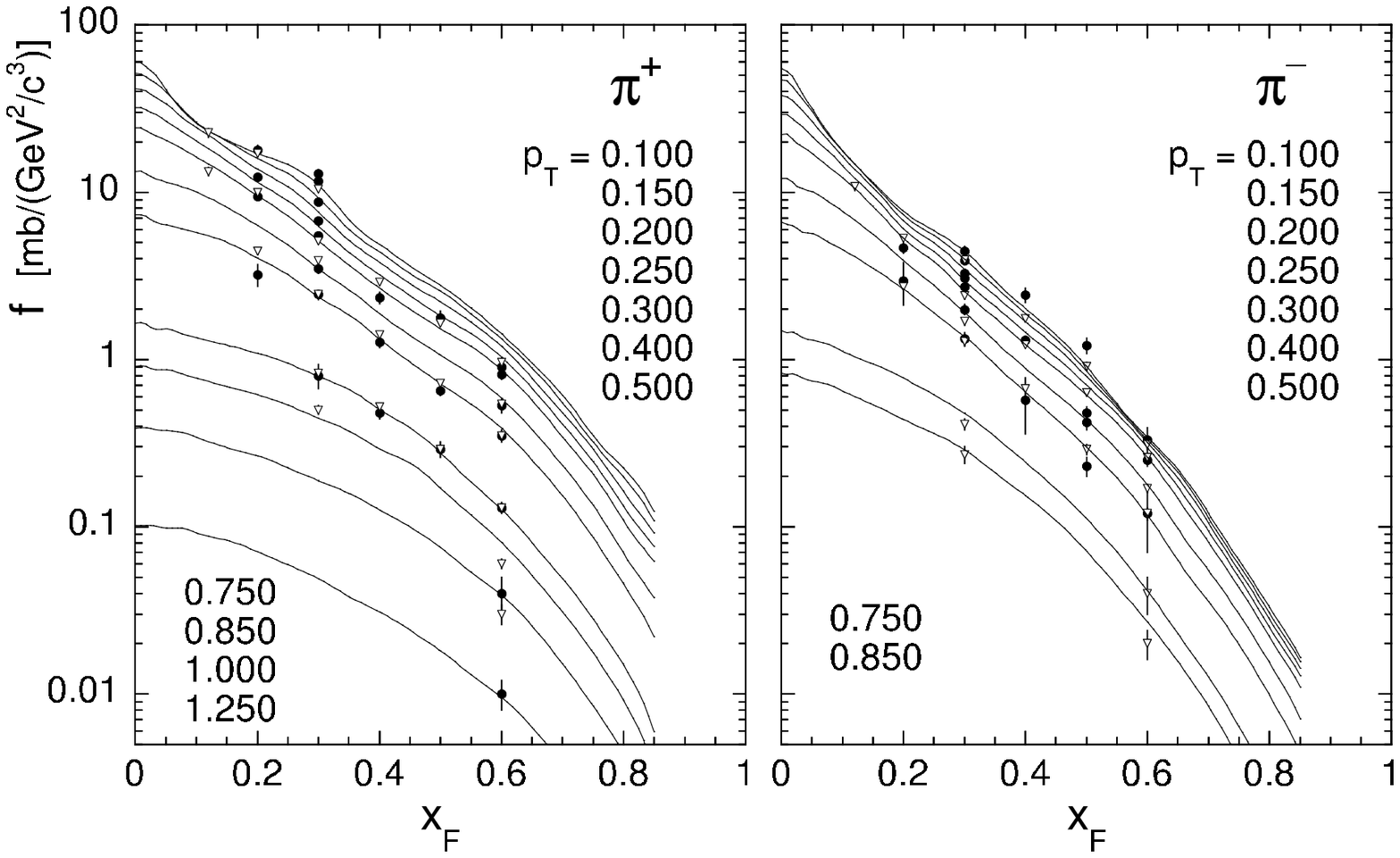,width=16cm}
\caption{Comparison of invariant cross section as a function of $x_F$ at fixed $p_T$ from
NA49 (lines) with measurements from \cite{bre} at 100 (full circles) and 175~GeV/c (open triangles).
}
\label{pi-bre-xf}
\end{figure}

\begin{figure}
\centering
\epsfig{file=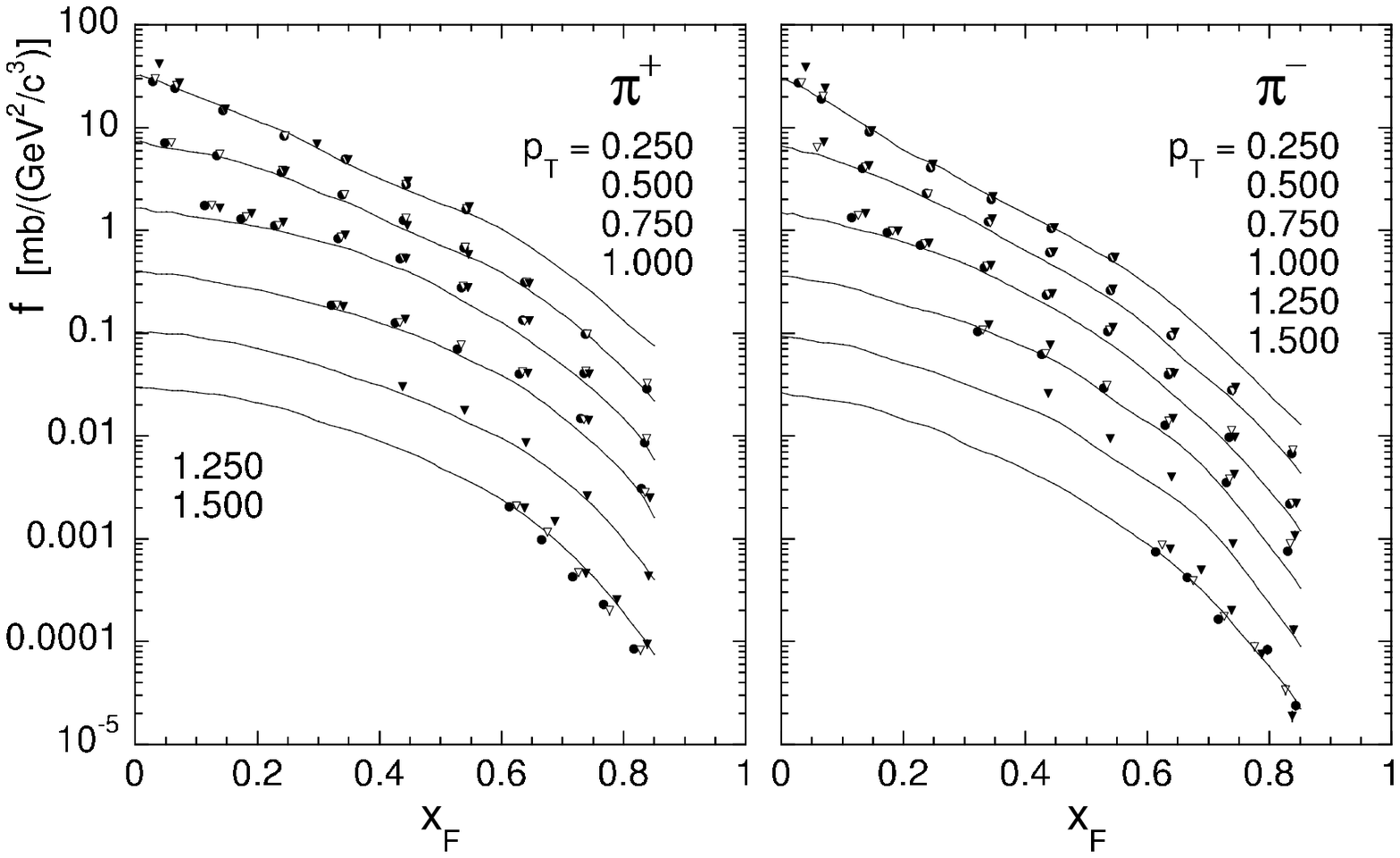,width=16cm}
\caption{Comparison of invariant cross section as a function of $x_F$ at fixed $p_T$ measured
by NA49 (lines) with measurements from \cite{joh} at 100 (full circles), 200 (open triangles), 
and 400~GeV/c (full triangles).
}
\label{pi-joh-xf}
\end{figure}

\vspace{3mm}
As shown in Section~2 above, there are only two preceding experiments
which measured double differential cross sections for identified pions
in overlapping ranges of $\sqrt{s}$ allowing for direct comparison. 
The data of Brenner et al. \cite{bre} from the Fermilab SAS spectrometer
offer 107 data points at beam momenta of 100 and 175~GeV/c, the data 
of Johnson et al. \cite{joh} comprise 241 data points at the three beam momenta 
100, 200 and 400~GeV/c. The corresponding comparisons are presented 
in Figs.~\ref{pi-bre-pt}, \ref{pi-bre-xf} and \ref{pi-joh-xf} where the interpolated NA49 
data are used as reference and 
are shown as full lines at those values of $x_F$ and $p_T$ where direct comparison 
is feasible. As the Johnson data have not been obtained at fixed $x_F$ but  
with respect to the radial scaling variable $x_R = E^{cms}/E^{cms}_{max}$, their positions 
have to be retransformed into $x_F$ values depending both on $p_T$ and $\sqrt{s}$. 
Therefore in this case only $x_F$ distributions at fixed $p_T$ are given.

\begin{figure}[b]
\centering
\epsfig{file=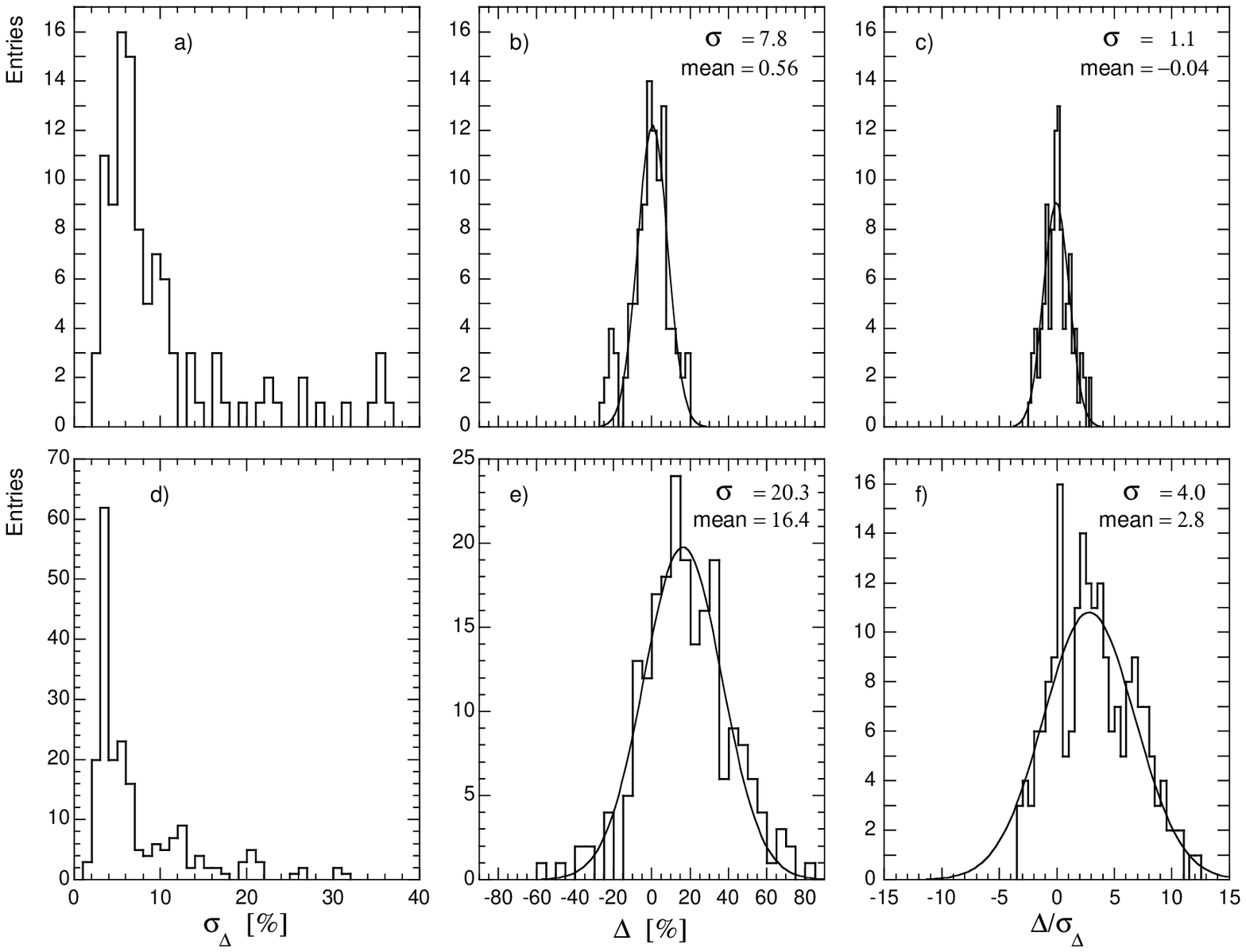,width=16cm}
\caption{Statistical analysis of the difference of the measurements of \cite{bre} 
(upper three pannels) and \cite{joh} (lower three pannels)
with respect to NA49: 
a) and d) error of the difference of the measurements; b) and e) difference
of the measurements; c) and f) difference divided by the error.
}
\label{stat-bre-joh}
\end{figure}

Inspection of Figs.~\ref{pi-bre-pt} and \ref{pi-bre-xf} reveals that the SAS data show a good overall
agreement with NA49, whereas very sizeable systematic deviations with
a complicated $p_T$ and $x_F$ dependence appear in the Johnson data (Fig.~\ref{pi-joh-xf}). 
This situation is quantified in the statistical analysis presented in Fig.~\ref{stat-bre-joh}.

In Fig.~\ref{stat-bre-joh}a and d the distribution of the point-by-point statistical error
of the cross section difference is given. These errors have most
probable values around 6 and 4\%, respectively, which are governed
by the statistics of the spectrometer experiments. The long upwards
tails are reflecting the decrease of cross section at high $x_F$ and/or
$p_T$. Here the NA49 errors become comparable or exceed the ones of the
comparison data. Fig.~\ref{stat-bre-joh}b and e shows the point-by-point cross section 
difference in percent with respect to the NA49 data. If this difference
is completely defined by statistical fluctuations, its ratio
to the statistical error, shown in
Fig.~\ref{stat-bre-joh}c and f, should obey a Gaussian distribution with variance unity. This
is indeed true for the SAS data as demonstrated by the Gaussian
fits superimposed to the histogrammes. In fact the mean difference
over the 107 points is less than 1\% with an error of about 1\%.
In view of the overall systematic errors of 2\% and 7\% given for
NA49 and SAS respectively, this agreement has to be regarded as
fortuitous. More importantly however, the agreement of the data
within purely statistical fluctuations helps to exclude systematic
effects between the two measurements as well as $s$-dependences
beyond the level of about 5\%.

The experimental situation is less favourable with respect to the
data of Johnson et al. Although their statistical errors are smaller
than the ones of SAS (see Fig.~\ref{stat-bre-joh}d) already the cross section differences,
Fig.~\ref{stat-bre-joh}e, show a very broad and off-centered distribution. The 
normalized differences, Fig.~\ref{stat-bre-joh}f, are off by nearly 3 standard 
deviations whilst the variance is around 4 units. Even taking account
of a mean shift of the data by 16\%, which might be in agreement with
the normalization uncertainties mentioned in their publication, this
indicates major systematic effects which are also clearly visible
in the direct data comparison of Fig.~\ref{pi-joh-xf}.  

\subsection{Comparison with Forward Production at ISR Energies and Extension
            of Data Interpolation for $\pi^+$
}
\vspace{3mm}
As the identification of $\pi^+$ suffers from the preponderant proton
component for $x_F > 0.6$, it is of interest to try to make use of the
large set of ISR data in the forward region in order to extend the
NA49 results on $\pi^+$ to the same phase space coverage as the $\pi^-$ data.
The CHLM Collaboration, Albrow et al. \cite{alb-72, alb-73, alb-74, sin} has 
published sizeable sets
of cross sections in the region $0.3<x_F<0.85$ and $0.3<p_T<1.2$~GeV/c. 
These data are more abundant for $\pi^+$ production where samples
at $\sqrt{s}= 31$, 45 and 53~GeV \cite{alb-74} as well as independent measurements 
at $\sqrt{s}= 45$~GeV \cite{sin} are available. For $\pi^-$ only measurements at 
$\sqrt{s}= 45$~GeV \cite{sin} and a set of data at one fixed angle \cite{alb-73} 
have been published.

\begin{figure}[h]
\centering
\epsfig{file=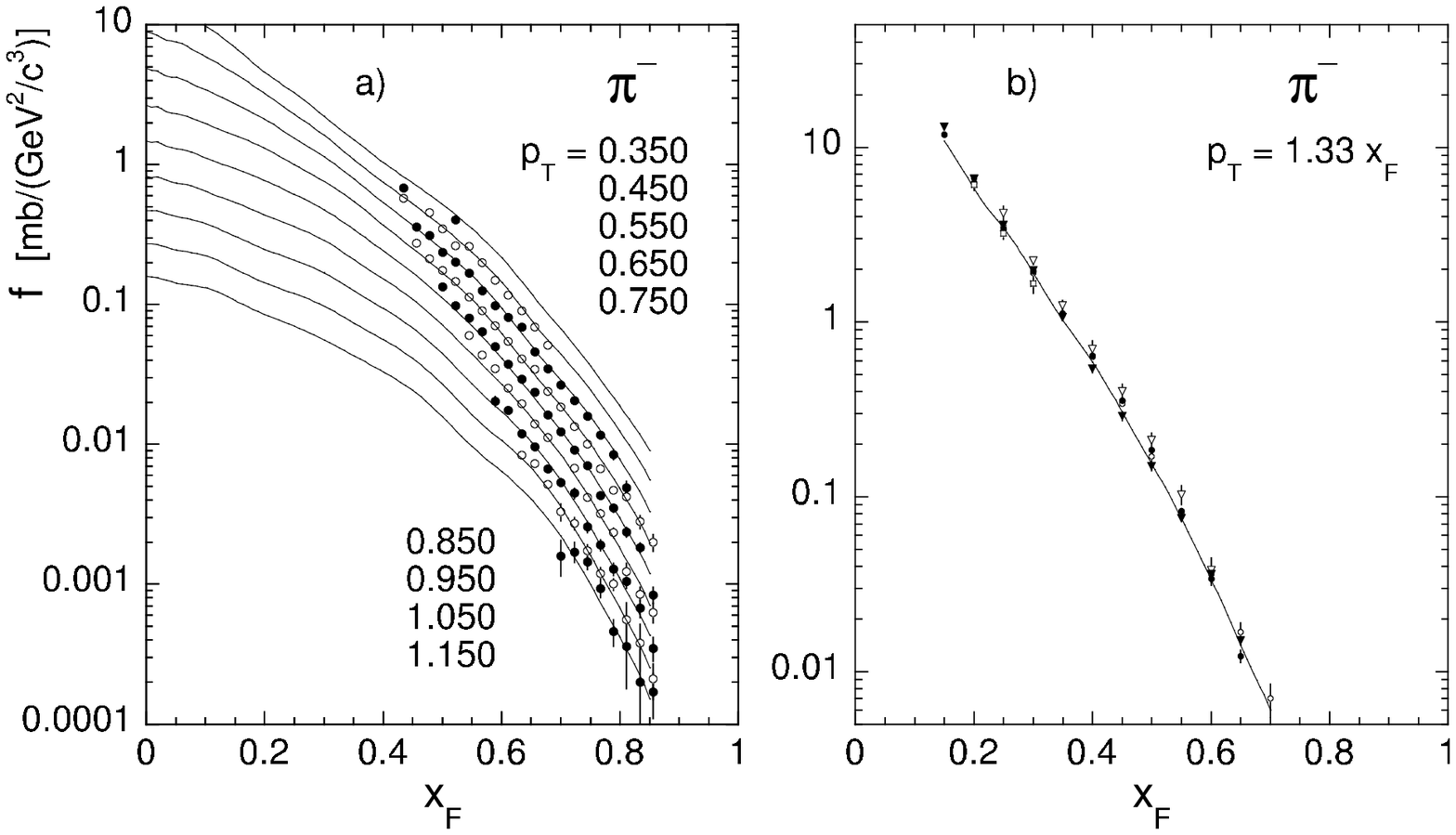,width=16cm}
\caption{Comparison of invariant cross section as a function of $x_F$ at fixed $p_T$ measured
by NA49 (lines) with measurements at a) $\sqrt{s}=45$~GeV \cite{sin} and 
b) $\sqrt{s}=23, 31, 45, 53, 63$~GeV \cite{alb-73}.
}
\label{pim-alb}
\end{figure}

\begin{figure}
\centering
\epsfig{file=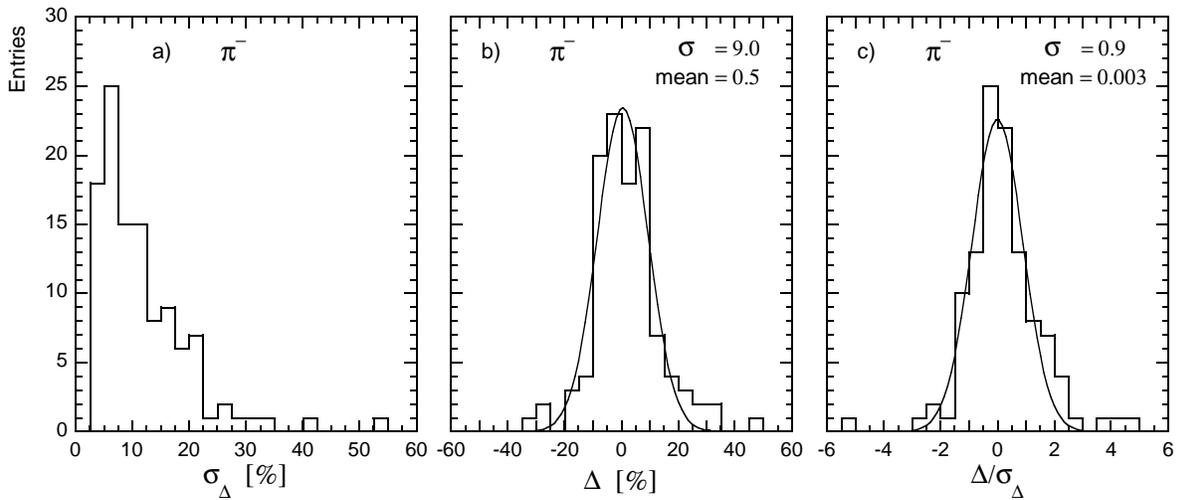,width=16cm}
\caption{Statistical analysis of the difference of the measurements of \cite{sin}
with respect to NA49: a) error of the difference of the measurements; b) difference
of the measurements; c) difference divided by the error.
}
\label{stat-alb-pim}
\end{figure}

The ISR data can be compared directly to the NA49
results in the region of overlapping measurements. For $\pi^-$ this is presented
in Fig.~\ref{pim-alb} showing various $x_F$ distributions at fixed $p_T$. A
detailed statistical analysis making use of a total of 111 data points is
presented in Fig.~\ref{stat-alb-pim}.
There is good agreement over the complete range of measurements
with the exception of a series of points obtained at 
$\sqrt{s}=23$~GeV \cite{alb-73} which are clearly above the NA49 data by about 25\%
(open triangles in Fig.~\ref{pim-alb}).
 
\begin{figure}
\centering
\epsfig{file=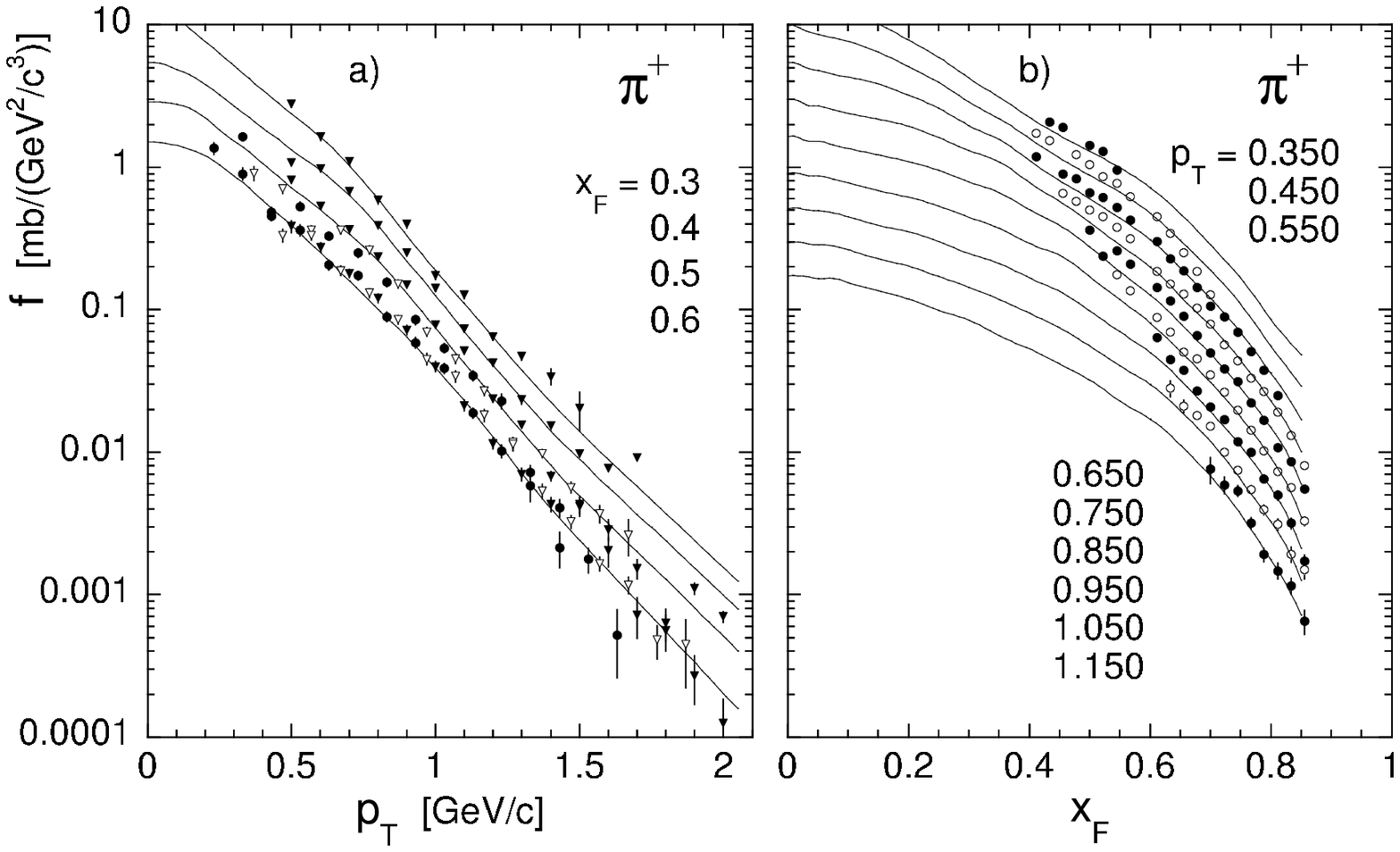,width=16cm}
\caption{Comparison of invariant cross section as a function of 
a) $p_T$ at fixed $x_F$ published by \cite{alb-74} at $\sqrt{s}=31, 45, 53$~GeV  and 
b) $x_F$ at fixed $p_T$ published by \cite{sin} at $\sqrt{s}=45$~GeV
to NA49 measurements represented as lines.
}
\label{pip-alb}
\end{figure}

\begin{figure}
\centering
\epsfig{file=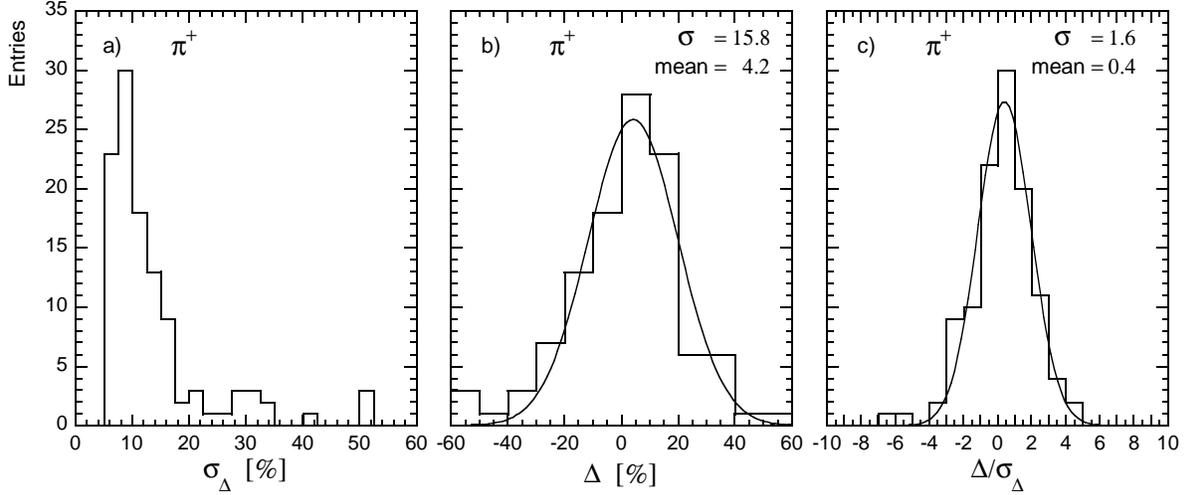,width=16cm}
\caption{Statistical analysis of the difference of the measurements of \cite{alb-74}
with respect to NA49: a) error of the difference of the measurements; b) difference
of the measurements; c) difference divided by the error.
}
\label{stat-alb-pip}
\end{figure}

For $\pi^+$ the situation is more complicated and summarized in Fig.~\ref{pip-alb}.
The ISR data of \cite{alb-74} 
obtained at $\sqrt{s}= 31$, 45 and 53~GeV show good agreement (Fig.~\ref{pip-alb}a)
whereas the data of \cite{sin} from the same collaboration show sizeable
upward shifts of about 20\% in the $x_F$ region where they overlap
with the NA49 measurements (Fig.~\ref{pip-alb}b). On the other hand this internal discrepancy 
vanishes for higher values of $x_F$, also seen in Fig.~\ref{pip-alb}b. It has therefore been
decided to use the combined data of \cite{alb-74} and \cite{sin} in the appropriate
$x_F$ ranges in order to extend the data parametrization of NA49 for $\pi^+$.
The statistical analysis of the \cite{alb-74} data  is presented in Fig.~\ref{stat-alb-pip}.

All in all, the data comparison over a range of $\sqrt{s}$ from 17 to 63~GeV
demonstrates Feynman scaling to the level of a few percent both for $\pi^+$
and $\pi^-$ in the range of $x_F > 0.4$. A more complete study of $s$-dependence
also in comparison to lower energies and for more central ranges of $x_F$
is outside the scope of this paper and will be addressed in a subsequent
publication. 

\subsection{Comparison of $\pi^+/\pi^-$ Ratios
}     
\begin{figure}
\centering
\epsfig{file=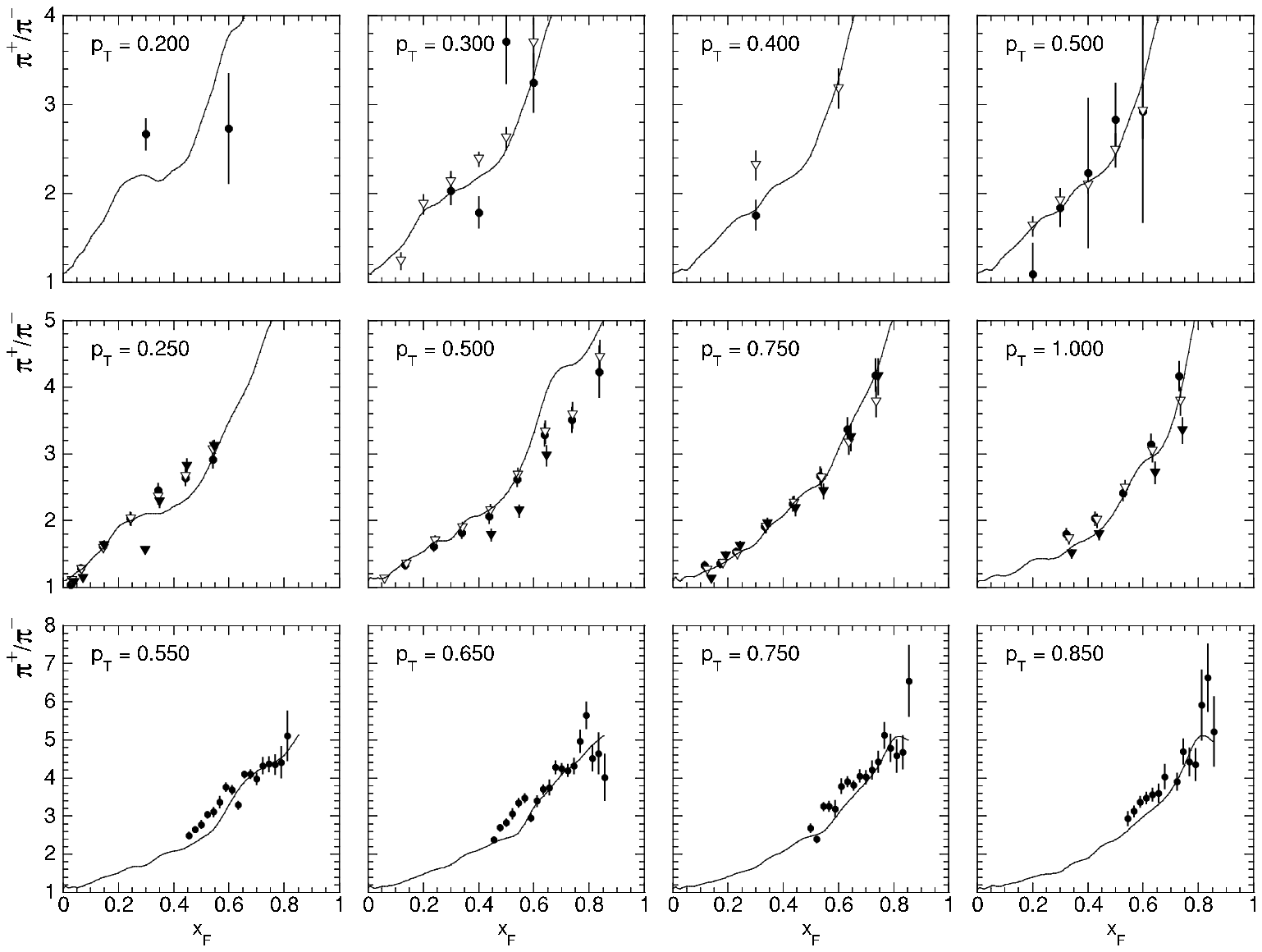,width=16cm}
\caption{Comparison of $\pi^+/\pi^-$ ratios as a function of $x_F$ for various $p_T$ values
measured by \cite{bre} (upper four pannels), \cite{joh} (middle four pannels), and
\cite{sin} (lower four pannels) to the NA49 results represented as lines.
}
\label{pirat-bja}
\end{figure}

\vspace{3mm}
A comparison of the $\pi^+/\pi^-$ ratios as a function of $x_F$ both in the 
Fermilab and ISR energy ranges is given in Fig.~\ref{pirat-bja} for various $p_T$ values.
As expected from the comparison of the single particle cross
sections of \cite{bre, sin}, the general agreement is satisfactory
with the exception of the ISR data at $\sqrt{s}= 45$~GeV in the lower
range of $x_F$.

\begin{figure}
\centering
\epsfig{file=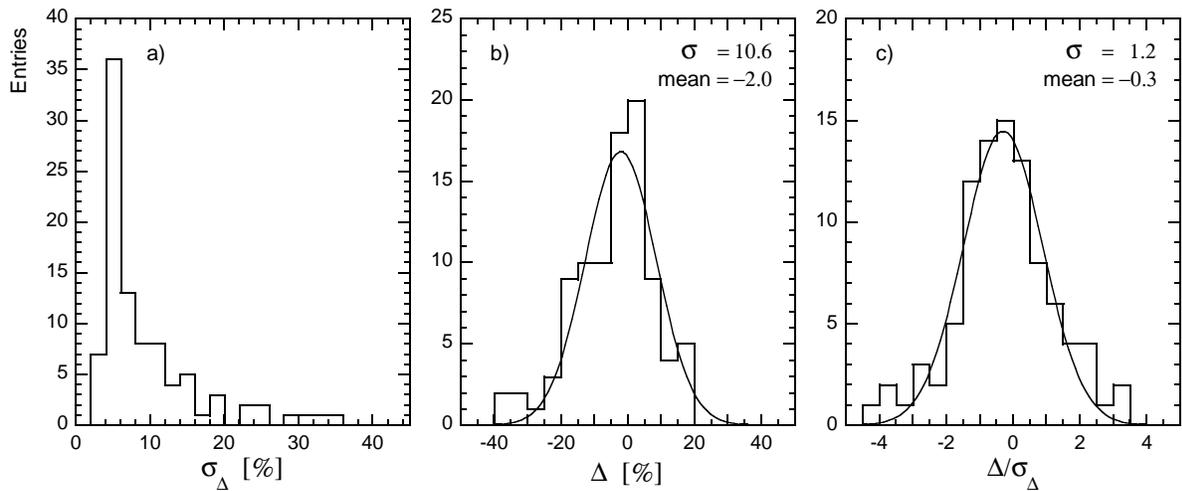,width=16cm}
\caption{Statistical analysis of the difference of the $\pi^+/\pi^-$ ratio of \cite{joh}
with respect to NA49: a) error of the difference of the ratio measurements; b) difference
of the ratio measurements; c) difference divided by the error.
}
\label{stat-johr}
\end{figure}

It is however interesting to note that also the data of Johnson et al.
\cite{joh} show reasonable agreement as far as the ratios are concerned, which is 
clearly confirmed by the statistical analysis presented in Fig.~\ref{stat-johr}. 
This indicates that the large systematic deviations found in the cross 
sections cancel out in the ratios thus allowing a further precise 
cross check of the NA49 data.

\section{Integrated Data
}
\vspace{3mm}
In addition to the double differential invariant cross sections discussed
above, integrated quantities such as invariant and non-invariant   
yields and particle ratios as a function of $x_F$ or $y$, 
first and second moments of the $p_T$ distributions and finally the total 
pion multiplicity are of interest. Such quantities are evaluated below
and compared to other available data. Integrations are generally performed
numerically applying Simpson´s parabolic approximation to the interpolated 
data. 

\clearpage
\subsection{$p_T$ Integrated Distributions
}
\begin{figure}
\centering
\epsfig{file=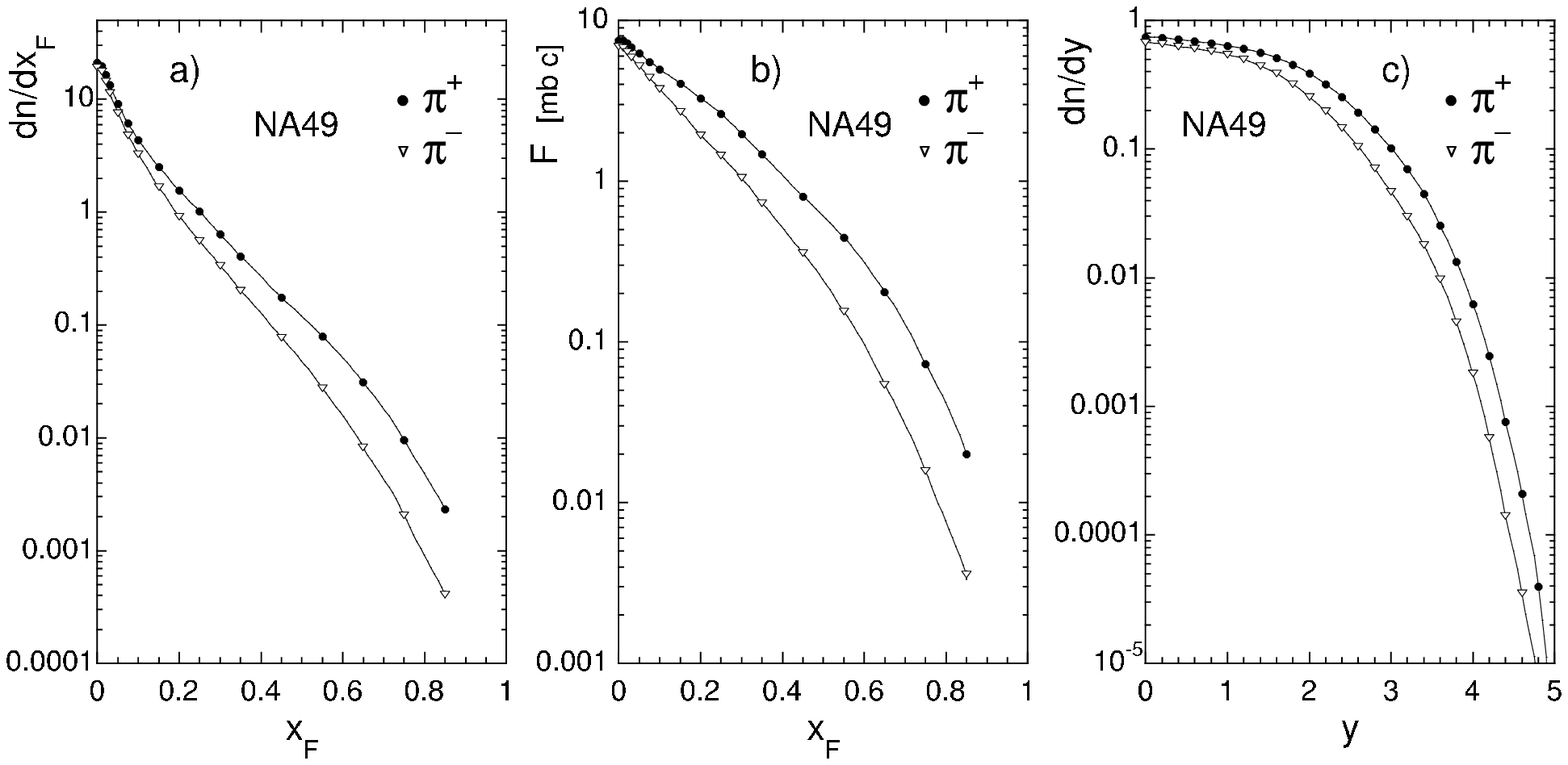,width=16cm}
\caption{Integrated distributions of $\pi^+$ and $\pi^-$ produced in p+p interactions at 
158~GeV/c: 
a) density distribution $dn/dx_F$ as a function of $x_F$; 
b) Integrated invariant cross section $F$ as a function of $x_F$; 
c) density distribution $dn/dy$ as a function of $y$.
}
\label{pi-xf-F-y}
\end{figure}

\vspace{3mm}
The distributions of the non-invariant and invariant yields are
defined as

\begin{equation}
dn/dx_F = \pi/\sigma_{inel} \cdot \sqrt{s}/2 \cdot \int{f/E \cdot dp_T^2}
\end{equation}
$$  F   = \int{f \cdot dp_T^2}  $$
$$  dn/dy = \pi/\sigma_{inel} \cdot \int{f \cdot dp_T^2}  $$
\noindent
with $f = E \cdot d^3\sigma/dp^3$, the invariant cross section.
These quantities are summarized in Table~\ref{ipi} and shown as a function of $x_F$ and $y$ 
in Fig.~\ref{pi-xf-F-y}.
 

\begin{table}[b]
\renewcommand{\tabcolsep}{0.10pc} 
\renewcommand{\arraystretch}{1.05} 
\begin{center}
{\scriptsize
\begin{tabular}{|l|cc|cc|cc|cc||cc|cc|cc|cc|||l|c||c|}
\hline
       &
\multicolumn{8}{|c||}{$\pi^+$} & \multicolumn{8}{|c|||}{$\pi^-$}  &
       & $\pi^+$   &   $\pi^-$                                            \\
\hline
~~$x_F$
& \multicolumn{2}{c|}{$F$~~~~~~~$\Delta$}   
& \multicolumn{2}{c|}{$dn/dx_F$~~$\Delta$}
& \multicolumn{2}{c|}{$\langle p_T \rangle $~~~$\Delta$}
& \multicolumn{2}{c||}{$\langle p_T^2 \rangle $~~~$\Delta$}  
& \multicolumn{2}{c|}{$F$~~~~~~~~$\Delta$}
& \multicolumn{2}{c|}{$dn/dx_F$~~$\Delta$}
& \multicolumn{2}{c|}{$\langle p_T \rangle $~~~$\Delta$} 
& \multicolumn{2}{c|||}{$\langle p_T^2 \rangle $~~~$\Delta$}
& ~~$y$    &     $dn/dy$  &     $dn/dy$                                     \\
\hline
0.0
&  7.505   &   0.13    & 20.959   &   0.14 &  0.2590  &   0.06 &  0.1015  &   0.13
&  6.832   &   0.13    & 19.136   &   0.13 &  0.2579  &   0.06 &  0.1009  &   0.13
&  0.0     & 0.7418    & 0.6713                                                     \\
\hline
0.01
&  7.500   &   0.12    & 19.640   &   0.13 &  0.2664  &   0.06 &  0.1062  &   0.13
&  6.704   &   0.12    & 17.586   &   0.13 &  0.2659  &   0.06 &  0.1055  &   0.13
&  0.2     & 0.7327    & 0.6567                                                     \\
\hline
0.02
&  7.171   &   0.12    & 16.348   &   0.12 &  0.2811  &   0.06 &  0.1165  &   0.12
&  6.323   &   0.12    & 14.334   &   0.13 &  0.2838  &   0.06 &  0.1183  &   0.13
&  0.4     & 0.7113    & 0.6281                                                     \\
\hline
0.03
&  6.819   &   0.12    & 13.344   &   0.12 &  0.2962  &   0.06 &  0.1274  &   0.12
&  5.865   &   0.12    & 11.377   &   0.13 &  0.3018  &   0.06 &  0.1312  &   0.13
&  0.6     & 0.6894    & 0.6022                                                     \\
\hline
0.05
&  6.216   &   0.12    &  9.134   &   0.12 &  0.3276  &   0.06 &  0.1520  &   0.12
&  5.180   &   0.12    &  7.550   &   0.13 &  0.3357  &   0.06 &  0.1579  &   0.13
&  0.8     & 0.6618    & 0.5760                                                     \\
\hline
0.075
&  5.519   &   0.12    &  6.084   &   0.12 &  0.3571  &   0.06 &  0.1776  &   0.14
&  4.391   &   0.13    &  4.807   &   0.14 &  0.3683  &   0.06 &  0.1868  &   0.15
&  1.0     & 0.6334    & 0.5425                                                     \\
\hline
0.1
&  4.963   &   0.12    &  4.353   &   0.12 &  0.3763  &   0.07 &  0.1955  &   0.14
&  3.734   &   0.14    &  3.254   &   0.14 &  0.3912  &   0.07 &  0.2093  &   0.18
&  1.2     & 0.6015    & 0.4973                                                     \\
\hline
0.15
&  4.050   &   0.11    &  2.502   &   0.12 &  0.4010  &   0.07 &  0.2214  &   0.15
&  2.707   &   0.14    &  1.663   &   0.14 &  0.4240  &   0.07 &  0.2435  &   0.14
&  1.4     & 0.5614    & 0.4426                                                     \\
\hline
0.2
&  3.263   &   0.14    &  1.549   &   0.14 &  0.4170  &   0.08 &  0.2408  &   0.15
&  1.925   &   0.17    &  0.9091  &   0.17 &  0.4509  &   0.09 &  0.2728  &   0.17
&  1.6     & 0.5102    &  0.3841                                                    \\
\hline
0.25
&  2.619   &   0.20    &  1.007   &   0.20 &  0.4246  &   0.09 &  0.2520  &   0.19
&  1.437   &   0.24    &  0.5505  &   0.24 &  0.4625  &   0.12 &  0.2890  &   0.23
&  1.8     & 0.4509    &  0.3191                                                    \\
\hline
0.3
&  1.959   &   0.26    &  0.6325  &   0.26 &  0.4395  &   0.13 &  0.2697  &   0.24
&  1.042   &   0.35    &  0.3355  &   0.35 &  0.4750  &   0.17 &  0.3046  &   0.31
&  2.0     & 0.3863    &  0.2534                                                    \\
\hline
0.35
&  1.462   &   0.27    &  0.4061  &   0.27 &  0.4634  &   0.13 &  0.2932  &   0.23
&  0.7280  &   0.37    &  0.2020  &   0.37 &  0.4857  &   0.18 &  0.3184  &   0.31
&  2.2     & 0.3186    &  0.1964                                                    \\
\hline
0.45
&  0.8027  &   0.29    &  0.1745  &   0.29 &  0.4768  &   0.14 &  0.3112  &   0.24
&  0.3552  &   0.45    &  0.0772  &   0.45 &  0.4940  &   0.19 &  0.3312  &   0.33
&  2.4     & 0.2525    &  0.1462                                                    \\
\hline
0.55
&  0.4439  &   0.65    &  0.0792  &   0.65 &  0.4631  &   0.25 &  0.2964  &   0.50
&  0.1544  &   0.78    &  0.0275  &   0.78 &  0.4896  &   0.40 &  0.3232  &   0.74
&  2.6     & 0.1924    &  0.1034                                                    \\
\hline
0.65
&  0.2046  &   1.00    &  0.0309  &   1.05 &  0.4587  &   0.60 &  0.2915  &   1.15
&  0.0542  &   1.46    &  0.00820 &   1.46 &  0.4865  &   0.82 &  0.3269  &   1.58
&  2.8     & 0.1420    &  0.0703                                                    \\
\hline
0.75
&  0.0727  &   2.05    &  0.00955 &   2.10 &  0.4501  &   1.00 &  0.2804  &   1.75
&  0.0157  &   2.96    &  0.00205 &   2.96 &  0.4741  &   1.64 &  0.3068  &   3.05
&  3.0     & 0.1018    &  0.0463                                                    \\
\hline
0.85
&  0.0200  &   3.25    &  0.00232 &   3.35 &  0.4040  &   2.60 &  0.2314  &   3.50
&  0.00355 &   5.93    &  0.00041 &   5.93 &  0.4345  &   3.44 &  0.2602  &   6.25
&  3.2     & 0.0698    &  0.0298                                                    \\
\hline
&          &           &          &        &          &        &          &   
&          &           &          &        &          &        &          &   
&  3.4     & 0.0449    &  0.0179                                                    \\
\hline
&          &           &          &        &          &        &          &   
&          &           &          &        &          &        &          &   
&  3.6     & 0.0255    &  0.00966                                                   \\
\hline
&          &           &          &        &          &        &          &   
&          &           &          &        &          &        &          &   
&  3.8     & 0.0133    &  0.00446                                                   \\
\hline
&          &           &          &        &          &        &          &   
&          &           &          &        &          &        &          &   
&  4.0     & 0.00621   &  0.00179                                                   \\
\hline
&          &           &          &        &          &        &          &   
&          &           &          &        &          &        &          &   
&  4.2     & 0.00247   &  0.000566                                                  \\
\hline
&          &           &          &        &          &        &          &   
&          &           &          &        &          &        &          &   
&  4.4     & 0.000755  &  0.000141                                                  \\
\hline
&          &           &          &        &          &        &          &   
&          &           &          &        &          &        &          &   
&  4.6     & 0.000208  &  0.0000351                                                 \\
\hline
&          &           &          &        &          &        &          &   
&          &           &          &        &          &        &          &   
&  4.8     & 0.0000396 &  0.0000065                                                 \\
\hline
\end{tabular}
}
\end{center}
\vspace{-2mm}
\caption{$p_T$ integrated invariant cross section $F$ [mb$\cdot$c], 
density distribution $dn/dx_F$, mean transverse momentum $\langle p_T \rangle $ [GeV/c],
mean transverse momentum squared $\langle p_T^2 \rangle $ [(GeV/c)$^2$] as a function of
$x_F$, as well as density distribution $dn/dy$ as a function of $y$ for $\pi^+$ and $\pi^-$.
The statistical uncertainty $\Delta$ for each quantity is given in \%.
}
\label{ipi}
\end{table}

The statistical errors are in general below the percent level with the
exception of the $\pi^-$ data at $x_F > 0.55$. The overall experimental
uncertainties are therefore completely governed by the quoted systematic
errors. This is in particular true for the extrapolated data for $\pi^+$
in the phase space region above $x_F = 0.55$.

\begin{figure}
\centering
\epsfig{file=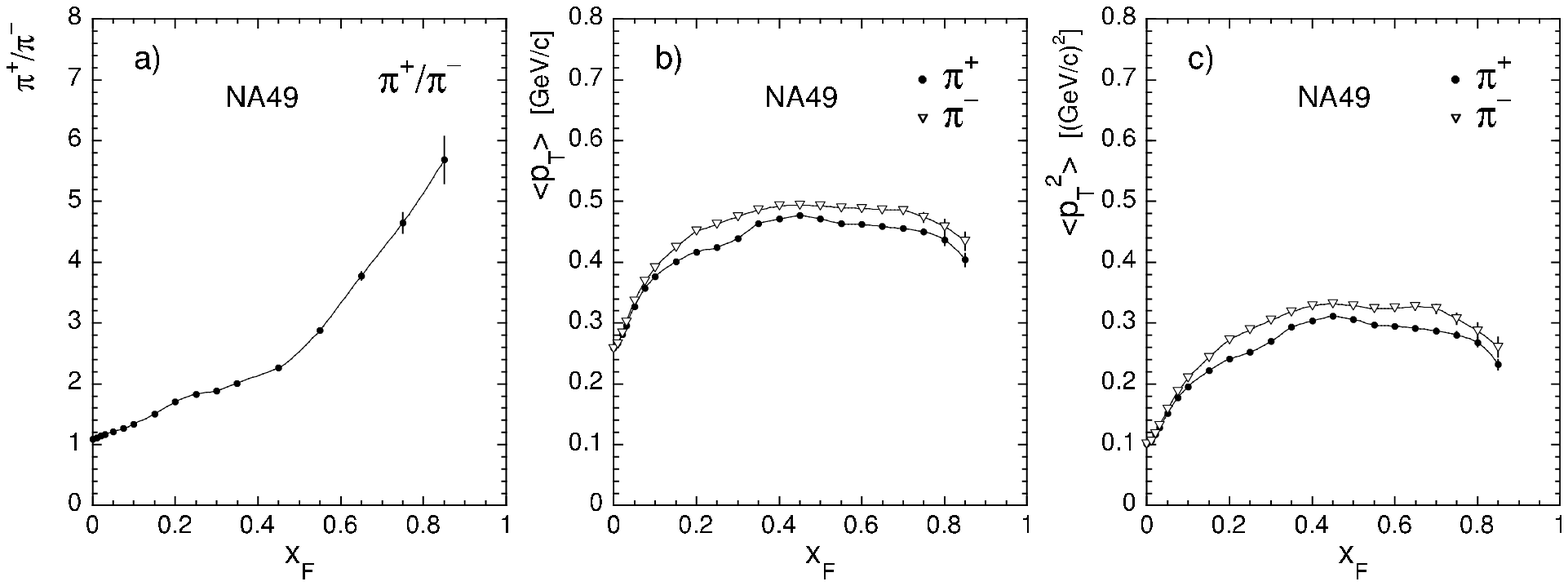,width=16cm}
\caption{a) $\pi^+/\pi^-$ ratio,
b) mean $p_T$, and
c) mean $p_T^2$ as a function of $x_F$ 
for $\pi^+$ and $\pi^-$ produced in p+p interactions at 158~GeV/c.
}
\label{pirat-mpt-mpt2}
\end{figure}

Further integrated quantities such as $\pi^+/\pi^-$ ratios, mean transverse momentum
$\langle p_T \rangle$ and mean transverse momentum squared $\langle p_T^2 \rangle$ are 
presented in Fig.~\ref{pirat-mpt-mpt2} again as a function of $x_F$.
The distinct structures and in particular the differences between $\pi^+$ and $\pi^-$ 
in the first and second moments of the $p_T$ distributions are noteworthy and
will be discussed in Section~10.

\subsection{Data Comparison
}
\begin{figure}[h]
\centering
\epsfig{file=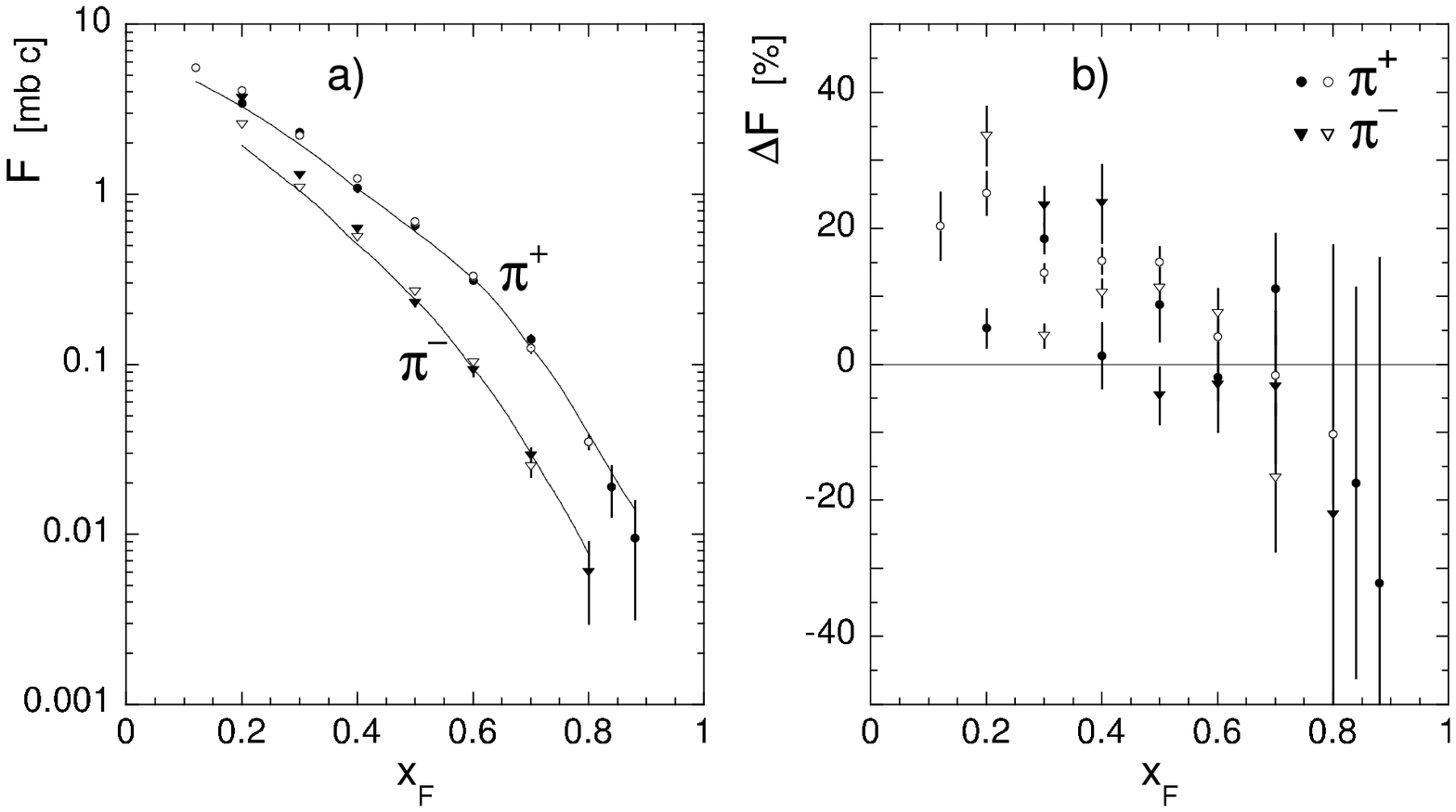,width=16cm}
\caption{  
a) Comparison of $p_T$ integrated invariant cross section $F$ as a function of 
$x_F$ for $\pi^+$ and $\pi^-$ measured by \cite{bre} to NA49 results (represented as lines); 
b) Deviation of the measurements of \cite{bre} from the NA49 results in percent.
}
\label{Bre-F-stat}
\end{figure}

\vspace{3mm}
Given the scarcity and generally incomplete phase space coverage of other
experiments the extraction of integrated quantities is liable to suffer
from sizeable systematic uncertainties. This becomes evident by comparing
the $p_T$ integrated pion yields from \cite{bre} to the NA49 data as shown in 
Fig.~\ref{Bre-F-stat}. Although the differential data are in good agreement with each 
other as shown in Section~8.1, the use of simple exponential or Gaussian parametrizations 
of the $p_T$ distributions for extrapolation into the unmeasured regions of $p_T$ 
\cite{bre} result in systematic deviations of the integrated values which reach many
standard deviations with respect to the given errors.

\begin{figure}
\centering
\epsfig{file=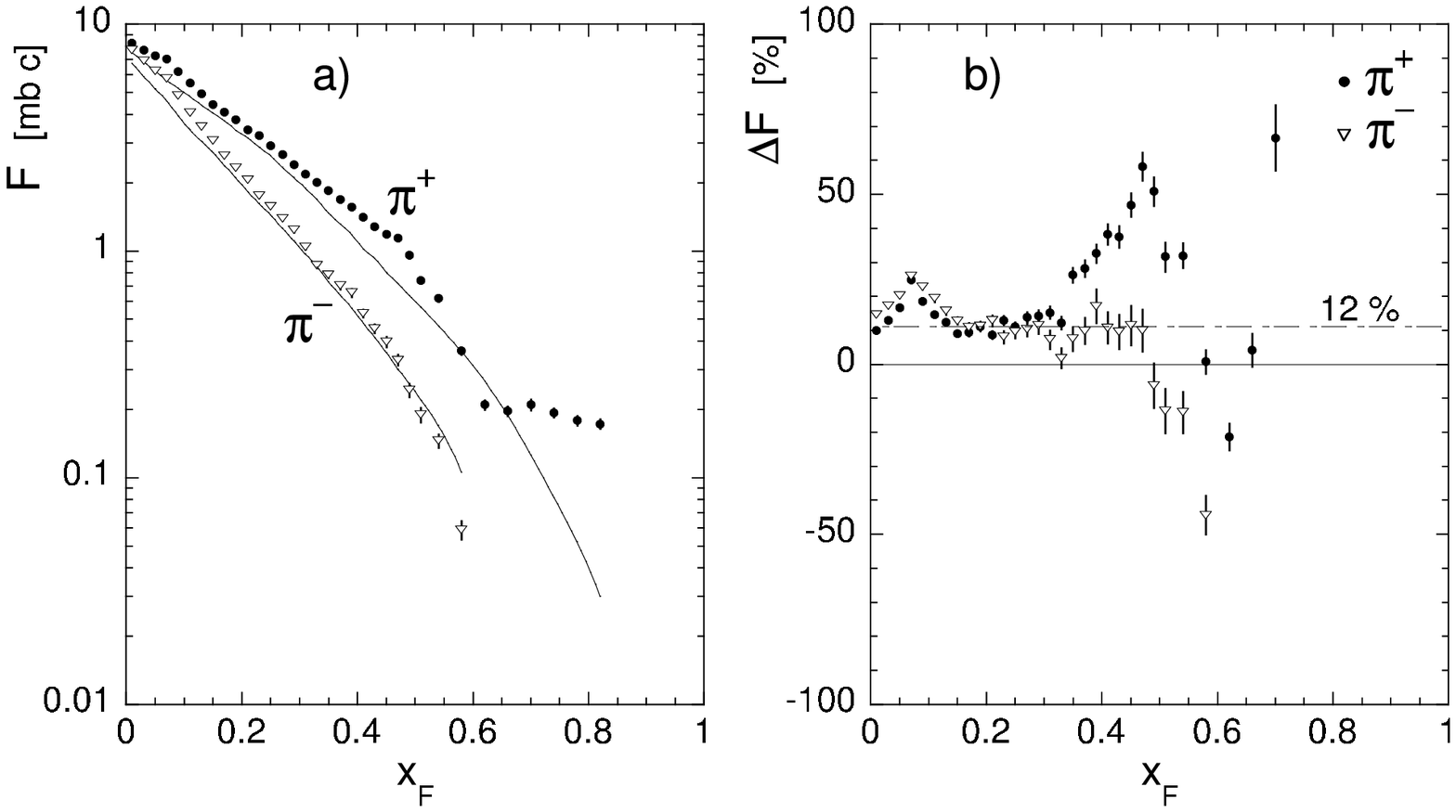,width=16cm}
\caption{  
a) Comparison of $p_T$ integrated invariant cross section $F$ as a function of 
$x_F$ for $\pi^+$ and $\pi^-$ measured by \cite{agu} to NA49 results (represented as lines); 
b) Deviation of the measurements of \cite{agu} from the NA49 results in percent.
}
\label{EHS-F-stat}
\end{figure}

The EHS experiment [10--12] which is directly comparable to NA49 in terms of phase 
space coverage and potential particle identification capability and which has accumulated 
a sizeable data sample of 470k~events at a beam momentum of 400~GeV/c has only
published $p_T$ integrated distributions of the invariant and non-invariant cross 
sections. The results are compared to the NA49 data in Fig.~\ref{EHS-F-stat}.
As shown by the percentage deviation of the invariant yields (Fig.~\ref{EHS-F-stat}b), 
a consistent 
upward shift of about 12\% is evident up to $x_F \simeq 0.35$. This shift is  
not compatible with the precise fulfillment of $x_F$ scaling exhibited by the 
ISR data for $x_F > 0.3$ in the same energy range as discussed in Section~8.2.
This indicates a normalization problem of the EHS data. Above $x_F = 0.35$, 
the EHS results show large and evidently unphysical systematic effects
especially for the $\pi^+$ yields. 

\begin{figure}
\centering
\epsfig{file=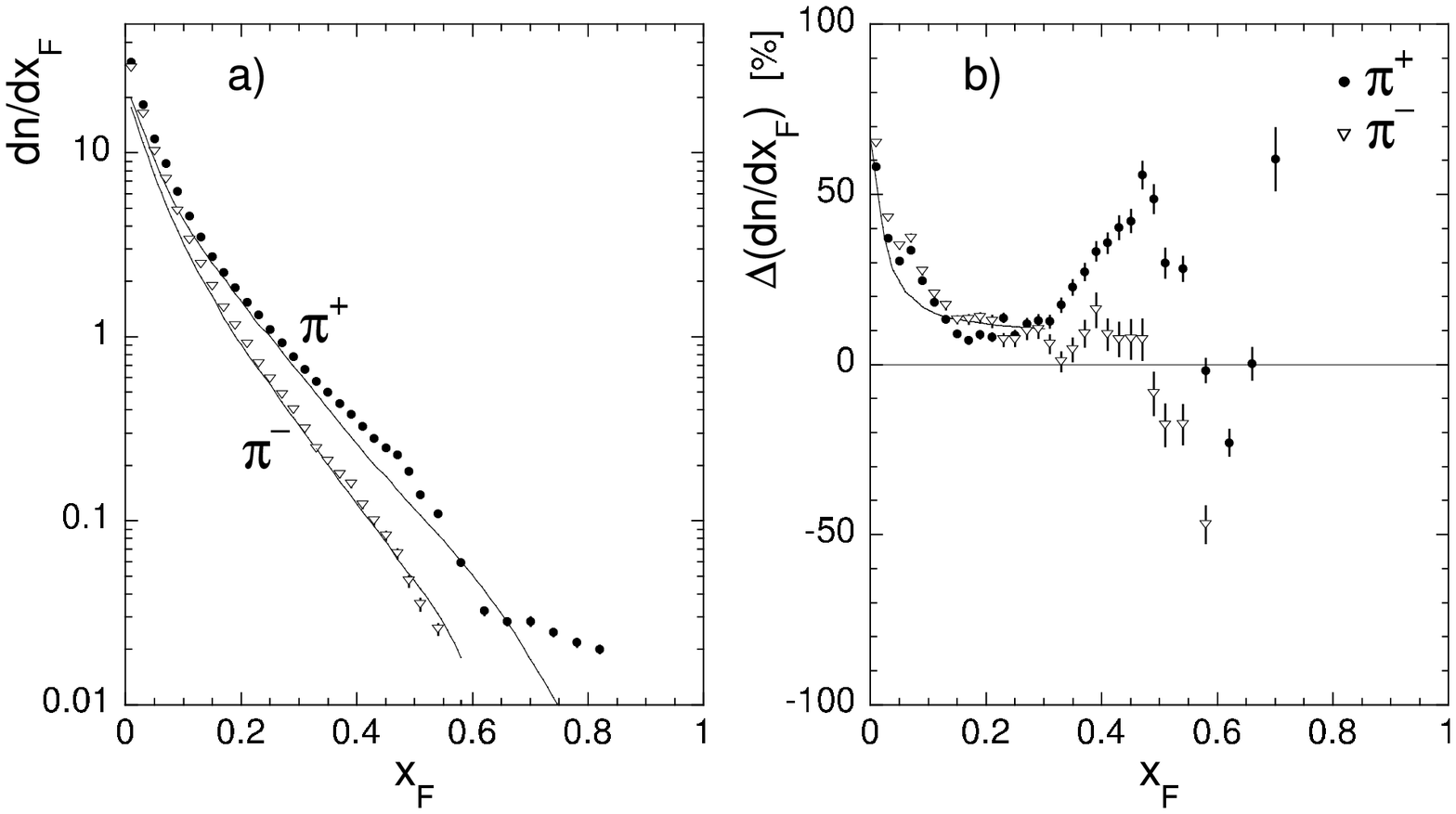,width=16cm}
\caption{  
a) Comparison of density distribution $dn/dx_F$ as a function of 
$x_F$ for $\pi^+$ and $\pi^-$ measured by \cite{agu} to NA49 results (represented as lines); 
b) Deviation of the measurements of \cite{agu} from the NA49 results in percent.
}
\label{EHS-dndxf-stat}
\end{figure}

It is interesting to also compare the non-invariant density distributions 
$dn/dx_F$, as presented in Fig.~\ref{EHS-dndxf-stat}. 
The increase of the percentage deviations below $x_F \simeq 0.15$ is entirely due to 
the energy dependence of the functional relation between non-invariant and invariant yields, see
equation (9), as shown by the superimposed line in Fig.~\ref{EHS-dndxf-stat}b. This relation 
predicts for $s$-independent invariant cross sections a linear increase
of particle density with $\sqrt{s}$ at $x_F=0$. This is 
evidently borne out by the data. A more detailed discussion of $s$-dependences 
will be carried out in a separate publication.

\begin{figure}[h]
\centering
\epsfig{file=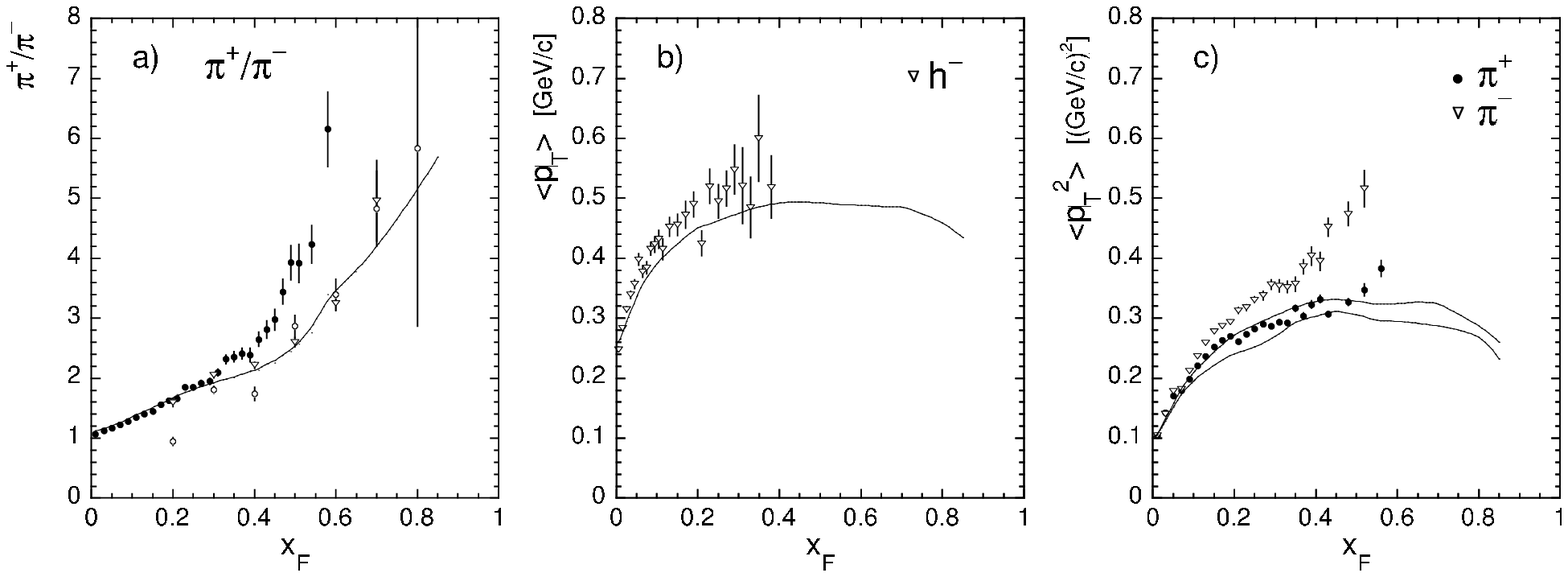,width=16cm}
\caption{  
Comparison as a function of $x_F$ of 
a) $\pi^+/\pi^-$ ratio  measured by \cite{agu} (full circles) and \cite{bre} (open symbols), 
b) mean $p_T$ of $h^-$ measured by \cite{bai}, and
c) mean $p_T^2$ for $\pi^+$ and $\pi^-$ measured by \cite{agu}
to NA49 results (represented as lines).
}
\label{EHS-pirat}
\end{figure}

A comparison of $p_T$ integrated $\pi^+/\pi^-$ ratios with NA49 is possible for the 
data of \cite{bre} and \cite{agu}, whereas mean $p_T$ and mean $p_T^2$ are only available from 
\cite{bai} and \cite{agu}. The summary of these comparisons is presented in 
Fig.~\ref{EHS-pirat}. 

The $\pi^+/\pi^-$ ratios of \cite{bre} are in fair agreement with the NA49 results whereas 
the data of \cite{agu} deviate above $x_F \simeq 0.3$. The first and second moments of the 
$p_T$ distributions show an upward trend also below this $x_F$ value which complies with the 
expected $s$-dependence of these quantities.


\subsection{Total Pion Multiplicity
}
\vspace{3mm}
The integration of the $dn/dx_F$ distribution presented in Table~\ref{ipi} yields 
the following total pion multiplicities:
\begin{center}
\begin{tabular}{ccr}
$\langle n_{\pi^+} \rangle$   &  =  &  3.018   \\
$\langle n_{\pi^-} \rangle$   &  =  &  2.360   \\
$\langle \pi^+/\pi^- \rangle$ &  =  &  1.279   \\
\end{tabular}
\end{center}
The statistical errors on these quantities are negligible compared to the overall 
normalization uncertainty of 1.5\% given in Table~\ref{tsyserr}.

\subsection{Availability of the Presented Data
}
\vspace{3mm}
The NA49 data are available in numerical form on the Web Site 
\cite{ows} as far as the tabulated values are concerned. In 
addition and in order to give access to the complete data interpolation 
developed in the context of this publication, two large sets of $\pi^+$ and 
$\pi^-$ momentum vectors ($5 \cdot 10^7$ each) can also be found on this site. 
These files are obtained by a Monte Carlo technique with importance sampling 
without weighting. By normalizing with the respective total pion 
multiplicities and the total inelastic cross section given above, 
distributions of cross sections in arbitrary coordinate representations 
and with arbitrary binning may readily be obtained. This might be 
found useful for comparison with production models using Monte 
Carlo methods with finite bin sizes especially in view of the 
importance of binning effects demonstrated in Section~5.7 above.

\section{Discussion
}
\vspace{3mm}
The purpose of this paper is the establishment of a complete 
and internally consistent set of inclusive pion cross sections at SPS energy.
As such, the data may serve as a basis of comparison with other hadronic
interactions, in particular with collisions involving nuclei in p+A
and A+A reactions. There is however, another important aspect of
this work which is more directly related to the status of the theoretical
understanding of soft hadronic collisions. This status might be described
as deeply unsatisfactory due to the lack of reliability and predictive
power of present attempts in the field of non-perturbative QCD. In the
presence of sufficiently consistent and precise experimental information,
both the multitude of ad-hoc microscopic multiparameter models and the
attempts at a statistical approach to hadronic interactions fail to
describe the detailed structure of the data even on the most primitive
level of single particle inclusive cross sections. It is outside the
scope of this publication to proceed with a detailed discussion of these
problems which has to be left to subsequent papers. Two aspects are
however worth mentioning here. The first concerns the important
local structure visible in the data, the second addresses the 
$s$-dependence.

\subsection{The Importance of Resonance Effects
}
\begin{figure}[h]
\centering
\epsfig{file=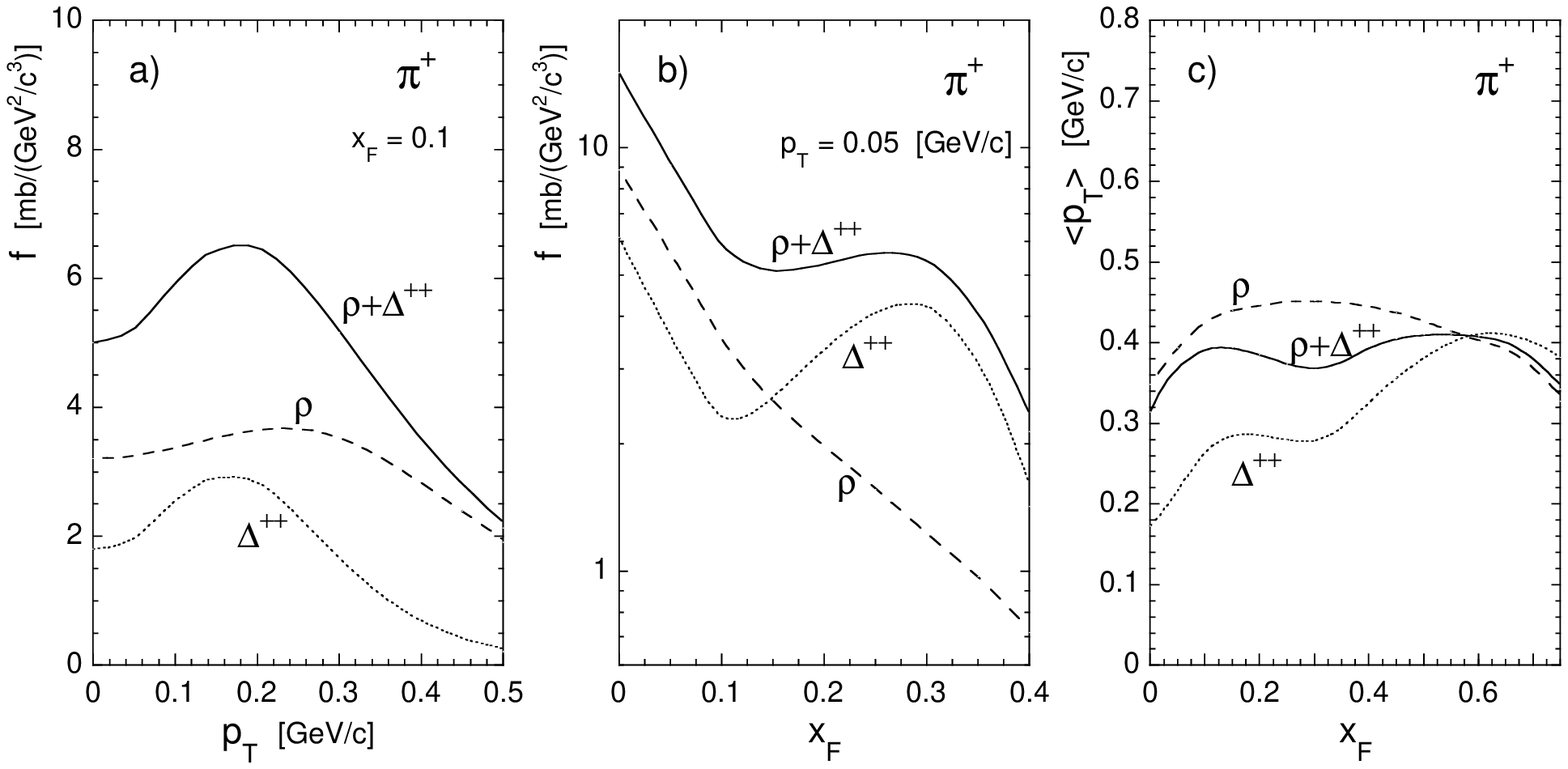,width=14cm}
\caption{Monte Carlo study of
a) $p_T$ distribution at fixed $x_F$,
b) $x_F$ distribution at fixed $p_T$,
c)~mean $p_T$ as a function of $x_F$ for $\pi^+$ 
resulting from $\rho$ and $\Delta^{++}$ decay.
}
\label{reson}
\end{figure}

\vspace{3mm}
Distinct local structures are visible in all inclusive distributions shown
in this paper, in the $x_F$ or $y$ as well as in the $p_T$ dependences,
in the $\pi^+/\pi^-$ ratios as well as in the moments of the $p_T$ distributions. 
These structures are significantly different for $\pi^+$ and $\pi^-$ production.
Similar structures have been first observed in a more restricted experiment 
at the CERN PS in 1974 \cite{ama} and discussed in relation to resonance decay. 
In fact, by using the measured inclusive cross sections for the low-mass
mesonic and baryonic resonances $\rho$ and $\Delta^{++}$ such structures are
readily predicted if the mass distribution of these resonances is properly
taken into account. This is illustrated in Fig.~\ref{reson}, where the 
$p_T$ dependence at fixed $x_F$, the $x_F$ dependence at fixed $p_T$, 
and the mean $p_T$ as a function of $x_F$ for $\pi^+$ originating from 
$\rho$ and $\Delta^{++}$ decays are shown. Clearly, the salient features of the 
data are predicted already from this very restricted set of resonances.
In addition, the known charge dependence of mesonic and hadronic 
resonance production in p+p collisions also implies a distinct difference
between $\pi^+$ and $\pi^-$ as observed in the data.

Evidently a number of questions are raised by this argumentation. It is
of course known since a long time \cite{gra} that the vast majority of all hadrons 
come from resonance decay. There has however been a surprising lack of 
attempts to assess the real consequences of this fact for all aspects 
of the inclusive distributions. In addition to the local structures 
discussed above also the full $x_F$ and $p_T$ dependences proper can be 
reproduced from resonance decay if an appropriate yield of higher mass 
resonances is allowed for.
The past estimates \cite{gra} have only taken account of a very limited 
number of mostly mesonic resonances in the low mass region.
There is however good reason to conclude that a much larger fraction of the almost 200
listed baryonic and mesonic resonances is produced in p+p interactions and 
contributes via cascading decay.
A measurement of inclusive resonance cross 
sections well beyond the lowest-lying states therefore becomes mandatory 
for any progress in understanding of both the inclusive sector and  
particle correlations.

\subsection{Energy Dependence
}
\vspace{3mm}
The problem of the $s$-dependence of the invariant cross sections is
touched upon in this paper only in the very limited forward region 
of $x_F > 0.3$ and in the range of $17 < \sqrt{s} < 53$~GeV. This extension       
in $\sqrt{s}$ is necessary due to the lack of comparison data in the more
immediate neighbourhood of interaction energy. It demonstrates
the $s$-independence of the inclusive cross section, to a few percent level,
in this area. This finding corresponds to the well-known hypotheses
of limiting fragmentation \cite{ben} and Feynman scaling \cite{fey}. From the physics
point of view these hypotheses have the dual aspect of the manifestation
of fragmenting excited objects or of a direct trace of the partonic
structure of the colliding hadrons. As such, a more detailed scrutiny 
also at lower and higher energies seems indicated. This study meets,
in the lower $\sqrt{s}$ region, with the usual problem of the scarcity of 
consistent data sets. At $\sqrt{s}$ values above the ISR energy range there
is in addition, the basic experimental problem of access to the forward
kinematic region. Already at RHIC energies, $\sqrt{s}=200$~GeV, the present 
experimental equipment is limited to $x_F < 0.15$ at the mean transverse momentum
of the produced particles. This limitation becomes more
and more restrictive with increasing energy at $\overline{\mbox{p}}$p colliders and 
at the LHC where at best some studies around $x_F=0$ and in the area of 
diffraction are feasible.
  
The zone at $x_F=0$ has, on the other hand, its proper interest due to the fact
that here the bulk of pionic multiplicity is produced and that the
invariant cross section shows a steady increase with $\sqrt{s}$, the
"rising rapidity plateau". These effects are not touched upon in
this paper due to the principal problem of feed-down corrections as
described in Section~5.6. As the ISR data for example are uncorrected
for strangeness feed-down, this correction has to be performed before
any reliable conclusions can be drawn. In a more general sense, the
subsequent crossing of flavour thresholds and the saturation of corresponding
particle yields with increasing $\sqrt{s}$, from SU(2) at the PS/AGS to SU(6)
at the LHC, poses considerable problems both on the experimental and
on the interpretation level. 

\section{Conclusion
}
\vspace{3mm}
A new and extensive set of inclusive cross sections for pion production
in p+p collisions at the CERN SPS is presented and compared in
detail to existing data in the corresponding energy range. The new data
cover the available phase space with a consistent set of 
double-differential cross sections and with systematic and
statistical uncertainties well below the 5\% level. This precision allows 
for the first time in the SPS energy range, the observation of a rich 
substructure in the data. This structure pervades all kinds of inclusive 
distributions and precludes any attempt at simple analytic parametrizations.
It is a direct manifestation of resonance decay. The importance
of this building-up of the inclusive particle distributions, over the
full phase space, from the decay of resonances is again stressed,
as it has profound consequences for the understanding of particle 
production in the non-perturbative sector of QCD. The new data may
also serve as a basis for comparison with the more complex hadronic
collisions involving nuclei in p+A and A+A interactions. Data in these
areas will be provided, with similar precision, in upcoming publications.

\section*{Acknowledgements
}
\vspace{3mm}
We wish to thank the crews of the CERN accelerator complex and the H2 beam line for the delivery 
of the high quality proton beam.
We are grateful to all groups at CERN and at the collaborating institutions involved in the design,
construction, installation, operation, and maintenance of the NA49 detector.
Special thanks goes to M.~Flammier and the groups of M.~Bosteels and C.~Ferrari for the continuous
support and very fruitful cooperation.
We furthermore acknowledge the effective operation of the Vertex Magnets by the CERN cryogenics and 
power supply groups.

This work was supported by 
the US Department of Energy Grant DE-FG03-97ER41020/A000,
the Bundesministerium f\"{u}r Bildung und Forschung, Germany, 
the Polish State Committee for Scientific Research (2 P03B 130 23, SPB/CERN/P-03/Dz 446/2002-2004, 
2 P03B 04123), 
the Hungarian Scientific Research Foundation (T032648, T032293, T043514),
the Hungarian National Science Foundation, OTKA, (F034707),
the Polish-German Foundation, 
the Korea Research Foundation Grant (KRF-2003-070-C00015),
and the Bulgarian National Science Fund (Ph-09/05).



\end{document}